\documentclass[a4paper,11pt]{article}

\pdfoutput=1 

\usepackage{jheppub} 
\newcommand{\la}{\langle}
\newcommand{\ra}{\rangle}
\usepackage[T1]{fontenc} 
\usepackage{tikz}
\usepackage[export]{adjustbox}
\allowdisplaybreaks

\usepackage{soul}

\newcommand \vev [1] {\langle{#1}\rangle}

\title{Nonperturbative Negative Geometries: Amplitudes at Strong Coupling and the Amplituhedron}

\author[a]{Nima Arkani-Hamed,}\emailAdd{arkani@ias.edu}
\author[b]{Johannes Henn,}\emailAdd{henn@mpp.mpg.de}
\author[c]{Jaroslav~Trnka}\emailAdd{trnka@ucdavis.edu}

\affiliation[a]{School of Natural Sciences, Institute for Advanced Study, Princeton, NJ 08540, USA}
\affiliation[b]{Max-Planck-Institut für Physik, Werner-Heisenberg-Institut, D-80805 München, Germany}
\affiliation[c]{Center for Quantum Mathematics and Physics (QMAP),\\Department of Physics, University of California, Davis, CA 95616, USA}

\abstract{The amplituhedron determines scattering amplitudes in planar ${\cal N}=4$ super Yang-Mills by  a single ``positive geometry'' in the space of kinematic and loop variables.  We study a closely related  definition of the amplituhedron for the simplest case of four-particle scattering,
given as a sum over complementary ``negative geometries'', which provides a natural geometric understanding of the exponentiation of infrared (IR) divergences, as well as a new geometric definition of an IR finite observable ${\cal F}(g,z)$ -- dually interpreted as the expectation value of the null polygonal Wilson loop with a single Lagrangian insertion -- which is directly determined by these negative geometries. This provides a long-sought direct link between canonical forms for positive (negative) geometries, and a  completely IR finite post-loop-integration observable depending on a single kinematical variable $z$, from which the cusp anomalous dimension $\Gamma_{\rm cusp}(g)$ can also be straightforwardly obtained. We study an especially simple class of negative geometries at all loop orders, associated with a ``tree'' structure in the negativity conditions, for which the contributions to ${\cal F}(g,z)$ and $\Gamma_{\rm cusp}$ can easily be determined by an interesting non-linear differential equation immediately following from the combinatorics of negative geometries. This lets us compute these ``tree'' contributions to ${\cal F}(g,z)$ and $\Gamma_{\rm cusp}$ for all values of the `t Hooft coupling. The result for $\Gamma_{\rm cusp}$ remarkably shares all main qualitative characteristics of the known exact results obtained using integrability.} 

\begin{document} 

\preprint{MPP-2021-199}

\maketitle

\section{Introduction and summary of results}

How does the miracle of AdS/CFT duality \cite{Maldacena:1997re} come about? 
Working at large $N$ in ${\cal N} = 4$ super Yang-Mills (sYM) with $SU(N)$ gauge group, and in the limit where the 't Hooft coupling $g^2 \equiv {g^2_{\rm YM} N}/{(16 \pi^2)}$ is small, we have a weakly coupled description in term of a theory of gluons, associated with a perturbative expansion in powers of $g^2$ with a finite radius of convergence, while when $g^2$ is large, we have a different weakly coupled description in terms of strings in AdS, giving an asymptotic expansion in powers of $1/g$. At the most basic level, one aspect of the miracle is simply understanding why physical observables have this qualitative property, of having a finite-radius-of-convergence expansion in powers of $g^2$ for $g \ll 1$ and an asymptotic expansion in $1/g$ for $g \gg 1$. 

The most famous observable where the transition from weak to strong coupling has been understood exactly is the cusp anomalous dimension $\Gamma_{\rm cusp}(g)$, which (amongst other things) determines the leading behavior of four-gluon scattering in the theory, 
\begin{align}\label{eqMdivintro}
{ A} \sim {A}_{\rm tree} \times  \exp \left[-  \Gamma_{\rm cusp}(g)  \log(\mu_{{\rm soft}})  \log (\mu_{{\rm collinear}})\right]\,.
\end{align}
Here and in the remainder of this paper we tacitly assume the `t Hooft limit.
In eq. (\ref{eqMdivintro}) we see the familiar fact that the amplitudes has soft-collinear divergences, so that the exclusive four-particle amplitude is exponentially suppressed.
 At infinite $N$, $\Gamma_{\rm cusp}$ has been computed exactly using the machinery of integrability for ${\cal N}=4$ sYM \cite{Beisert:2006ez}. 
 The exact answer takes the form of an integral representation, from which we can learn that it has a finite radius of convergence, $|g_*| = \frac{1}{4}$.
 The asymptotics of $\Gamma_{\rm cusp}(g)$ for small and large $g$ are given by \cite{Bern:2006ew,Benna:2006nd,Basso:2007wd}
\begin{equation}\label{formula_cusp_intro}
\Gamma_{\rm cusp}(g) \to \left\{  \begin{array}{cc} 4 g^2 - 8 \zeta_2 g^4 + \cdots & g \ll 1 \\  2 g - \frac{3 \log 2}{2 \pi}  + \cdots & g \gg 1 \end{array}  \right.
\end{equation}
The leading terms at weak coupling reflect the loop expansion for gluons, while the leading term of order $g$ at strong coupling is the area of the ``soap bubble in AdS'' anchored to the null-square Wilson loop on the boundary \cite{Kruczenski:2002fb}, with the $1/g$ expansion reflecting string loop corrections in AdS \cite{Roiban:2007dq}. 
Note the familiar characteristic feature that the expansion at small $g$ is in powers of $g^2$, but at large coupling, in powers of $1/g$.

This is a spectacular result, but it is natural to ask whether it is  possible to get non-perturbative results without using the heavy machinery of integrability. After all, the weak to strong coupling transition from gauge theories to gravity should be a general, robust phenomenon. 
There has long been an obvious fantasy for understanding this more directly, certainly in the planar limit of gauge theories: can we simply sum over all the diagrams? A nice toy example for what this might look like is seen in  the context of resumming certain infinite classes of loop diagrams, the off-shell ladder diagrams for four-point scattering in $\phi^3$ theory (further interesting infinite classes of diagrams are considered in \cite{Basso:2021omx}),
\begin{equation}
\centering{
\includegraphics[width=0.65\columnwidth]{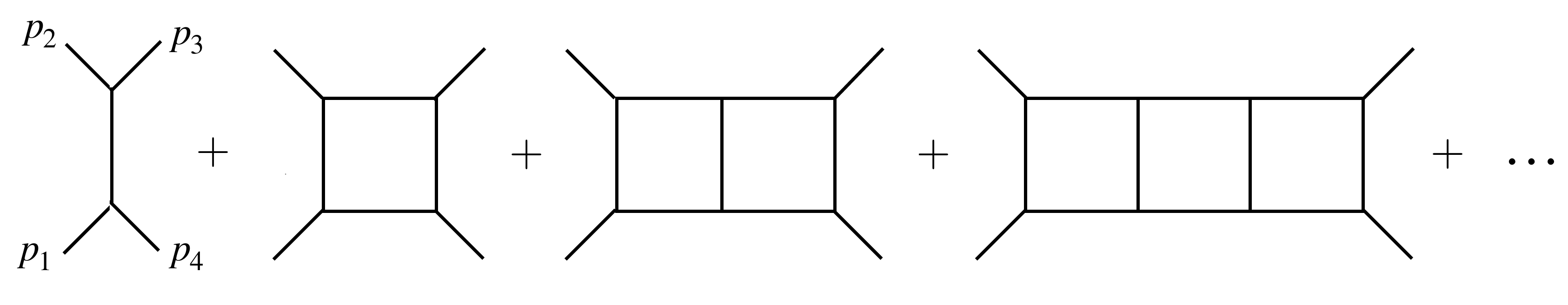} 
}
\end{equation}
The behaviour of this sum at strong coupling was analyzed by \cite{Broadhurst:2010ds} using a judicious integral representation.
Here we wish to mention an alternative approach that will generalize easily to the cases we discuss in this paper.
It is based on the fact that that the Laplace equation relates $L$- and $(L-1)$-loop ladder integrals  \cite{Drummond:2006rz,Drummond:2010cz}.
Denoting the kinematic variables by 
\begin{align}
\frac{p_1^2 p_3^2}{s t} = \frac{z \bar{z}}{(1-z)(1-\bar{z})} \,,\quad\quad
\frac{p_2^2 p_4^2}{s t} = \frac{1}{(1-z)(1-\bar{z})} \,,
\end{align}
and the normalized coupling by $\kappa^2 = -g_{\phi}^2/(16 \pi^2 s)$, where
$g_\phi$ is the $\phi^3$ coupling, it is easy to see that the appropriately normalized infinite sum satisfies the equation
\begin{equation}\label{DEladdersphi3}
z \partial_z  \bar{z} \partial_{\bar{z}}F(\kappa; z,\bar{z}) - \kappa^2 F(\kappa; z, \bar{z}) =0\,.
\end{equation}
It is interesting that while the structure of this equation is determined by inspection of the perturbative expansion of $F(\kappa)$, with the equation in hand, we can either solve this simple differential equation exactly and extrapolate to strong coupling, or failing that, set up a different perturbation theory around large $\kappa$.

Obviously the ladder diagrams do not remotely include all the interesting structure of planar theories, missing multi-particle cuts that would be present in generic planar diagrams, such as
\begin{equation}
\includegraphics[width=0.25\columnwidth]{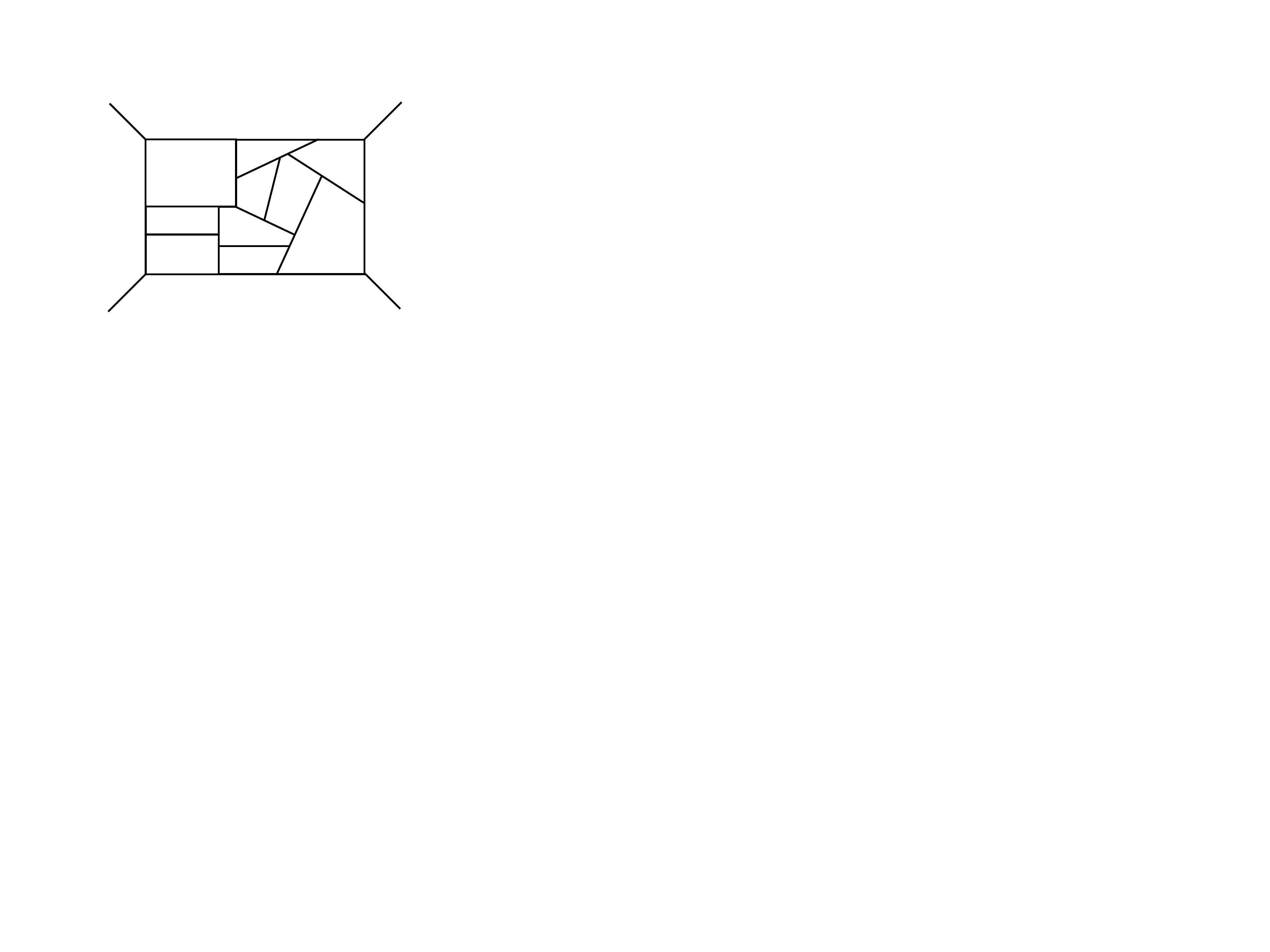} 
\end{equation}
Thus, if we wish to extend the direct resummation strategy to the full theory, there is a qualitative challenge to overcome: we need to find ``another direction'' to take into account the complexity of general planar diagrams; moreover, in gauge theories, there is the challenge of identifying meaningful classes of gauge-invariant diagrams to resum.

In this paper, we will introduce such a new expansion direction, based on a novel geometric representation of amplitudes as a sum over a collection of ``negative geometries'' which are suggested by the single ``positive geometry'' of the amplituhedron. 
But before introducing these ideas, let us first discuss the simple, completely finite observable that will be our main object of study in this paper. 

The on-shell four-particle amplitude in $\mathcal{N}=4$ sYM is infrared divergent; it is useful to instead deal with a closely related simple observable that is completely finite, from which $\Gamma_{\rm cusp}$ can be easily extracted. This observable is the expectation value of the four-point Wilson loop defined on a null polygon of points $x_1,x_2,x_3,x_4$, with a Lagrangian insertion at a point $x_0$ (and normalized by the Wilson loop itself) \cite{Alday:2011ga,Engelund:2011fg,Engelund:2012re},
\begin{equation}
\includegraphics[width=0.5\columnwidth]{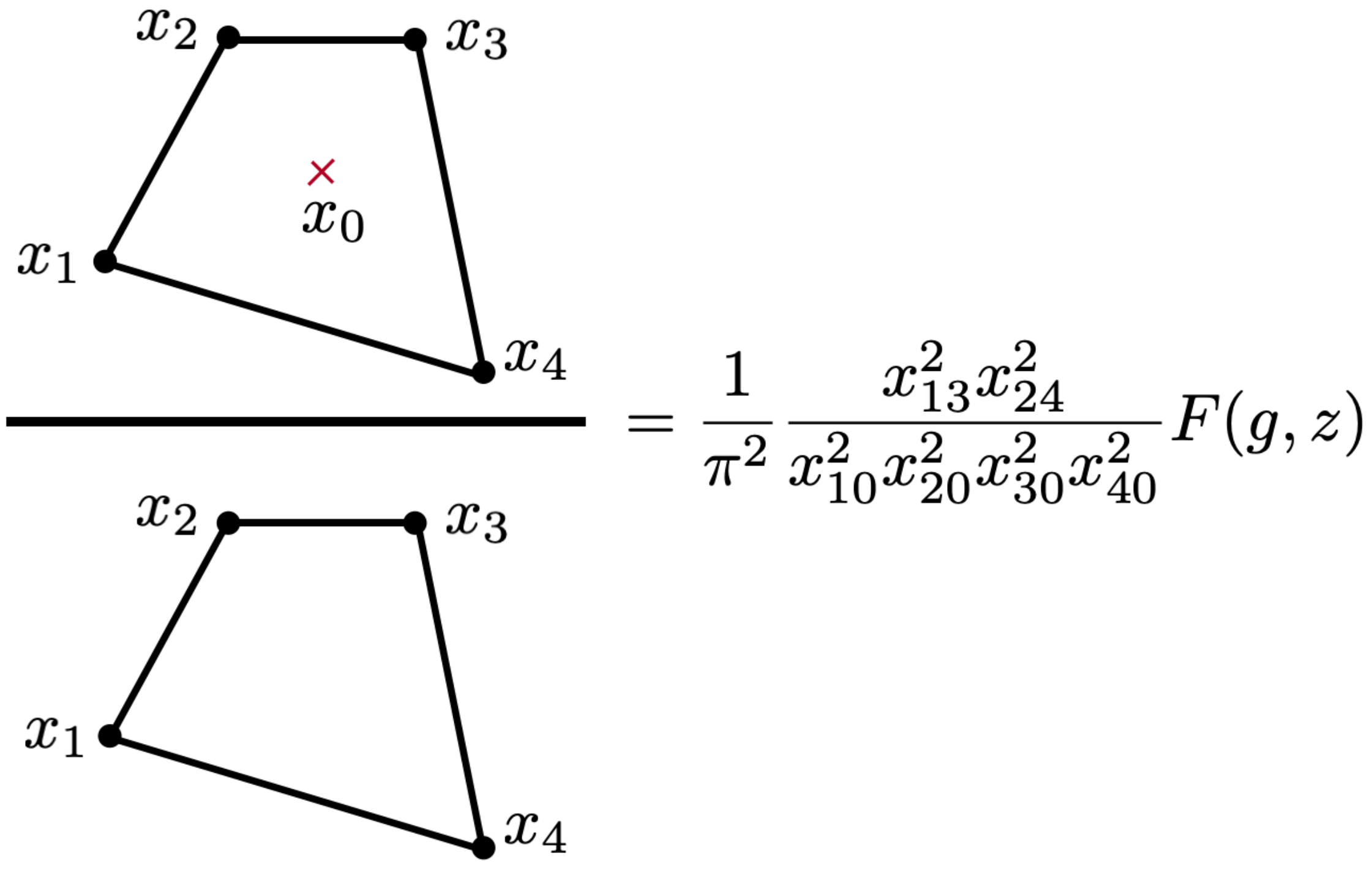}
\end{equation}
Thanks to the conformal symmetry of the Wilson loop, $F$ is a function of a {\it single} kinematical variable, built from the cusp points of the Wilson loop, and the Lagrangian insertion point, namely
\begin{align}
z = \frac{x_{20}^2 x_{40}^2 x_{13}^2}{x_{10}^2 x_{30}^2 x_{24}^2} \,,
\end{align}
where $x_{ij}^2 = (x_i-x_j)^2$.

This object is closely related to the logarithm of the four-point Wilson loop. 
Indeed, taking a derivative in the coupling $g \partial_{g}$ of the logarithm of the Wilson loop leads to a Wilson loop with a Lagrangian insertion.
More precisely, one obtains the ratio $F(z)$, but integrated over the insertion point $x_{0}$. In this sense $F(z)$ can be considered as an especially simple {\it integrand} of the logarithm of the Wilson loop, depending on a single variable. 

This connection helps to understand why $F(z)$ is finite in four dimensions. The Wilson loop at $L$ loops has $1/\epsilon^{2 L}$ (ultraviolet) divergences, due to cusps formed by adjacent light-like segments. It is well-known that the logarithm of the Wilson loop can be expressed in terms of certain ``maximally non-Abelian'' diagrams, which only have $1/\epsilon^2$ divergences \cite{Gatheral:1983cz,Frenkel:1984pz,Korchemskaya:1992je}. This is due to the fact that divergences only arise when probing integration regions where entire ``web'' diagrams come close to the cusps. In other words, there are no lower loop subdivergences, but all $L$ integrations are required to produce the divergence. This last observation makes it clear why $F(z)$ is finite: it is the last integration, over the Lagrangian insertion point $x_0$, that would produce the $1/\epsilon^2$ divergence.
Moreover, given the function $F$, we can extract $\Gamma_{\rm cusp}$ by doing the final loop integration. Since only the leading divergent term is needed, the latter integration simplifies and turns into a certain functional operation on $F$ \cite{Alday:2013ip}.

Thanks to the duality between scattering amplitudes and Wilson loops \cite{Alday:2007hr,Drummond:2007aua,Brandhuber:2007yx,Drummond:2007cf}, this simple interpretation can be equally formulated in terms of (integrands of) scattering amplitudes.
Consider the loop integrand for the {\it logarithm} of the amplitude, $\log { M} $, thought of at $L$ loops as a function of $L$ 
dual integration variables.
Physically we know that while the amplitude itself has ${1}/{\epsilon^{2L}}$ infrared divergences in dimensional regularization, the logarithm of the amplitude is only ${1}/{\epsilon^2}$ divergent. In practice, this means that if we leave one of the loop variables un-integrated, the integral over the remaining loop variables is completely finite; it is only the integral over the final variable, say $x_0$, that produces the divergence. 

In this paper we will formulate the problem of solving for the function $F(g^2;z)$ by examining the positive geometry of the amplituhedron \cite{Arkani-Hamed:2013jha,Arkani-Hamed:2013kca,Franco:2014csa,Arkani-Hamed:2017vfh,Arkani-Hamed:2018rsk,Damgaard:2019ztj}. Let us begin by quickly recollecting the definition of the four-point amplitude given by the amplituhedron. 
In a nutshell, the amplituhedron defines a geometric region in the kinematics of the scattering process. This includes the external kinematics, as well as conditions on certain scalar products involving loop integration variables and external kinematics.
The most novel aspect of the positive geometry, however, is in the interaction between the loops -- given two loop integration variables, we have to have ``mutual positivity'' conditions. 
The integrand for the $L$-loop amplitude is the ``canonical form'' of this positive geometry -- i.e. the unique $4L$-form with logarithmic singularities on, and only on, all the boundaries  of the positive geometry. 

We can represent {\it any} positive geometry of mutual positivities between integration variables $i$ and $j$ graphically, by drawing a graph with $L$ vertices--one for each loop variables $i$--and connecting any pair of vertices with a wiggly line, denoting the positivity condition. 
Thus with this notation, the positive geometry of the amplituhedron just corresponds to the complete graph on $L$ vertices, where all vertices are connected to each other via wiggly lines. Having done this, we can also define any ``negative'' geometries, by again drawing a graph, but now connecting two nodes $i,j$ with a solid line, denoting the mutual {\it negativity} condition:
\begin{equation}
\includegraphics[width=0.37\columnwidth]{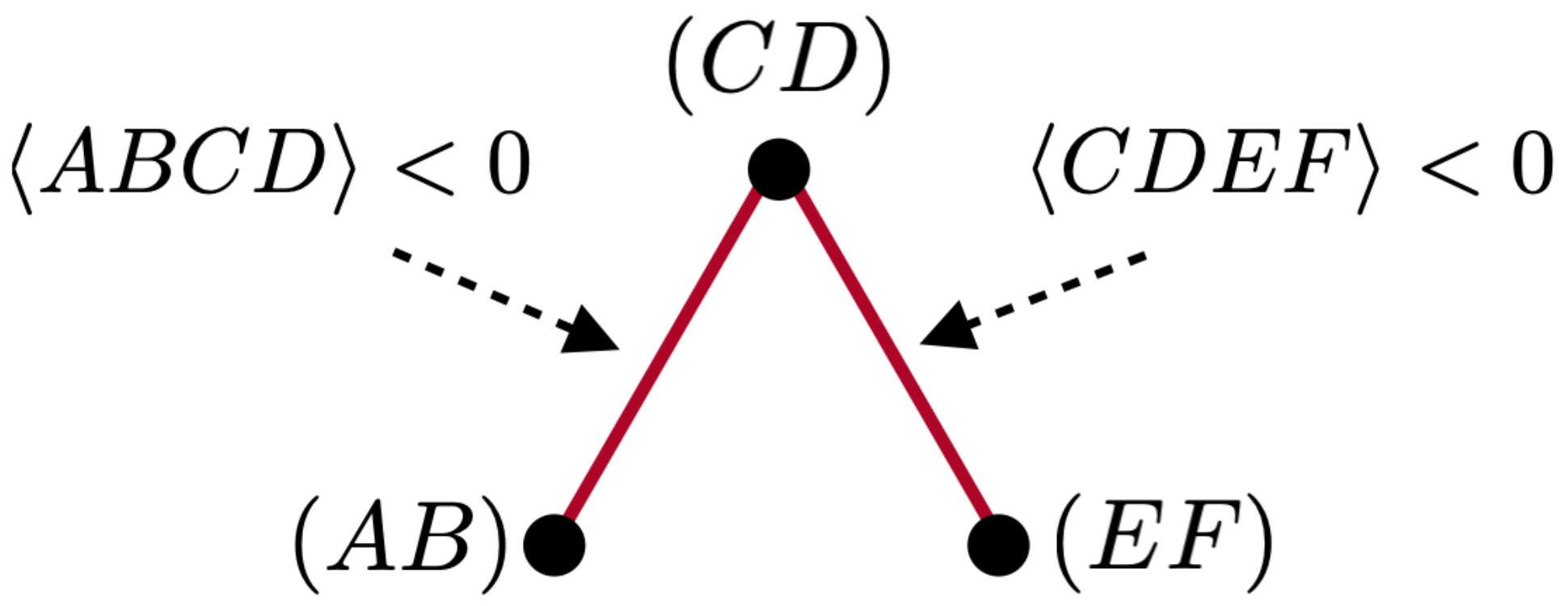} 
\end{equation}
Starting from this basic definition of the amplituhedron for four-point scattering, in this note we will see two things:

(1) We will discover that, quite apart from any physical motivation about the exponentiation of infrared divergence, the positive geometry of the loop amplituhedron itself tells us that the logarithm of the amplitude is a natural object to define. Indeed, the integrand of the logarithm of the amplitude will be given as 
\begin{equation}
{\rm log} M = \sum_{{\rm {\bf connected}} \,  {\rm graphs} \, G} (-1)^{E(G)} (g^2)^L F_{G}
\end{equation}
where $E(G)$ is the number of edges of the graph $G$. 
Quite beautifully, integrating the canonical form for {\it any one} of the ``negative geometries'' $G$ gives us something that is only $1/\epsilon^2$ divergent; and leaving a single loop $x_0$ un-integrated, associates a {\it completely finite} function $F_G(g^2;z)$.
This gives us, for the first time, a direct link between integrating the canonical form of positive geometries, and completely finite polylogarithmic functions, which are as simple as possible, depending only on a single variable $z$. We will give the explicit expressions for these polylogarithmic through to $(L-1) =2$ integrated loops. We stress that while this expansion of log $M$ in terms of ``negative geometries'' is naturally handed to us by the amplituhedron, it is an entirely different sort of expansion than then Feynman diagram expansion. While the integrand for each graph $G$ has only local poles, it is in general not related to sums/products of planar Feynman diagrams.

(2) Purely from the perspective of the geometry, there is a clear separation in the complexity involved with understanding these ``negative'' geometries. Graphs that are ``trees'' are especially simple, since the mutual negativity can be imposed recursively one loop at a time; and indeed we will find it easy to determine the integrand here explicitly (even though we can get by without seeing it explicitly). But ``closed loops'' are clearly more complicated, since they impose ``global'' negativity constraints that cannot be solved one loop at a time.

This motivates us to try and sum the results for all the negative geometries associated with the ``tree'' graphs, for which the geometry is most easily understood. We will find that this can easily be done: the canonical forms for all the ``tree'' negative geometries can be computed in a simple recursive way, and precisely the same ``Laplace equation'' structure mentioned above in the context of the planar $\phi^3$ ladders allows the loop integrations to be easily performed, leaving us with a simple non-linear differential equation that can straightforwardly be solved non-perturbatively for all values of the 't Hooft coupling. 

We denote this object by $F_{\rm tree}(g,z)$, and find that it is given as  
\begin{equation}
 F_{\rm tree}(g,z) = -{A^2} \frac{z^A}{(z^A + 1)^2}\,,\quad {\rm where} \quad \frac{A}{2 g {\rm cos} \frac{\pi A}{2}} = 1\,.
\end{equation}
From here, by performing the final loop integration and extracting the coefficient of the $1/\epsilon^2$ divergence, we can easily extract the contribution to $\Gamma_{\rm cusp}$ at ``tree-level'', finding
\begin{equation}
 \Gamma_{\rm tree}(g) = A \left(\frac{4}{\pi}  {\rm tan} \frac{\pi A}{2} - A\right) \,.
\end{equation}
Remarkably, $\Gamma_{\rm tree}(g)$ has all the same qualitative properties as the known exact expression for $\Gamma_{\rm cusp}(g)$. 
The asymptotics have exactly the same form, with an expansion in powers of $g^2$ for small $g$, while asymptoting to be linear in $g$ at large $g$, with an asymptotic expansion in powers of $1/g$,
The analytic structure in the $g$ plane is also the same, with a branch cut on the imaginary axis, starting at $|g_{*,{\rm tree}}| = 0.21\cdots$. Amusingly, this shows that the radius of convergence of 
$\Gamma_{\rm tree}$ is even {\it smaller} than for the exact result which is at $g_*  = 0.25$; in this sense $\Gamma_{\rm tree}$ incorporates even {\it larger} non-perturbative effects than the the full result from $\Gamma_{\rm cusp}$.

We can also sum a subset of the tree-like negative geometries, corresponding to a single chain; since these are roughly analogous to the ``ladder'' diagrams we call the corresponding contributions $F_{{\rm ladder}}$ and the associated contribution to cusp anomalous dimension $\Gamma_{{\rm ladder}}$, for which there is a closed form analytic expression:
\begin{equation}
   F_{{\rm ladder}}(g,z)=  \frac{z^{i \sqrt{2} g} + z^{-i \sqrt{2} g}}{2 {\rm cosh}(\sqrt{2} g \pi)}\,,
 \end{equation}
 \begin{equation}
\Gamma_{{\rm ladder}}(g)= \frac{4}{\pi} {\rm log \, cosh} (\sqrt{2} \pi g)\,.
\end{equation}
$\Gamma_{\rm ladder}$ also has an expansion in powers of $g^2$ at small coupling, and asymptotes to being linear in $g$ for large coupling, but has corrections that are powers of of $\exp(-g)$, and thus is missing ``string loop corrections'' that are a series in $1/g$. The analytic structure is also different than the exact result, with an infinite series of poles on the imaginary $g$ axis rather than a branch cut.
This $\Gamma_{{\rm ladder}}$ is also amusingly closely related to another sort of anomalous dimension, ``$\Gamma_{{\rm octagon}}$'' \cite{Coronado:2018cxj,Belitsky:2020qrm}, which controls the remainder function for six-particle scattering \cite{Basso:2020xts}, and also appears in scattering amplitudes on the Coulomb branch, in particular limits \cite{Caron-Huot:2021usw}: 
\begin{equation}
    \Gamma_{{\rm ladder}}(g) = 2 \Gamma_{{\rm octagon}}(\frac{g}{\sqrt{2}})\,.
\end{equation}
It is fascinating that the truncation of the full negative geometry problem to trees and even further to ladders, produces expressions for $\Gamma$ that are so close to the full exact result, so much so that one might wonder, what is the ``point'' of the rest of the negative regions with ``loops'', necessary for the full result? The answer can readily be seen by looking at the functions $F(g,z)$, which {\it do} have qualitative differences from the known exact result. The latter is asymptotic to being linear in $g$ at strong coupling, with as usual string loop corrections in powers of $1/g$. Instead $F_{{\rm tree}}$ asymptotes to $g^0$ at large $g$, with corrections that are a series in $1/g$, while $F_{{\rm ladder}}$ has a more unusual behavior, oscillating wildly with exponentially falling envelope at large coupling. These differences are washed away in performing the loop integration over $z$ to extract $\Gamma$, but clearly the full negative geometries are needed to correctly capture all aspects of the physics at strong coupling. 

The outline of this paper is as follows. In section \ref{sec:Wilsonloop}, we briefly review the definition of the Wilson loop with a Lagrangian insertion, and its relationship to scattering amplitudes and the cusp anomalous dimension. In section \ref{sec:geometry}, we introduce the idea of negative geometries, and explain how they naturally lead us to the logarithm of the amplitude, and to the infrared finite quantity $F$. We then show, in section \ref{section-examples-forms-and-functions}, explicit examples of infrared finite functions associated to negative geometries. In section \ref{sec:resummation}, we use differential equations in order to sum, to all orders in the coupling, contributions from ``tree'' and ``ladder'' geometries and discuss their contribution to the cusp anomalous dimension. We close by giving an outlook in section \ref{sec:outlook}.

\section{Wilson loop with Lagrangian insertion and cusp anomalous dimension}
\label{sec:Wilsonloop}

\subsection{Definition of the Wilson loop observable $F$ and review of known results}

We follow the conventions of \cite{Henn:2019swt}, except that we write perturbative expansions in terms of the expansion parameter $g^2$, while in that reference $g^2/4$ was used, and that we define Wilson loops in the fundamental representation.
Let us quickly review the relevant equations.
We define Wilson loops in the fundamental representation as follows,
\begin{align}\label{defcuspedWL}
\vev{W_F(x_1,x_2,x_3,x_4)}
 = \frac{1}{N_F} 
 \vev{{\rm tr}_F P \exp 
 \left( {i \oint_C dx\cdot A(x)} \right)
 }
 \,,
\end{align}
where the contour $C$ is a rectangle with vertices located at four points $x_i$ that are light-like separated, $x_{i,i+1}^2 = 0$ (with $i=1,\dots,4$ and $i+4\equiv i$),
and where we normalized the Wilson loop by 
$N_{F} = N$,
so that its perturbative expansion starts with $1$.

$W_F$ has ultraviolet divergences, which in dimensional regularization, with $D=4-2\epsilon$, take the form 
\begin{align}\label{W-div}
\log \vev{W_F(x_1,x_2,x_3,x_4)} = - \sum_{L\ge 1}  \frac{1}{( L \epsilon)^2}  g^{2 L} \Gamma_{{\rm cusp}}^{(L)}   + {\cal O}(1/\epsilon)\,,
\end{align}
where $\Gamma_{{\rm cusp}}^{(L)}$ are expansion coefficients of the  cusp anomalous dimension
\begin{align}
 \Gamma_{{\rm cusp}} =\sum_{L\ge 1}  g^{2 L} \Gamma_{{\rm cusp}}^{(L)}\,,
\end{align}
that was already given in eq. (\ref{formula_cusp_intro}).

Let us now consider the ratio mentioned in the introduction \cite{Alday:2011ga,Engelund:2011fg,Engelund:2012re}
\begin{align}\label{defF}
\frac{ \vev{W_{F}(x_1,x_2,x_3,x_4) \mathcal{L}(x_0)}}{\vev{W_{F}(x_1,x_2,x_3,x_4) }}  = \frac{1}{\pi^2} \frac{x_{13}^2 x_{24}^2}{x_{10}^2 x_{20}^2 x_{30}^2 x_{40}^2} F(g ;z) \,.
\end{align}

The fact that the kinematics depends on a single variable follows from conformal symmetry of the Wilson loop (which is dual conformal symmetry as seen from the amplitudes vantage point).
Moreover,  invariance of the Wilson loop under cyclic permutation of points $x_1,\dots,x_4$ implies $F(z)=F(1/z)$. 

In addition, $F$ depends on the rank of the gauge group $N$, and on
the Yang-Mills coupling $g_{\rm YM}^2$. In this note we will take the `t Hooft limit, so hat we only have the dependence on $g^2$.
We expand $F$ in powers of $g^2$,
\begin{equation}\label{defexpansionF}
F(z) = g^2  F^{(0)}(z) 
+ g^4  F^{(1)}(z) 
+g^6  F^{(2)}(z)    +{\cal O}(g^8) \,.
\end{equation}
Note that the expansion starts at order $g^2$, with $F^{(0)}(z)=-1$ being the Born-level contribution. 

The first four perturbative orders of $F(z)$ are known (including a non-planar contribution at three loops) \cite{Alday:2013ip,Henn:2019swt}, as well as the leading term at strong coupling \cite{Alday:2011ga}.
We reproduce the first few terms in order to establish our conventions
\begin{align}
F^{(0)}(z)=& -1 \,,\label{Fweak0} \\
F^{(1)}(z)=& \log^2 z + \pi^2 \,, \label{Fweak1}\\
F^{(2)}(z)=& -\frac{1}{2} \log^4 z  
+ \log^2 z \left[ \frac{2}{3} \text{Li}_2\left(\frac{1}{z+1}\right)+\frac{2}{3} \text{Li}_2\left(\frac{z}{z+1}\right)-\frac{19 \pi ^2}{9} \right]  \label{Fweak2}\\
&\hspace{-1cm} + \log z \left[4 \text{Li}_3\left(\frac{1}{z+1}\right)-4 \text{Li}_3\left(\frac{z}{z+1}\right) \right] 
+ \frac{2}{3} \left[\text{Li}_2\left(\frac{1}{z+1}\right)+\text{Li}_2\left(\frac{z}{z+1}\right)-\frac{\pi ^2}{6}\right]^2 \nonumber \\
&\hspace{-1cm} +\frac{8}{3} \pi ^2 \left[\text{Li}_2\left(\frac{1}{z+1}\right)+\text{Li}_2\left(\frac{z}{z+1}\right)-\frac{\pi ^2}{6}\right]
+ 8 \text{Li}_4\left(\frac{1}{z+1}\right)+8 \text{Li}_4\left(\frac{z}{z+1}\right)-\frac{23 \pi ^4}{18}\,. \nonumber
\end{align}
At strong coupling, we have \cite{Alday:2011ga}
\begin{align}\label{Fstrong}
F(z) \stackrel{g \gg 1}{=} g \frac{z}{(z-1)^3} \left[ 2 (1-z) + (z+1) \log z \right] + \ldots \,.
\end{align}
 As mentioned in the introduction, integration over the insertion point $x_{0}$ yields the cusp anomalous dimension. As explained in refs. \cite{Alday:2013ip,Henn:2019swt}, this relationship can be formulated as a functional that maps a function $F(z)$ to a number $\Gamma$. The precise relation is as follows,
\begin{align}\label{extractgammacusp}
g \frac{\partial}{\partial_g} {\Gamma}_{\rm cusp}(g) = -2 \, \mathcal{I}[F(g; z)]   \,,
\end{align}
where $\mathcal{I}$ acts on individual terms as in
\begin{align}\label{rulefunctional-intro}
\mathcal{I}[z^p] = \frac{\sin (\pi p)}{\pi p} \,. 
\end{align}
For self-contained completeness, we give a slightly different derivation of this fact, using the methods of \cite{Progress1}.  Let us begin with a usual massless box integral (in terms of the dual notation, $s=x_{13}^2, t=x_{24}^2$), 
\begin{equation}
I = s t \int \frac{d^4 x_{0}}{i\pi^2}
\frac{1}{x_{10}^2 x_{20}^2 x_{30}^2 x_{40}^2}\,,
\end{equation}
and see why it has a double-logarithmic IR divergence. After Feynman parametrizing each loop propagator as 
\begin{equation}\label{eq:Schwingerparam}
\frac{1}{(x_{i0}^2)^{a}} =  \frac{1}{\Gamma(a)} \int_0^\infty d \alpha_i \alpha_{i}^{a-1} e^{-\alpha_i x_{i0}^2},
\end{equation}
(with $a=1$) and carrying out the loop integration, we are left with the familiar 
\begin{equation}
s t \int \frac{d^4 \alpha}{{\rm GL(1)}} \frac{1}{(s \alpha_2 \alpha_4 + t \alpha_1 \alpha_3)^2}.
\end{equation}
We can as usual fix the GL(1) by setting e.g. $\alpha_4 \to 1$. Why is the resulting integral log$^2$ divergent? The reason is is simple: the integrand has two {\it additional} GL(1) invariances, under $\alpha_1 \to a \alpha_1, \alpha_3 \to b \alpha_3, \alpha_2 \to a b \alpha_2$. This pulls out a factor of $\int \frac{da}{a} \frac{db}{b}$ from the integral that gives the log$^2$ divergence. To get the {\it coefficient} of this divergence, we simply further mod-out by these two new GL(1)'s. We can do this for instance by setting $\alpha_{1,3} \to 1$, leaving us with
\begin{equation}
   s t \int_0^\infty d \alpha_2/(s \alpha_2 + t)^2 = 1
\end{equation}
for the coefficient of the log$^2$ divergence. We can easily repeat this exercise if the original loop integrand has a factor of 
\begin{equation}
z^p = \left(\frac{s x_{20}^2 x_{40}^2}{t x_{10}^2 x_{30}^2} \right)^p \,,
\end{equation}
for any power $p$. 
We now Feynman parametrize again, using eq. (\ref{eq:Schwingerparam}) for $a=1-p$ and $a=1+p$, depending on the propagator.
Note these expressions are formally valid for $|p| < 1$ but our final expression can be extended to general $p$ by analytic continuation. Performing the loop integration, we have 
\begin{equation}
   s t \left(\frac{s}{t} \right)^p \frac{1}{\Gamma(1-p)^2 \Gamma(1+p)^2} \int \frac{d^4 \alpha}{{\rm GL(1)}} \frac{1}{(\alpha_2 \alpha_4 s + \alpha_1 \alpha_3 t)^2} \left(\frac{\alpha_1 \alpha_3}{\alpha_2 \alpha_4}\right)^p.
\end{equation}
Again, we can gauge-fix the GL(1) by setting $\alpha_4 \to 1$ as usual, and the integral continues to have the two additional accidental GL(1) invariances giving the log$^2$ divergence. The coefficient of the divergence is again obtained simply by modding out by these GL(1)'s, setting $\alpha_1 \to 1, \alpha_3 \to 1$ we are left with 
\begin{equation}
    s t \left( \frac{s}{t}\right)^p \frac{1}{\Gamma(1 - p)^2 \Gamma(1 + p)^2} \int_0^\infty \frac{d\alpha_2 \alpha_2^{-p}}{(s \alpha_2 + t)^2}\,.
\end{equation} 
Performing the integral and using $\Gamma(1 + p) \Gamma(1 - p) = \frac{\pi p}{{\rm sin} \pi p}$ yields $\frac{{\rm sin} \pi p}{\pi p}$ for the coefficient of the log$^2$ divergence. 

The action of ${\cal I}$ has been specified on a power $z^p$, but we can identify $\frac{{\rm sin} \pi p}{\pi p}$ =  $\frac{1}{2 \pi i} \int_{{\cal C}} \frac{dz}{z} z^p$, where ${\cal C}$ is a contour that starts just below the real axis at $z = -1$, goes around the origin and comes back just above the real axis at $z=-1$. Thus, for a general function $F(z)$, we can write 
\begin{equation}\label{eq:functionalform2}
    {\cal I}[F] = \frac{1}{2 \pi i} \int_{{\cal C}} \frac{dz}{z} F(z)\,.
\end{equation}
If one further assumes that $F$ has a branch cut on the negative real axis only, and is analytic elsewhere (this is the case for the functions $F$ we encounter in this paper), then one can deform the contour $\cal C$ to be a unit circle around the origin, with $z=\exp(i \phi)$. In this way we obtain a further representation of the functional ${\cal I}$,
\begin{equation}\label{eq:functionalform3}
    {\cal I}[F] = \frac{1}{2 \pi } \int_{-\pi}^{\pi}  d\phi F(e^{i \phi})  \,.
\end{equation}
One may verify that upon using eq. (\ref{extractgammacusp}) in combination with (\ref{eq:functionalform3}) on eqs. (\ref{Fweak0}),(\ref{Fweak1}),(\ref{Fweak2}) and (\ref{Fstrong}), one obtains the value quoted for the cusp anomalous dimension in eq. (\ref{formula_cusp_intro}).

\subsection{Integrand for $F$ from planar scattering amplitudes}

There are several ways of computing $F$ in perturbation theory. While it is possible to start directly from the definition of the Wilson loops on the LHS of eqs. (\ref{defF}) and evaluate the necessary Wilson loop integrals, we can profit from the known duality between Wilson loops and scattering amplitudes. (Alternatively, a duality to correlation functions, in the light-like limit, can also be used, as was done in ref. \cite{Henn:2019swt}).

Let $A_4$ be the four-point color-ordered scattering amplitude in planar $\mathcal{N}=4$ sYM theory.
Then we define the loop factor $M$ via 
\begin{align}
M = A_{4} / A_{4}^{(0)}\,,
\end{align}
by dividing by the tree-level contribution $A_{4}^{(0)}$.
The duality relation states that
\begin{align}\label{dualityWM}
\log W_{F} \sim \log M \,.
\end{align}
We wrote $\sim$ because eq. (\ref{dualityWM}) is rather formal. The reason is that both objects have divergences. The Wilson loop has ultraviolet (short distance, cusp) divergences, cf. eq. (\ref{W-div}),
while the scattering amplitude has infrared (soft and collinear) divergences,\footnote{Recall that in this paper we work in the `t Hooft limit. The first non-planar corrections to eq. (\ref{M-div}) occur at four loops, and have been evaluated in \cite{Henn:2019swt}.}
\begin{equation}\label{M-div}
    \log M =     - \sum_{L\ge 1} g^{2L} \frac{\Gamma_{\rm cusp}^{(L)}}{(L \epsilon)^2} + {\cal{O}}\left({1}/{\epsilon}\right)\,.
\end{equation}
One possibility to make the relation exact is to turn to well-defined (and scheme-independent) finite parts, or ratio functions (in the case of more general helicity configurations). This has the disadvantage that it requires at least six points. Another possibility is to consider eq. (\ref{dualityWM}) at the level of the finite {\it integrand} of both quantities. This is the route we will take in this note. 

How does one formulate eq. (\ref{dualityWM}) at the level of loop integrands? For the scattering amplitude $M$, 
one writes the loop contributions in terms of dual (or region) coordinates, which eliminate the issue of ambiguity under momentum rerouting. 
Moreover, requiring the correct pole structure fixes a potential total derivative ambiguity, so that the integrand is well defined, and can be computed, for example, using the loop-level BCFW recursion relations \cite{Arkani-Hamed:2010zjl}.
For the Wilson loop, one uses the well-known Lagrangian insertion method - roughly speaking, the Wilson loop with $L$ Lagrangian insertions at Born level corresponds to the $L$-loop integrand.
The equivalence of the integrand of $W_F$ and $M$ (and to the integrand of correlation functions) has been well studied in the literature, see for example  references \cite{Caron-Huot:2010ryg,Eden:2010zz}.

Therefore we can profit from the fact that advanced techniques exist to obtain integrands for planar scattering amplitudes $M$. In this note our starting point will be the powerful amplituhedron construction, as reviewed and developed further in the following two sections.

Given the definition (\ref{defF}) it is clear that the knowledge of the Wilson loop with $L$ Lagrangian insertions at Born level, or equivalently the $L$-loop integrand for $M$, allows us to define the integrand of $F^{(L-1)}$.
More concretely, let us introduce the notation $\Omega$ for the integrand of $ \log M$,
\begin{align}
 \log M   = 1 + \sum_{L\ge1} g^{2 L} \int \frac{d^{D}x_{5} \ldots d^{D}x_{4+L}}{(i \pi^{D/2})^L} \Omega^{(L)}(x_1, \ldots x_4 ; x_{5} , \ldots x_{4+L}) \,,
\end{align}
where we assume that $\Omega$ is explicitly symmetrized between all loop integration variables $x_{5} \,,\ldots x_{4+L}$.
Then, taking into account that at integrand level $M$ and $W_{F}$ are equivalent, as well as the Lagrangian insertion formula $g^2 \partial_{g^2} \langle {\cal O} \rangle = - i \int d^{D}x_{0} \langle {\cal O} {\cal L}(x_0) \rangle$, as well as the definition (\ref{defF}), one finds that $F^{(L-1)}$ is given by
\begin{align}\label{defFfromintegrandM}
F^{(L-1)}(z) = L\,  \frac{x_{10}^2 x_{20}^2 x_{30}^2 x_{40}^2}{x_{13}^2 x_{24}^2} \int \frac{d^{4}x_{6} \ldots d^{4}x_{4+L}}{(i \pi^{2})^{L-1}} \Omega^{(L)}(x_1, \ldots x_4 ; x_{0} , x_{6}, \ldots x_{4+L}) \,.
\end{align}
Notice that in eq. (\ref{defFfromintegrandM}), one of the integration points, $x_{6}$ has been ``frozen'' to be $x_0$, and is left unintegrated. As a result, $F$ does not require regularization and can be computed, in principle, in four dimensions.

Here we show, for illustration and to set our conventions, the first few loop orders.
The four-point amplitude has the following expansion 
\begin{align}
M = 1+ g^2 M^{(1)} + g^4 M^{(2)} + {\cal O}(g^6) \,.
\end{align}
The one- and two-loop contributions are given in terms of box integrals. In dual notation, they read
\begin{align}\label{expressionM1integrand}
M^{(1)}  = - \int \frac{d^{D}x_{5}}{i \pi^{D/2}} \frac{x_{13}^2 x_{24}^2}{x_{15}^2 x_{25}^2 x_{35}^2 x_{45}^2} \,,
\end{align}
and
\begin{align}\label{expressionM2integrand}
M^{(2)}  =&\int \frac{d^{D}x_{5} d^{D}x_{6}}{(i \pi^{D/2})^2} \frac{1}{2}  \left[ \frac{(x_{13}^2)^2 x_{24}^2}{ x_{15}^2 x_{25}^2 x_{35}^2 x_{56}^2 x_{26}^2 x_{36}^2 x_{46}^2 } +  \frac{x_{13}^2 (x_{24}^2 )^2 }{ x_{25}^2 x_{35}^2 x_{45}^2 x_{56}^2 x_{36}^2 x_{46}^2 x_{16}^2 } + \right.  \nonumber \\
& \quad\quad\quad\quad\quad\quad\quad \left. +  \frac{(x_{13}^2)^2 x_{24}^2}{ x_{16}^2 x_{26}^2 x_{36}^2 x_{56}^2 x_{25}^2 x_{35}^2 x_{45}^2 } +  \frac{x_{13}^2 (x_{24}^2 )^2 }{ x_{26}^2 x_{36}^2 x_{46}^2 x_{56}^2 x_{35}^2 x_{45}^2 x_{15}^2 }  \right]  \,.
\end{align}
The first two summands in the first line correspond to the (normalized) $s$-channel and $t$-channel ladder integrals. The second line corresponds to the same integrals, symmetrized in the integration variables $x_{5},x_{6}$.

Taking into account the duality with the Wilson loop, and eqs. (\ref{defF}) and (\ref{defexpansionF}), one finds \cite{Alday:2012hy}
\begin{align}
F^{(0)} =& -1 \,, \\
F^{(1)} =&  \int \frac{d^{4}x_{6}}{i \pi^{2}}   \frac{\left[x_{16}^2 x_{24}^2 x_{30}^2 + x_{26}^2 x_{13}^2 x_{40}^2  + x_{36}^2 x_{24}^2 x_{10}^2+x_{46}^2 x_{13}^2 x_{20}^2  - x_{06}^2 x_{13}^2 x_{24}^2   \right] }{x_{16}^2 x_{26}^2 x_{36}^2 x_{46}^2 x_{06}^2}  \,.\label{F1integrand}
\end{align} 
Here, the last term in the second line comes from the contribution of $-1/2 (M^{(1)})^2$ to $\log M$.
The linear combination appearing in the numerator of eq. (\ref{F1integrand}) is exactly (up to an overall sign) that appearing in the finite combination of box and triangle integrals discussed in section 6 of \cite{Henn:2020omi}. As discussed there, the integral is manifestly finite in four dimensions.
Upon integration, this is in agreement with eq. (\ref{Fweak1}).

A comment is due on the nature of the integrals contributing to $F$. By definition, one could always express $\log M$, and hence $F$, as sums of products of planar integrals. In general, the latter will be individually divergence (such as the box and triangle integrals contributing to eq. (\ref{F1integrand}), and it is only their sum that is finite in four dimensions. 

What is more, we will introduce a geometric decomposition of $\log M$ that expresses the answer in terms of finite contributions, just like eq. (\ref{F1integrand}), which however are not expressible (in general) in terms of standard planar integrals. Instead, the geometric decomposition is a completely new representation of the answer. We will introduce methods for computing certain classes of these diagrams directly in four dimensions, without needing the standard machinery of loop integration \cite{Henn:2019swt}.

\subsection{Momentum twistor notation}

In the following we will use momentum twistor variables \cite{Hodges:2009hk}, as they considerably streamline the geometric analysis and the discussion of infrared finiteness. 
In this language, the external kinematics is described by four momentum twistors $Z_{1}, Z_{2},Z_{3},Z_{4}$. 
The Lagrangian insertion point corresponds to a line $Z_{A} Z_{B}$, and loop integration points are denoted by $Z_{C} Z_{D}$, $Z_{E} Z_{F}$, etc. at low loop orders, or by $(Z_{A} Z_{B})_{L}$ in general.
Then we have
\begin{align}
z =  \frac{\vev{12AB} \vev{34AB}}{\vev{14AB}\vev{23AB}} \,. \label{ratioz}
\end{align}
Likewise, we can write the loop integrals in momentum twistor space (see e.g. \cite{Arkani-Hamed:2010pyv}).
Following closely section 3 of \cite{Alday:2013ip}, the expression in eq. (\ref{F1integrand}) becomes 
\begin{align}
F^{(1)} =& \int_{CD}   \frac{{\cal N}}{\vev{41CD}\vev{12CD}\vev{23CD}\vev{34CD}\vev{ABCD}}\,,\label{F1}
\end{align}
with
\begin{align}
{\cal N} =& \langle 1 2 C D \rangle \langle 3 4 A B \rangle \langle 4 1 2 3 \rangle
\langle1 2 A B\rangle \langle3 4 C D\rangle \langle4 1 2 3\rangle +  \langle1 2 3 4\rangle \langle2 3 C D\rangle \langle4 1 A B\rangle \nonumber \\
& + 
 \langle1 2 3 4\rangle \langle2 3 A B\rangle \langle4 1 C D\rangle - 
 \langle1 2 3 4\rangle \langle4 1 2 3\rangle \langle C D A B\rangle\,.
\end{align}
This expression can be rewritten using a twistor identity, 
\begin{align}
{\cal N} =& - \langle 1 2 3 4 \rangle \left[  \langle A B 1 3 \rangle \langle C D 2 4 \rangle 
+\langle A B 2 4 \rangle \langle C D 1 3 \rangle 
\right]\,.
\end{align}
In this form, the infrared finiteness of the integral is obvious, cf. \cite{Alday:2013ip}, and upon integration we recover eq. (\ref{Fweak1}).

In the next section, we will present a novel method to obtain the integrand for $\log M$, and hence for $F$, directly. 
It will be given in a form that makes the infrared properties of the objects manifest, i.e. $\log M$ will be given in terms of contributions that are only $1/\epsilon^2$ divergent each, and $F$ will be given as a sum of finite contributions.

\section{Geometry of integrands}
\label{sec:geometry}

\subsection{Review of four-point amplituhedron}

We start with reviewing the amplituhedron geometry \cite{Arkani-Hamed:2013jha,Arkani-Hamed:2017vfh} for MHV four-point amplitudes. The external kinematical data is given by four momentum twistors $Z_1$, $Z_2$, $Z_3$, $Z_4$ with $\la 1234\ra \equiv \epsilon_{abcd}Z_1^aZ_2^bZ_3^cZ_4^d>0$. The one-loop amplituhedron ${\cal A}_1(AB)$ is a collection of all lines $(AB)_{IJ}\equiv Z_A^IZ_B^J$ in $\mathbb{P}^3$, where $Z_A$ and $Z_B$ are two arbitrary points on that line, with the following conditions
\begin{equation}
   {\cal A}_1(AB):\,\{\la AB12\ra, \la AB23\ra, \la AB34\ra, \la AB14\ra >0,\qquad \la AB13\ra,\la AB24\ra<0\} \,. \label{oneloop_pos}
\end{equation}
We can choose a convenient parametrization of the line 
\begin{equation}
    Z_A = Z_1 + x Z_2 + y Z_4,\qquad Z_B = Z_3 - z Z_2 + w Z_4 \,. \label{param}
\end{equation}
The conditions (\ref{oneloop_pos}) then force all parameters (\ref{param}) to be positive,
\begin{equation}
    x,y,z,w > 0\,.
\end{equation}
There is a unique form $\Omega_1(AB)$ with logarithmic singularities on the boundaries of ${\cal A}_1(AB)$. This form reproduces the four point one-loop integrand, and the one-loop amplitude $M^{(1)}$ is given by integrating over all lines $(AB)$,
\begin{equation}
    M^{(1)} = \int_{AB} \Omega_1(AB)  \,.
\end{equation}
The contour is chosen such that the corresponding loop momentum $\ell$ is real. 

At two-loops, the amplituhedron geometry is a configuration of two lines $AB$ and $CD$ (for lucidity we use this notation rather than $(AB)_1$, $(AB)_2$) where each line is in the one-loop amplituhedron with additional mutual condition $\la ABCD\ra = \epsilon_{abcd}Z_A^aZ_B^bZ_C^cZ_D^d>0$,
\begin{equation}
     {\cal A}_2(AB,CD) = \{AB,CD \in {\cal A}_1,\,\,\mbox{and}\,\, \la ABCD\ra > 0\}\,.
\end{equation}
We again parametrize the respective lines $AB$, $CD$ as in (\ref{param}) and the one-loop amplituhedron inequalities (\ref{oneloop_pos}) force $x,y,z,w>0$. The mutual positivity condition is a non-trivial relation between all eight parameters,
\begin{equation}
    - (x_1-x_2)(z_1-z_2) - (w_1-w_2)(y_1-y_2) > 0\,. \label{mutual2}
\end{equation}
Here we introduce a graphic notation: for each loop line we draw a dot and mutual positive condition is represented by blue dashed line. The figure then represents a dlog form on the positive geometry,
\begin{equation}
\includegraphics[width=0.23\columnwidth]{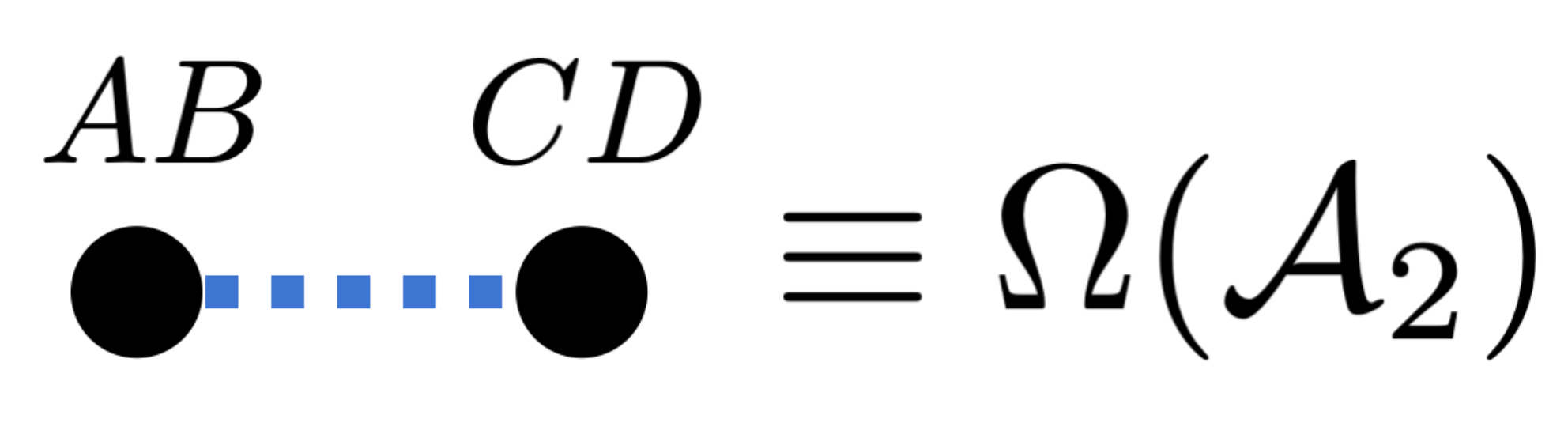}
\end{equation}
At $L$-loops we demand that each line $(AB)_i$ lives in the one-loop amplituhedron and each pair of lines satisfies mutual positivity condition, $\la (AB)_i (AB)_j\ra >0$,
\begin{equation}
  {\cal A}_L = \{(AB)_k\in {\cal A}_1,\,\mbox{and}\,\la (AB)_i(AB)_j\ra>0\,\mbox{for each pair $i,j$}\}\,.
\end{equation}
Finding the dlog form $\Omega_L$ on ${\cal A}_L$ for general $L$ (and hence the four point $L$-loop integrand) is a huge task and very complicated challenge. While some progress has been made for certain lower dimensional boundaries (cuts of the amplitude) \cite{Arkani-Hamed:2013kca,Arkani-Hamed:2018rsk}, the general strategy for the triangulation is still missing. 

In the graphical representation, $\Omega_L$ corresponds to a completely connected graph, 
\begin{equation}
\includegraphics[width=0.27\columnwidth]{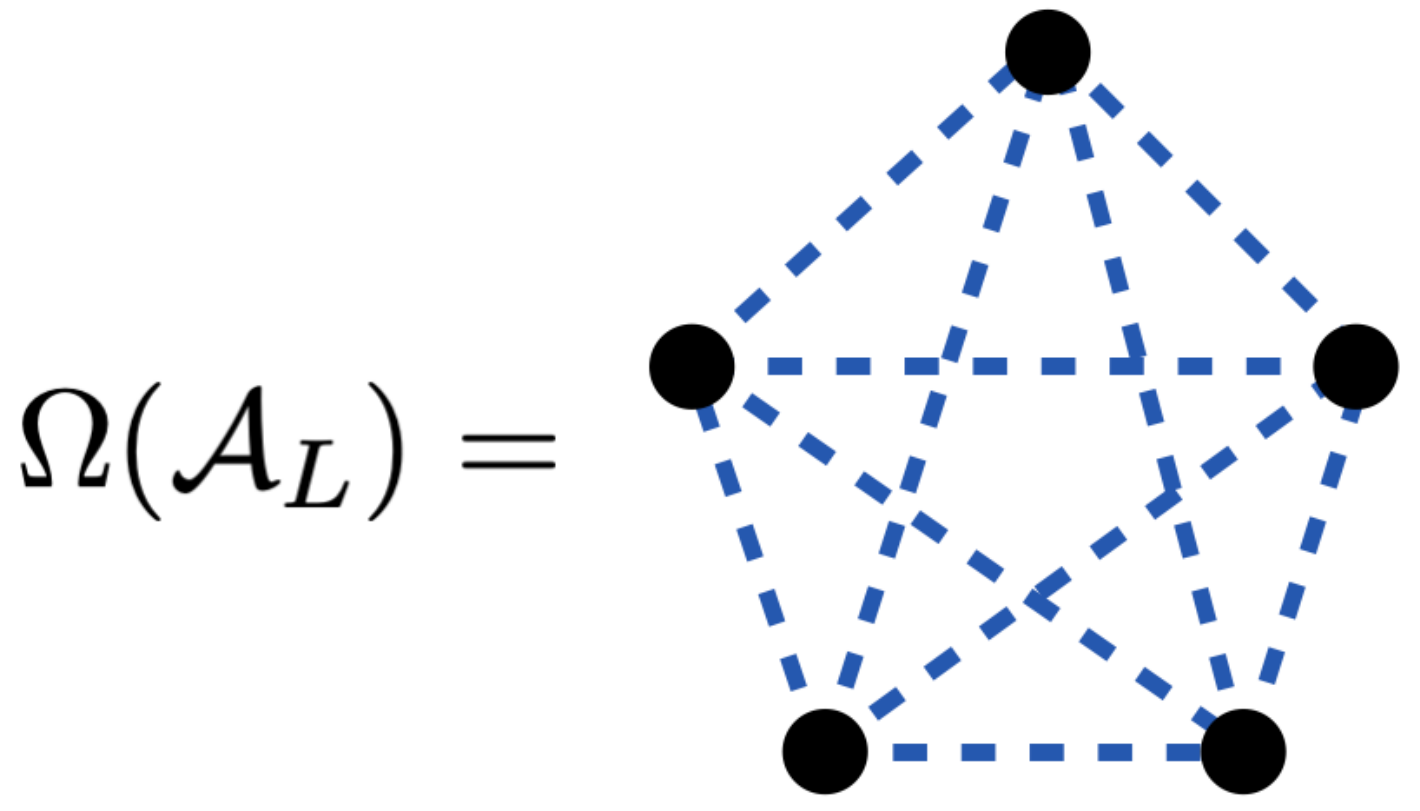} \label{eq:graphnegativegeometries}
\end{equation}
Each dot represents loop $(AB)_i$ and each edge is a negativity condition $\la (AB)_i(AB)_j\ra<0$.

\subsection{Negative geometries }

The mutual positivity conditions $\la (AB)_i(AB)_j\ra>0$ are essential to the definition of the amplituhedron. From the geometry point of view, an equally interesting case is when $\la (AB)_i(AB)_j\ra<0$. 

Let us start at two loops. We can define the positive geometry of two lines $AB$, $CD$, each of them being in the one-loop amplituhedron, further satisfying $\la ABCD\ra<0$,
\begin{equation}
    \widetilde{{\cal A}}_2 = \{AB,CD\in {\cal A}_1,\,\mbox{and}\,\la ABCD\ra<0\}
\end{equation}
We denote by $\widetilde{\Omega}_2$ the canonical dlog form for this space. We again use the graphic notation where the vertices represent the loop lines, and the ``negative link" $\la ABCD\ra<0$ is represented by a red thick line. The connection between $\Omega_2$, $\widetilde{\Omega}_2$ and $\Omega_{1,1}$ is then
\begin{equation}
\includegraphics[width=0.4\columnwidth]{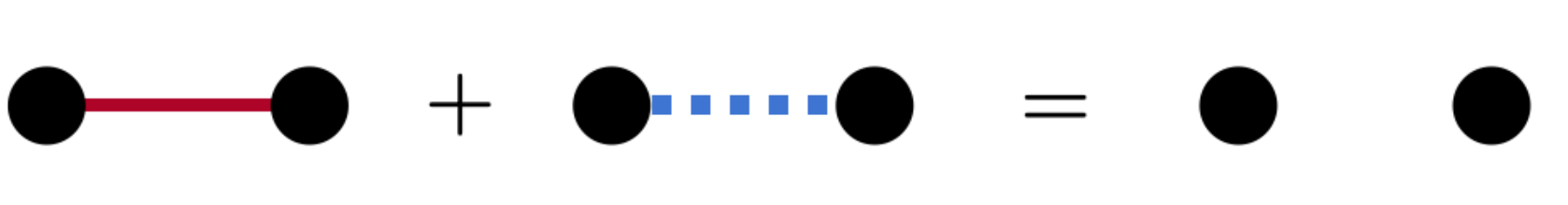} \label{rel}
\end{equation}
There is a natural question if this form has any physical significance. Using the relation (\ref{rel}) we see that at the level of integrands of amplitudes we have 
\begin{equation}
\widetilde{\Omega}_{2} \sim M^{(2)} - \frac12 (M^{(1)})^2\,,
\end{equation}
where the relative factor of $1/2$ comes from symmetrization in lines $AB$ and $CD$. This is nothing else than the two-loop logarithm of the four point amplitude. In the next subsections, we will see that this is not an accidental connection. 

Finally, we can use the relation (\ref{rel}) multiple times on the general $L$-loop connected graph (\ref{eq:graphnegativegeometries}), and rewrite all positive links (blue dashed lines) in terms of negative links (red thick lines) and no links at all. As a result, we get a sum over all graphs with red links only,
\begin{equation}
\includegraphics[width=0.7\columnwidth]{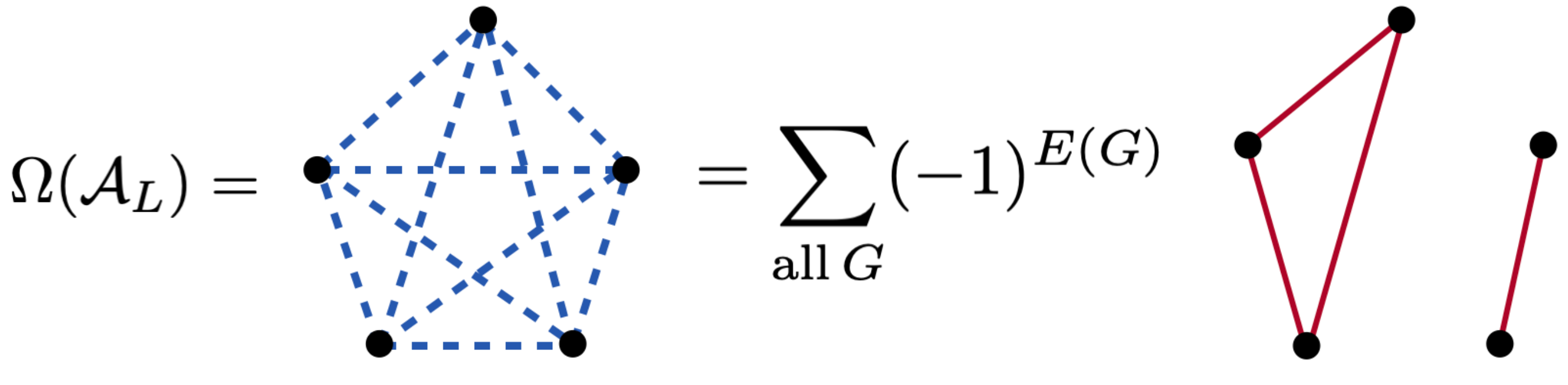} \label{negexp}
\end{equation}
Here the graph on the right hand side is just a representative, we sum over all graphs $G$. As an example, we give the $L=3$ expansion,
\begin{equation}
\includegraphics[width=0.6\columnwidth]{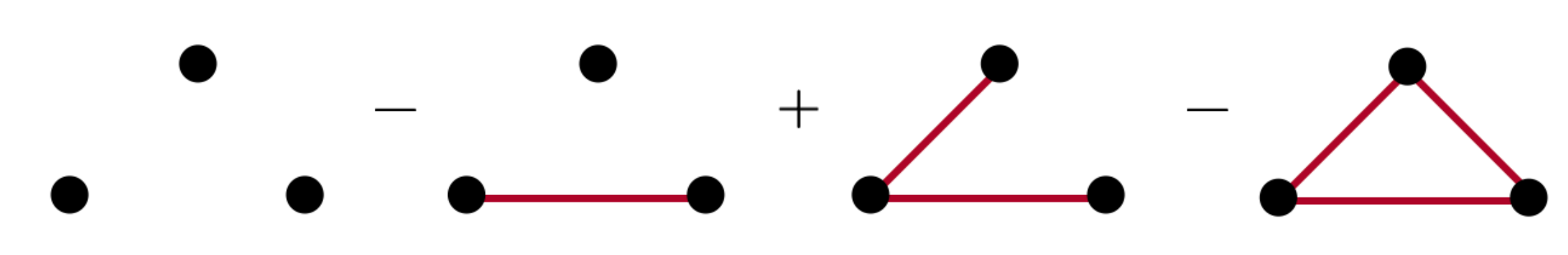} \label{L3}
\end{equation}
Each diagram is implicitly symmetrized in all lines $(AB)_i$. In particular, for the three-loop case we denote these lines by $AB$, $CD$, $EF$ for convenience, and each unlabeled diagram in (\ref{L3}) then represents    
\begin{equation}
\includegraphics[width=0.4\columnwidth]{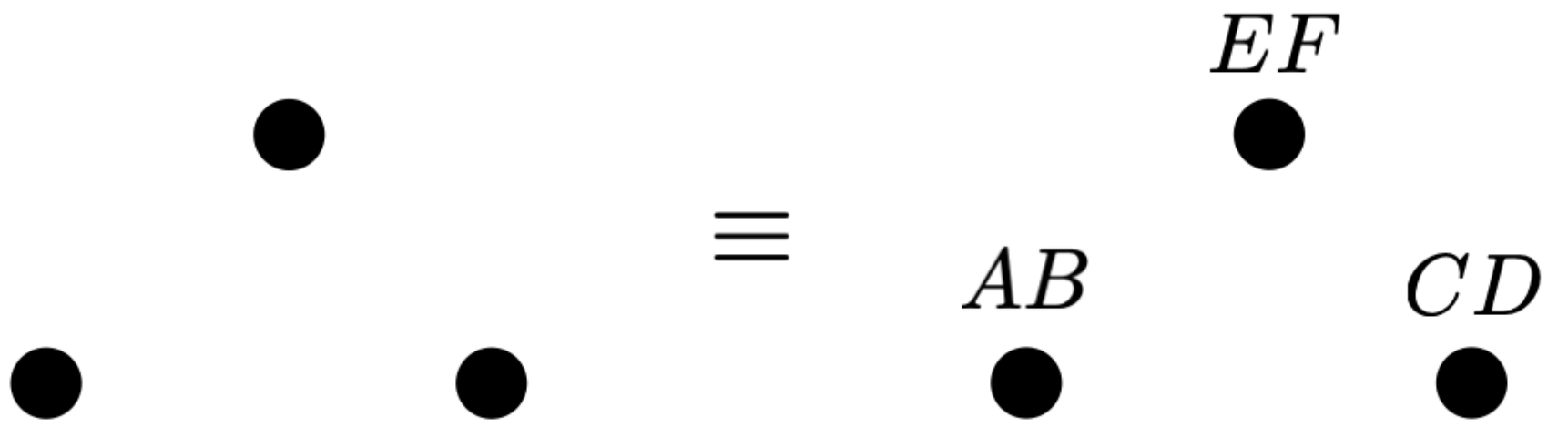} \label{L3a}
\end{equation}
\begin{equation}
\includegraphics[width=0.65\columnwidth]{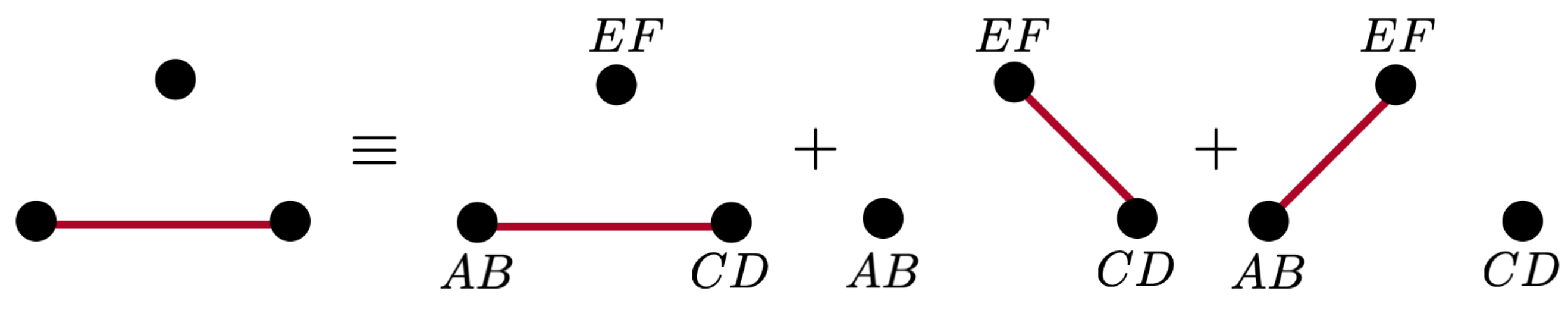} \label{L3b}
\end{equation}
\begin{equation}
\includegraphics[width=0.65\columnwidth]{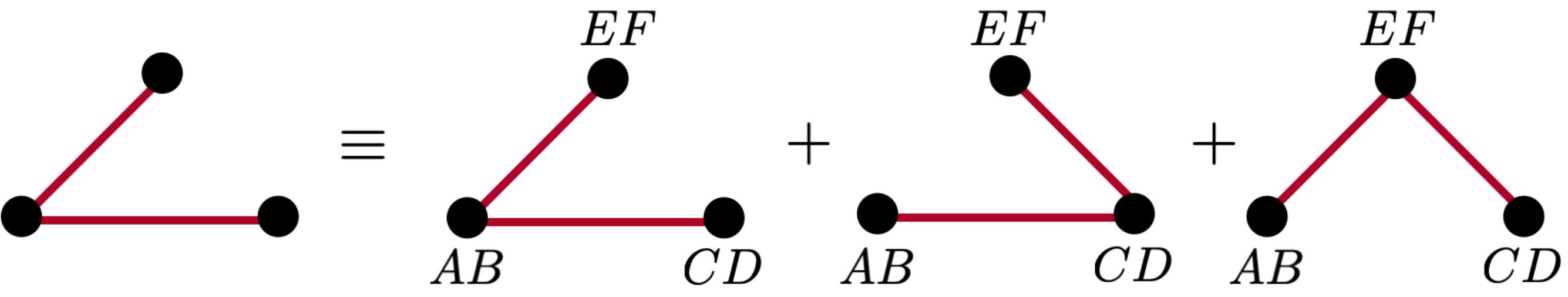} \label{L3c}
\end{equation}
\begin{equation}
\includegraphics[width=0.35\columnwidth]{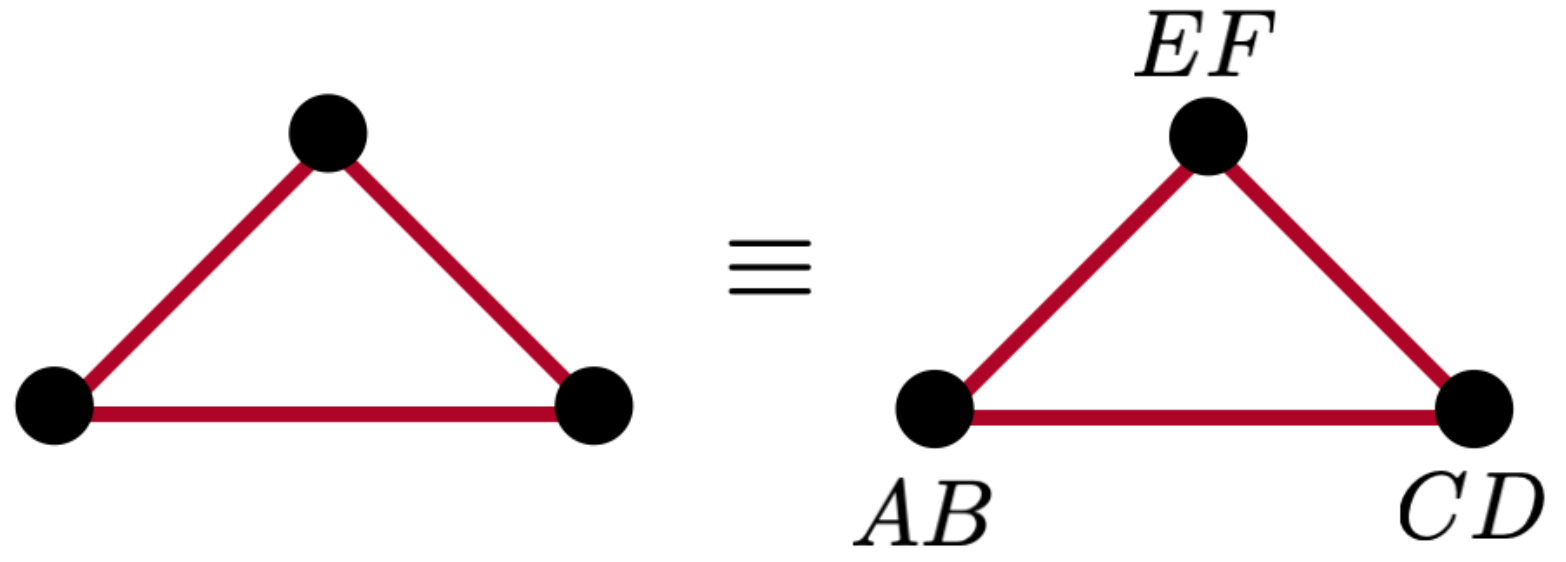} \label{L3d}
\end{equation}
where each diagram represents a positive geometry with certain set of inequalities
\begin{equation}
\includegraphics[width=0.28\columnwidth]{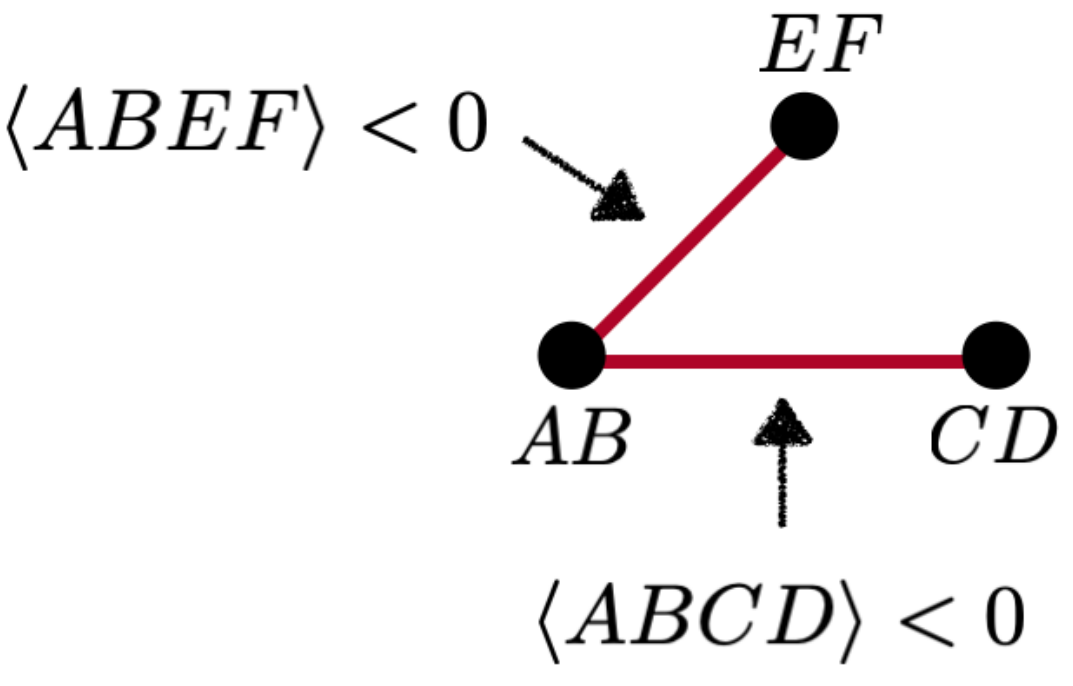}
\end{equation}
Now we decorate the relation (\ref{negexp}) with $(g^2)^L$ and sum over all $L$,
\begin{equation}
    \Omega(g) = \sum_{L=0}^\infty (-g^2)^L \Omega({\cal A}_L)\,,
\end{equation}
where $\Omega({\cal A}_0) = 1$. Now, as familiar in many settings, the generating function for a sum over {\it all} graphs, is the exponential of the generating function for {\it connected} graphs. Thus using (\ref{negexp}) we find the expansion for $\Omega(g)$ as the exponential of the sum over connected graphs,
\begin{equation}
\includegraphics[width=0.61\columnwidth]{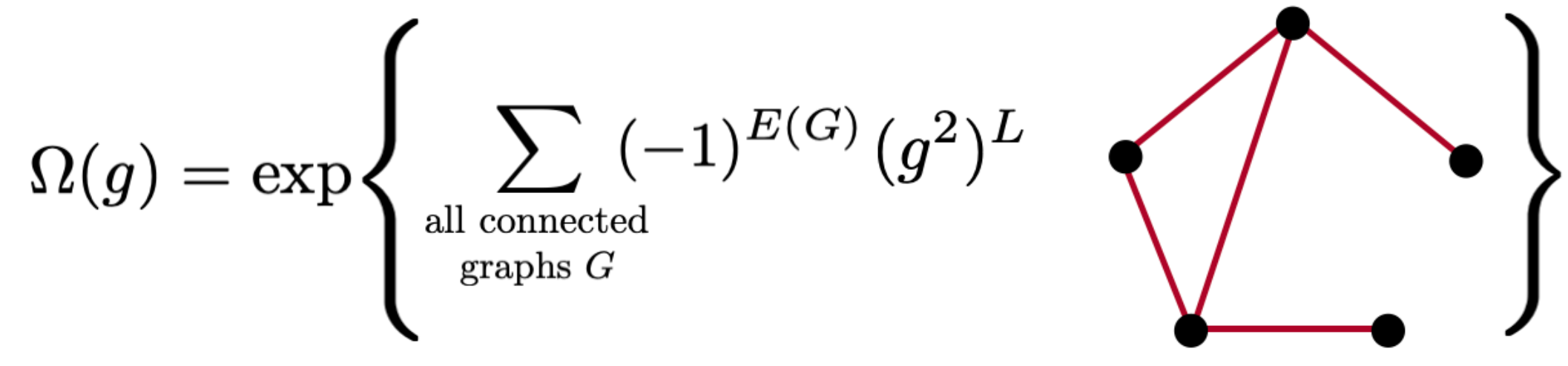} \label{exp1}
\end{equation}
Again, the graph in the equation above is just one representative, we have to sum over all connected graphs. Note that the graph with $L$ vertices and $E(G)$ edges is decorated with the coefficient $(-1)^{E(G)}(g^2)^L$. Now we can take the logarithm of this expression,
\begin{equation}
\includegraphics[width=0.56\columnwidth]{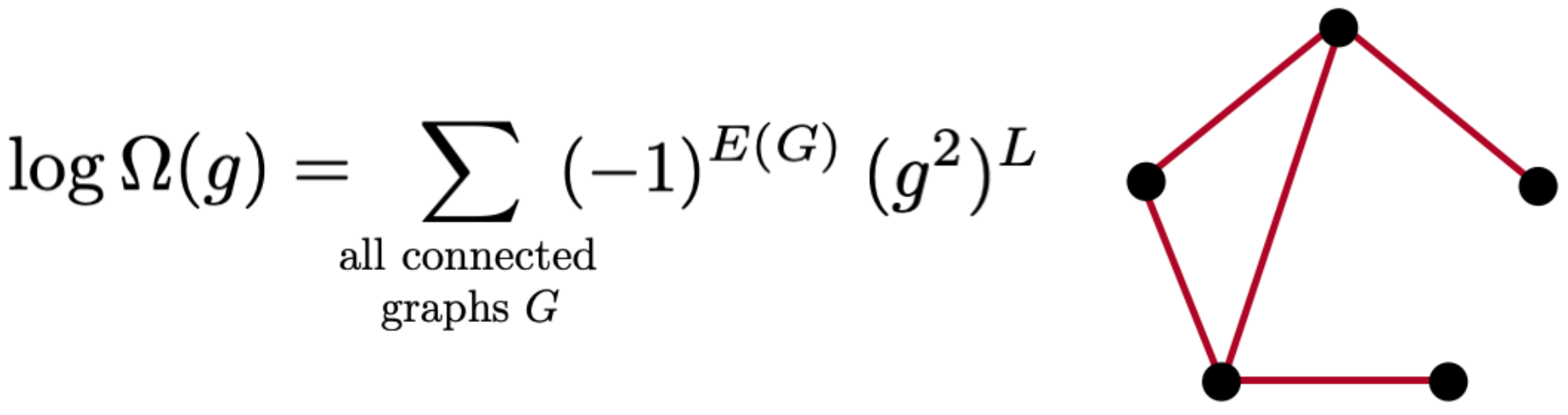} \label{log1}
\end{equation}
We can now expand $\log\Omega(g)$ in coupling $g$, and write the graphic expansion for the coefficient of $(g^2)^L$. We denote this form $\widetilde{\Omega}_L$,
\begin{equation}
\includegraphics[width=0.4\columnwidth]{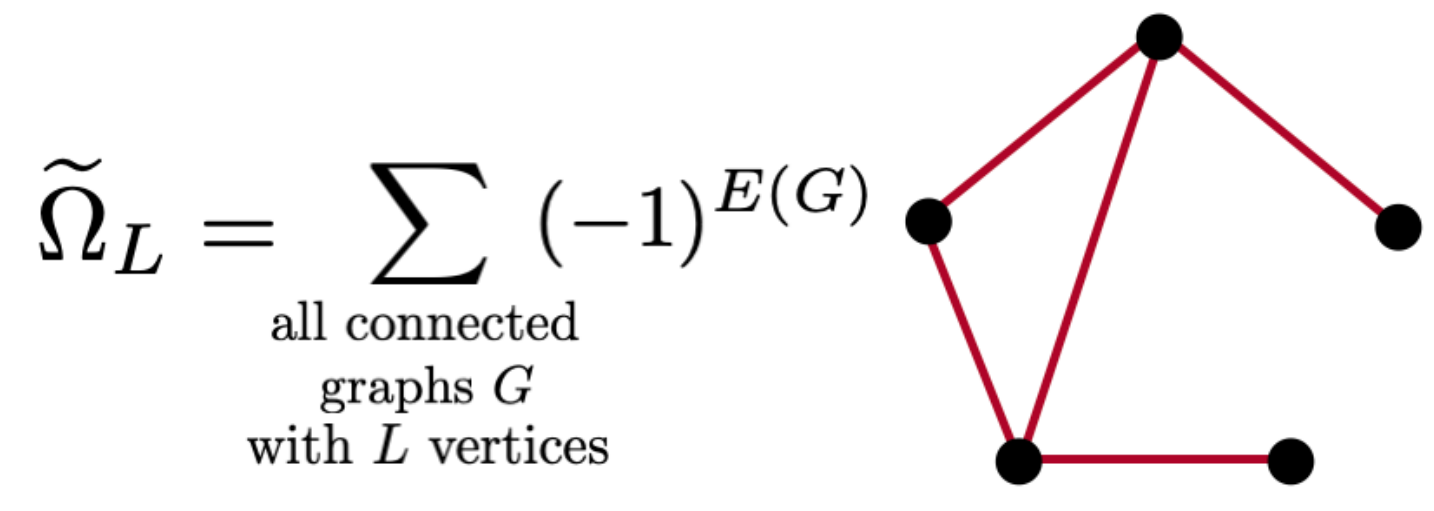} \label{log2}
\end{equation}
We succeeded in formulating the problem of finding the $L$-loop integrand for the logarithm of the amplitude in a purely geometric language. Each term in (\ref{log2}) represents a dlog form on a positive geometry given by the inequalities $\la (AB)_i(AB)_j\ra<0$ encoded by the graph. This is an analogue of the expansion (\ref{negexp}) for the $L$-loop amplitude, with the difference that in (\ref{negexp}) we sum over all graphs and in (\ref{log2}) only over all connected graphs. 

This raises the question: What is special about connected graphs with negative links? We will see in the following discussion that the answer is related to collinear regions and IR divergences.

\subsection{Collinear safety}

The negativity condition $\la ABCD\ra<0$ (or generally $\la (AB)_i(AB)_j\ra<0$) has an important implication on the allowed boundaries of the positive geometry. In particular, we look at the collinear configuration when the line $AB$ passes through point $Z_2$ and lies in the plane $(Z_1Z_2Z_3)$, 
\begin{equation}
\includegraphics[width=0.28\columnwidth]{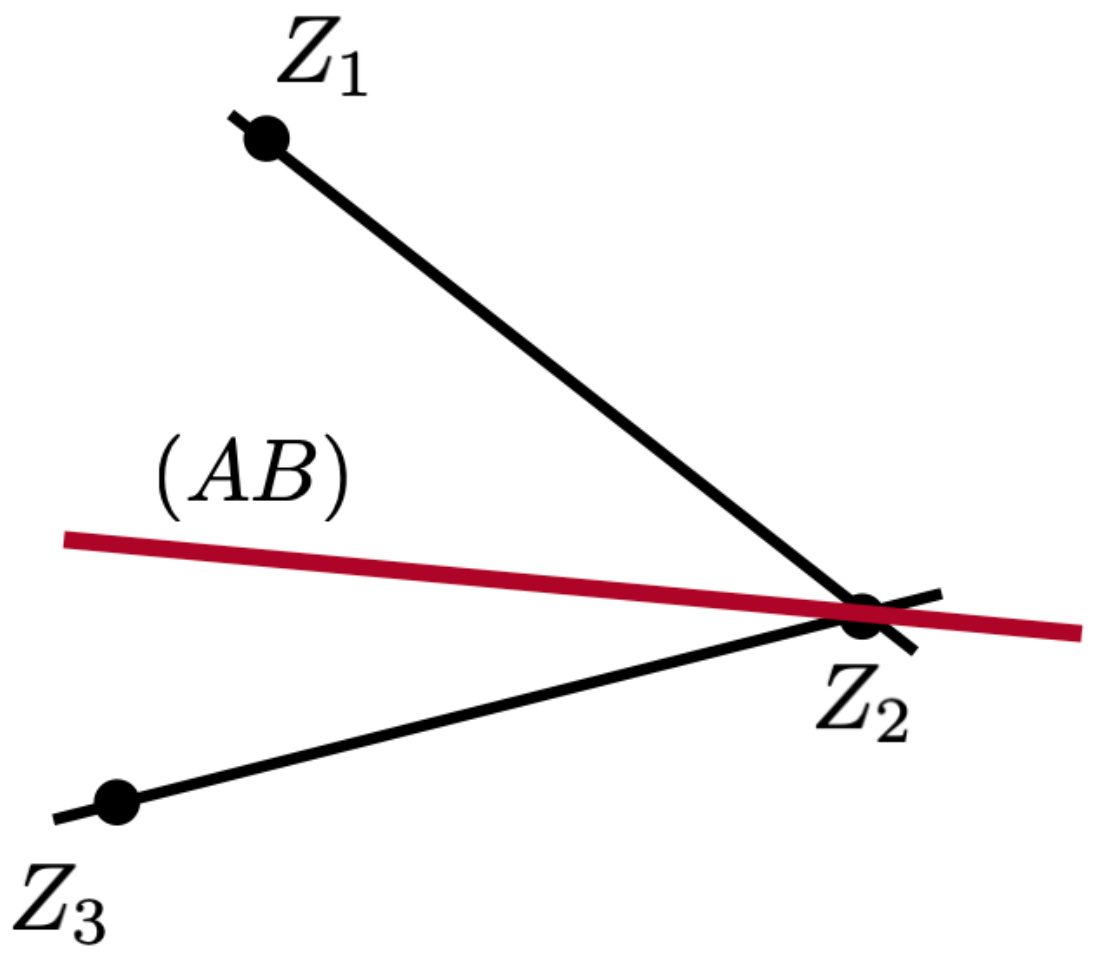}
\end{equation}
We can choose $Z_A=Z_2$ and $Z_B=-Z_1+\alpha Z_3$, where $\alpha>0$ due to the one-loop positivity conditions. The mutual positivity condition with $CD$ is then
\begin{equation}
    \la ABCD \ra = \la CD 12\ra + \alpha \la CD 23\ra  < 0
\end{equation}
Because $\la CD12\ra$, $\la CD23\ra$, $\alpha$ are all positive, this forces $\la ABCD\ra>0$ which is in conflict with the negativity condition. 
Therefore this configuration is not allowed.

This extends to higher loops and more general positive geometries. Consider a line $(AB)_i$ which has a negative link with another line $(AB)_j$, i.e. $\la (AB)_i(AB)_j\ra<0$. Placing either $(AB)_i$ or $(AB)_j$ into a collinear region violates this condition, and therefore such configurations are inconsistent with the positive geometry. As a result, if a graph with negative links is connected, then none of the lines $(AB)_i$ can be placed into a collinear region.

There is a more involved situation when some subset of all lines $(AB)_1,\dots,(AB)_k$ are placed into a collinear region at the same time, while the remaining lines $(AB)_{k{+}1},\dots,(AB)_L$ are kept generic,
\begin{equation}
\includegraphics[width=0.5\columnwidth]{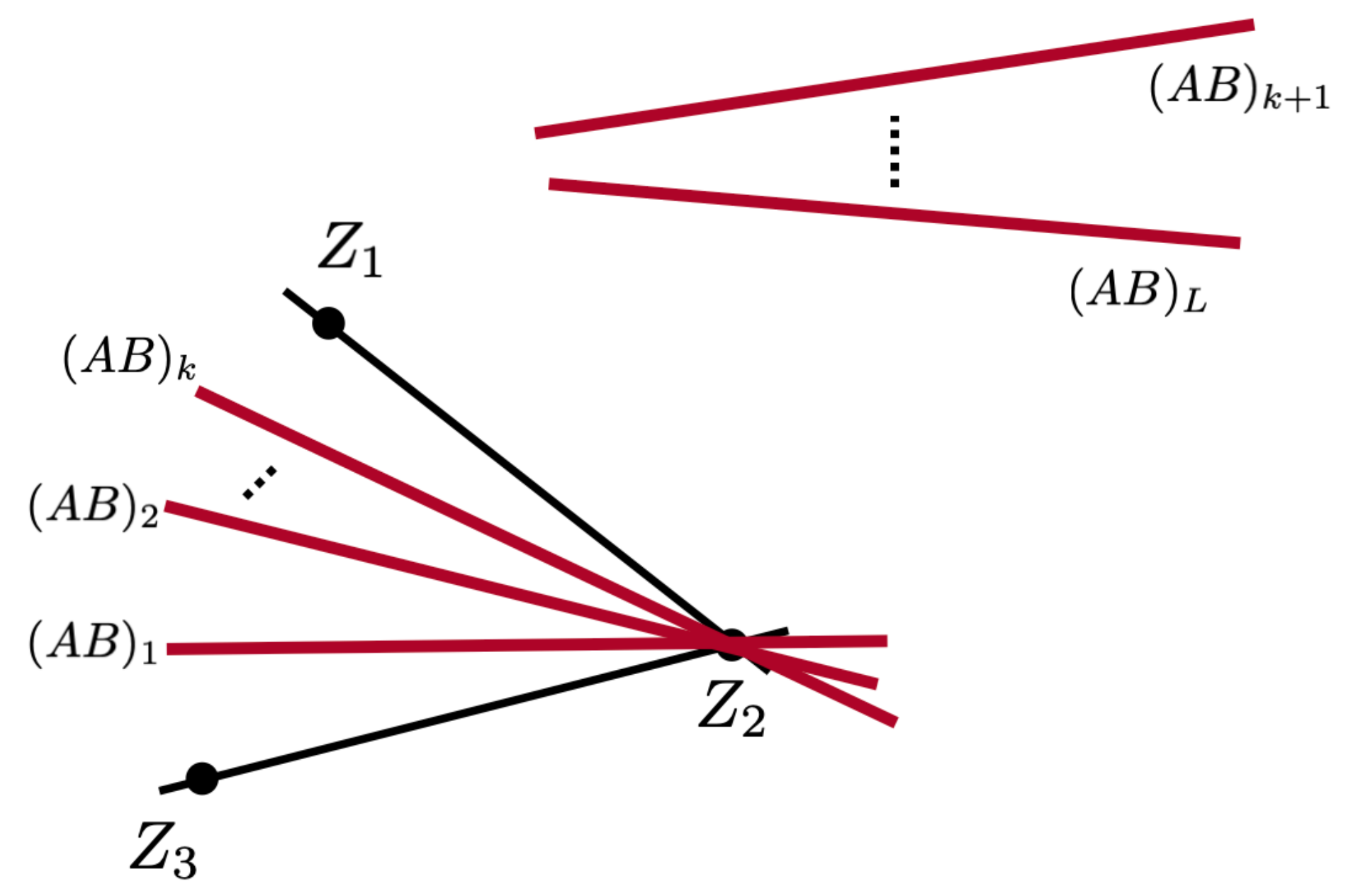}
\end{equation}
In that case all mutual negative conditions of lines in collinea region are zero,
\begin{equation}
    \la (AB)_i(AB)_j\ra = 0 \quad\mbox{i,j$\in$\,{\rm collinear}}
\end{equation}
However, if the graph is connected there is at least one inequality which includes one line in the collinear region and one generic line. This bracket is positive and the negativity condition is violated,
\begin{equation}
    \la (AB)_i(AB)_j\ra > 0 \quad\mbox{i$\in$\,{\rm collinear}},\,\,\,\mbox{j$\in$\,{\rm generic}}
\end{equation}
Consequently, the only way how to preserve the negativity conditions and send some lines to the collinear region is to send \emph{all} lines to the collinear region at the same time. 
\begin{equation}
\includegraphics[width=0.32\columnwidth]{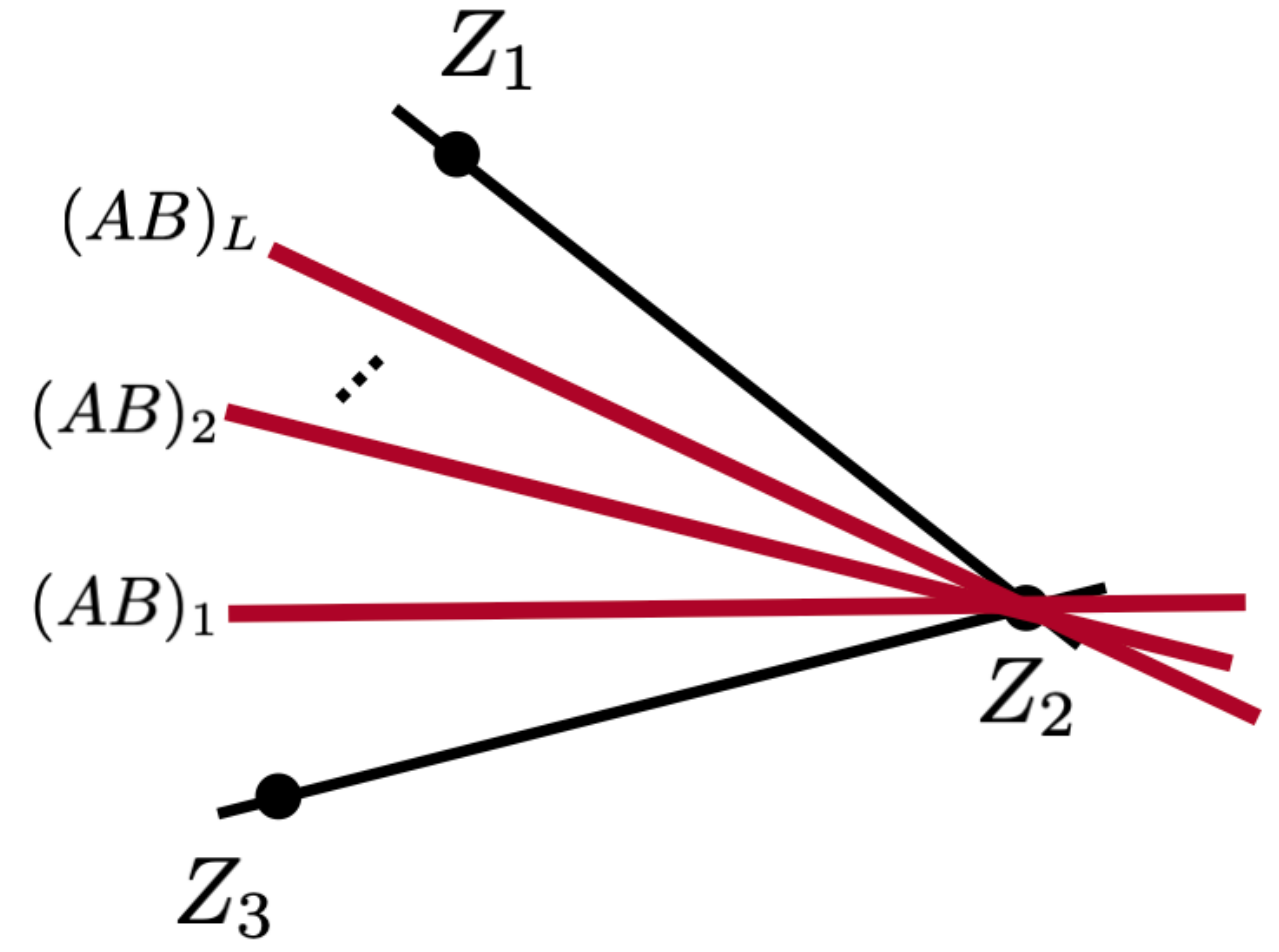}
\end{equation}
In that case all mutual brackets are zero, $\la (AB)_i(AB)_j\ra=0$, $i,j=1,\dots,L$ and nothing is violated. This is indeed an allowed boundary of the positive geometry. 

It is worth mentioning an important aspect of the mutual \emph{positivity} conditions $\la (AB)_i(AB)_j\ra>0$: they enforce ``planarity'' in the sense of the expansion in terms of planar diagrams. Let us put $(AB)_1$ and $(AB)_2$ in one-loop amplituhedra, respectively. We have to ensure that any lower-dimensional boundary inconsistent with planarity is forbidden by imposing $\la (AB)_1(AB)_2\ra>0$. For example, we can localize $(AB)_1$ to intersect lines $(12)$, $(34)$ and $(AB)_2$ to intersect lines $(23)$, $(14)$. This is indeed cut inconsistent with planarity,
\begin{equation}
\includegraphics[width=0.32\columnwidth]{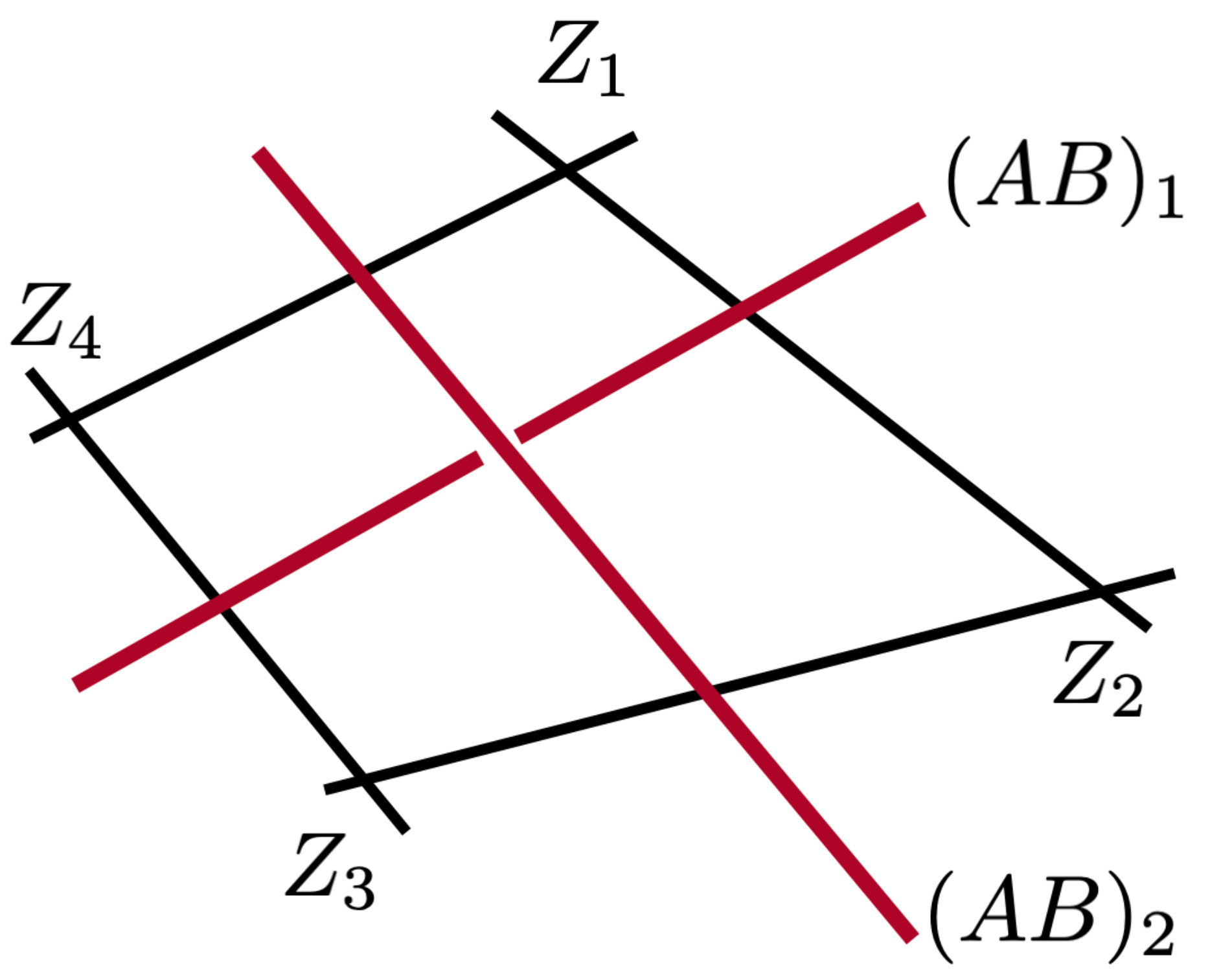}
\end{equation}
In the parametrization (\ref{param}) this sets $y_1=z_1=x_2=w_2=0$ and reduces the quadratic inequality (\ref{mutual2}) to
\begin{equation}
    -x_1w_1 - y_2z_2 > 0
\end{equation}
which is inconsistent with all parameters being positive. One can check more intricate planarity-violating configurations at higher loops but always the mutual positivity conditions come to rescue. 

Let us summarize the results of this part: 
\begin{itemize}
    \item The positivity condition $\la (AB)_i(AB)_j\ra>0$ is related to the planarity of the integrand, in a sense of the expansion in terms of planar diagrams. The object is planar only if all $\la (AB)_i(AB)_j\ra>0$ are enforced, which is the case of the $L$-loop amplituhedron.
    \item The negativity condition $\la (AB)_i(AB)_j\ra<0$ ensures that neither line $(AB)_i$ nor $(AB)_j$ can access the collinear region individually. If the graph with negative links is connected only the collection of \emph{all} lines can access the collinear region at the same time.
\end{itemize}

\subsection{Infrared finite observable}

The collinear region of the integrand is closely related to the presence of IR divergences post-integration. In the context of planar ${\cal N}=4$ SYM theory, if the integrand does not have support when one of the loops is localized in the collinear region, then the corresponding integral is IR finite. This can happen with special numerators for the integrand, which vanish in collinear regions, rendering the integrals finite. A simple example is the integral (\ref{F1}), where the collinear regions are present, as demonstrated by the presence of relevant poles in the denominator, but the numerator vanishes whenever the line $(AB)$ enters any of the collinear regions. 
As discussed earlier, our situation is different: for each connected graph we can not send individually any line $(AB)_j$ to the collinear region (as it would violate negativity conditions), but we can send all lines collectively there. As a result, this generates a mild $1/\epsilon^2$ divergence after integration,
\begin{equation}
\int \Omega_\Gamma = \frac{1}{\epsilon^2}(\dots) + {\cal O}\left(\frac{1}{\epsilon}\right)\,.
\end{equation}
As the $L$-loop logarithm is the sum over all connected graphs with $L$ vertices, the integration reveals the same degree of divergence. This is in agreement with eq. (\ref{M-div}). 

Consider now a situation when we freeze one of the loops in the connected graph and integrate over all others. As noted earlier, the integration region now does not contain the collinear region and such an object is IR finite. We denote this object ${\cal F}_\Gamma$ for each graph and the frozen loop $AB_0\equiv AB_L$, and we integrate over all other loops $AB_1,\dots,AB_{L-1}$. The function ${\cal F}_{\Gamma}$ depends on a single cross-ratio $z$ defined in (\ref{ratioz}),
\begin{equation}
    z = \frac{\la AB_012\ra\la AB_034\ra}{\la AB_023\ra\la AB_014\ra}\,,
\end{equation}
where now the frozen loop $AB_0$ appears in $z$. Then we define
\begin{equation}
 {\cal F}_{\Gamma}(z) = N_0\int_{AB_1}\int_{AB_2}\dots \int_{AB_{L{-}1}} \Omega_\Gamma \,, \label{F1}
\end{equation}
where $N_{0}$ is a simple normalization factor
\begin{equation}
    N_0 = \left(\frac{\la1234\ra^2}{\la AB_012\ra\la AB_023\ra\la AB_034\ra\la AB_014\ra}\right)^{-1} \,.
\end{equation}
which ensures that the ${\cal F}_\Gamma(z)$ associated with a single dot (no integration to be done) is equal to 1. 
We use the graphic notation for ${\cal F}_{\Gamma}$ where the graph represents an integral (\ref{F1}) with line $AB_0$ left frozen (denoted by white vertex with cross),
\begin{equation}
\includegraphics[width=0.29\columnwidth]{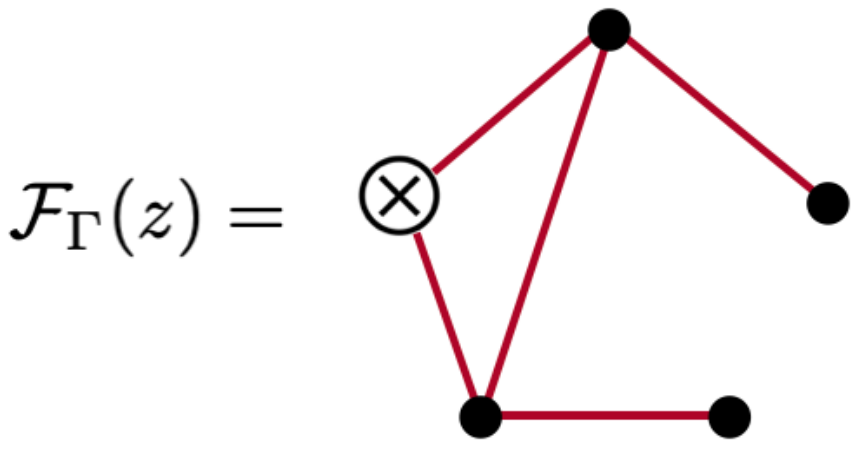}
\end{equation}
For the $L$-loop logarithm of the amplitude we define the IR finite function ${\cal F}_L(z)$ as an integral of $\widetilde{\Omega}_L$ over $AB_1,\dots,AB_{L{-}1}$, which is also the sum of ${\cal F}_\Gamma$ over all connected diagrams with $L$ vertices,
\begin{equation}
    {\cal F}_L(g,z) = N_0\int_{AB_1}\int_{AB_2}\dots \int_{AB_{L{-}1}} \widetilde{\Omega}_L \,.
\end{equation}
Once we label this diagram we have to symmetrize over all loop lines $AB_1,\dots,AB_{L{-}1}$. However, all these loops are integrated, so there is no need to label them in the diagram.

Using this definition we find the graphic expansion for ${\cal F}(g,z)$ starting from $\log\Omega(g)$. For $L=2$ there is only one connected graph,
\begin{equation}
\includegraphics[width=0.27\columnwidth]{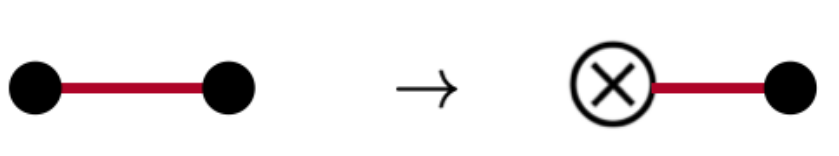} 
\end{equation}
while for $L=3$ we get two connected graphs, 
\begin{equation}
\includegraphics[width=0.5\columnwidth]{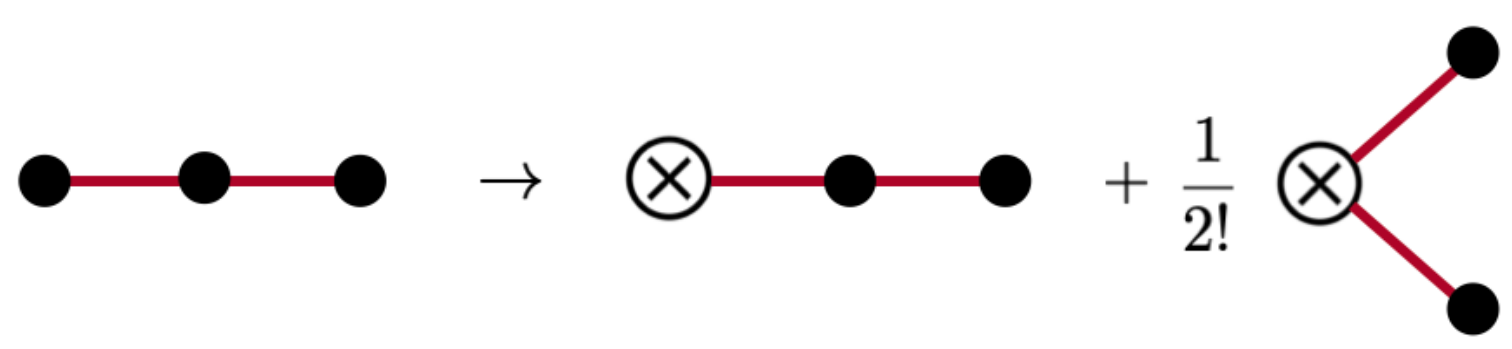}
\end{equation}
\begin{equation}
\includegraphics[width=0.3\columnwidth]{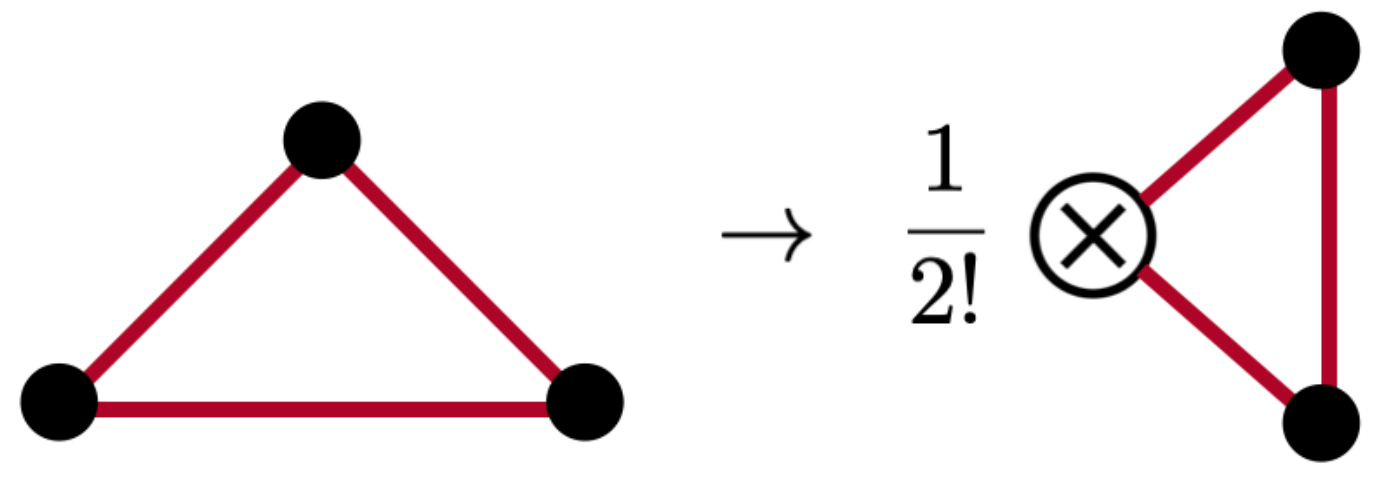}
\end{equation}
In the first figure, we see the symmetry factor $1/2$ which comes from the fact that both the graph on the left and the first term on the right represent two permutations, while the second term on the right is only one permutation. For $L=4$ we have multiple graphs, two of them are
\begin{equation}
\includegraphics[width=0.6\columnwidth]{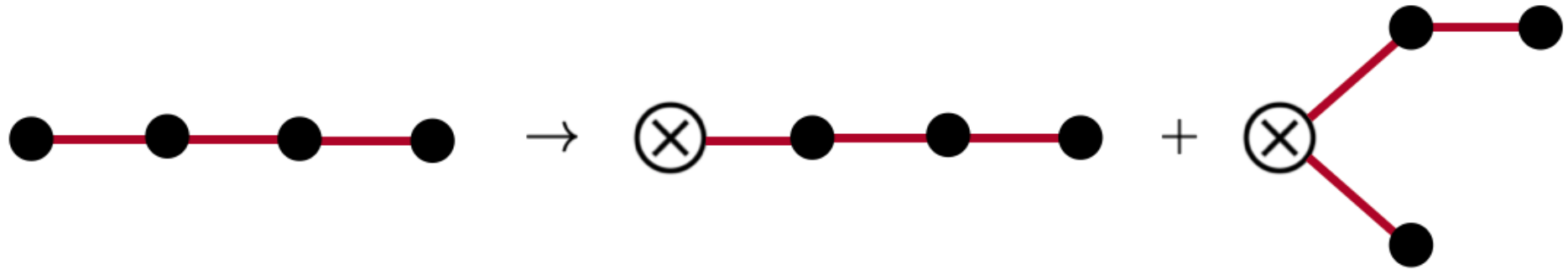}
\end{equation}
\begin{equation}
\includegraphics[width=0.5\columnwidth]{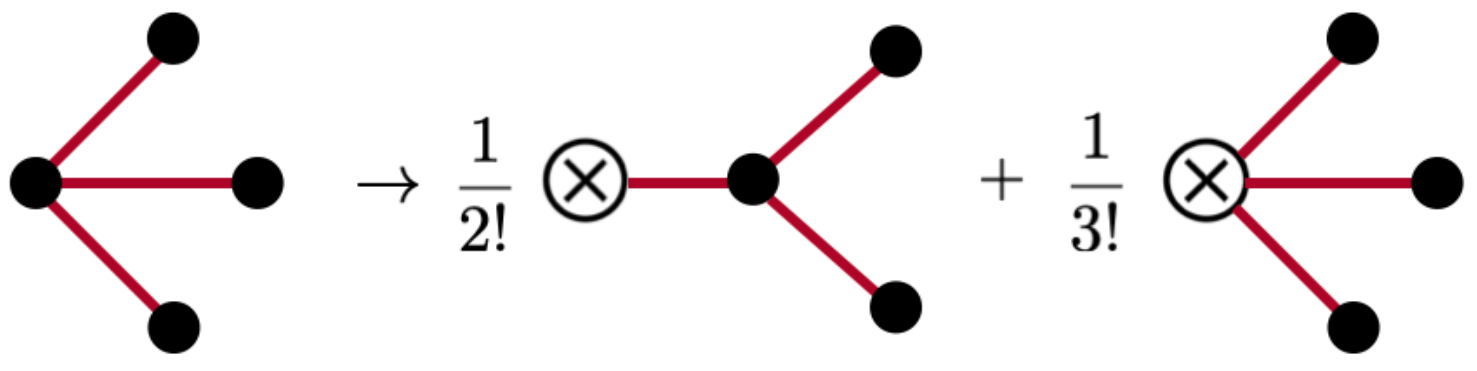}
\end{equation}
As before we can now dress each term with $(g^2)^L$ and sum over $L$ giving us
\begin{equation}
    {\cal F}(g,z) = \sum_{L=1}^\infty (g^2)^L(-1)^{E(G)} {\cal F}_L(g,z)
\end{equation}
In the graphic notation we get,
\begin{equation}
\includegraphics[width=0.8\columnwidth]{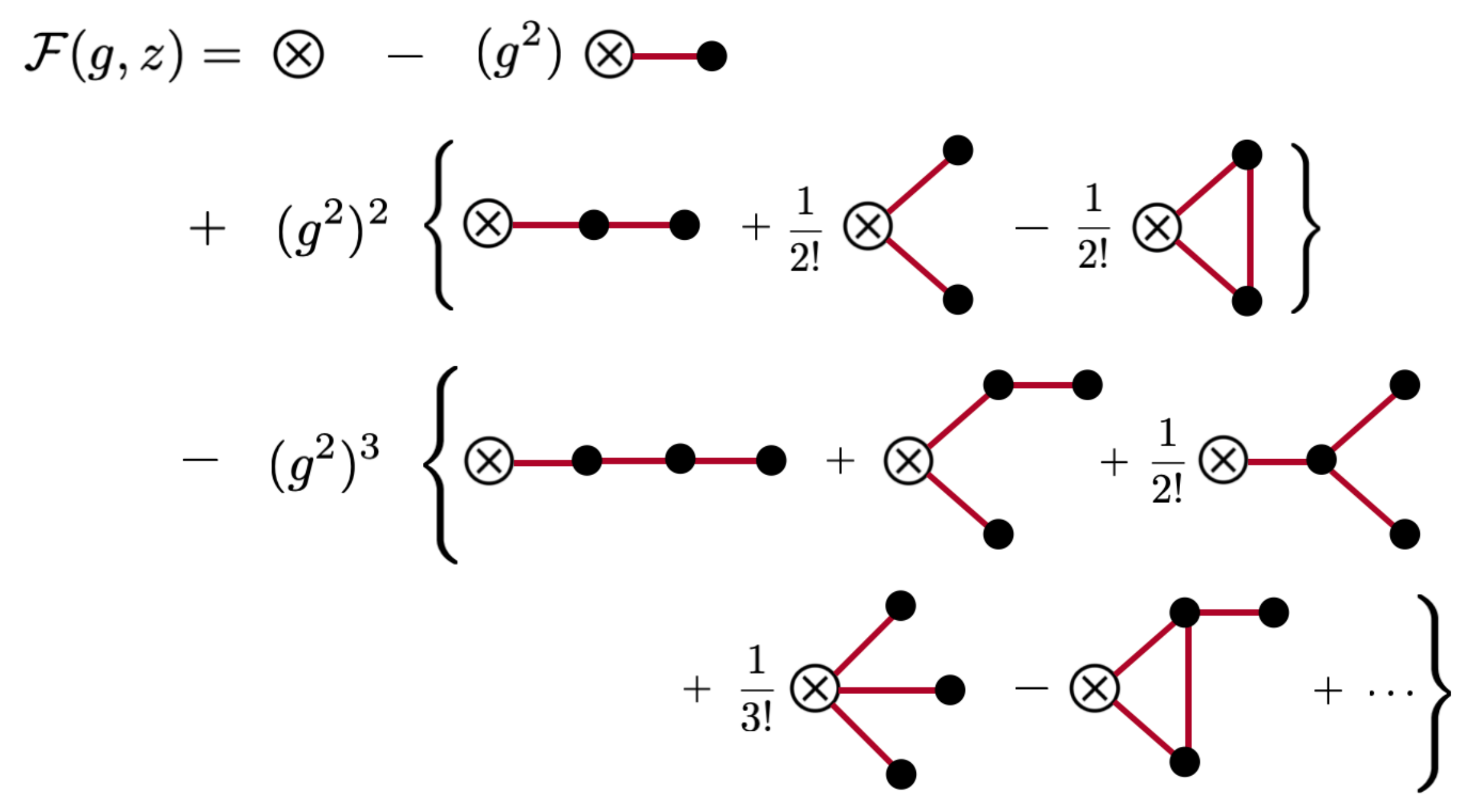}\label{calF}
\end{equation}
The function ${\cal F}(g,z)$ defined by this formula is directly related to $F(g,z)$ defined in the previous section as
\begin{equation}
  F(g,z)   = - g^2 {\cal F}(g,z) \,.
\end{equation}
so these are the same physical quantities (up to some rescaling). The representation (\ref{calF}) is a purely geometric expansion in terms of integrated dlog forms on negative geometries, it has no physical analogue.

\section{Examples of canonical forms and functions associated to negative geometries}
\label{section-examples-forms-and-functions}

The positive geometry is defined by the set of inequalities, positive or negative conditions on the lines $(AB)_j$. The next step is to find the ``dlog'' form $\Omega$ which has logarithmic singularities on the boundaries of the positive geometry. While we have many powerful tools to find these forms \cite{Arkani-Hamed:2010zjl,Arkani-Hamed:2010pyv,Arkani-Hamed:2012zlh,Bourjaily:2013mma,Arkani-Hamed:2014via,Bourjaily:2015jna,Herrmann:2020qlt,Herrmann:2020oud,Henn:2020lye,He:2020uxy,He:2018okq,Arkani-Hamed:2017tmz}, it is hard to find the solution for the general case.

The most natural approach is the triangulation of the space, ie. dividing the full positive space into subregions with trivial dlog forms. This method was followed in \cite{Arkani-Hamed:2013kca} with many non-trivial results. Here our objective is to find the dlog forms for the graphs with negative links. We will see that, surprisingly, for a class of tree graphs we can solve this problem in complete generality.

\subsection{One- and two-loop results}

In this subsection we give some examples of dlog forms associated with graphs $\Omega_\Gamma$ and functions ${\cal F}(z)$. Let us start with the simplest case, $L=1$. The dlog form $\Omega_\Gamma$ is represented by a single dot,
\begin{equation}
\includegraphics[height=1cm,valign=c]{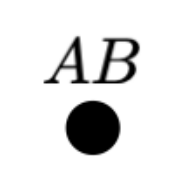}= \frac{d\mu_{AB}\la1234\ra^2}{\la AB12\ra\la AB23\ra\la AB34\ra\la AB14\ra} \,.
\end{equation}
and the associated function ${\cal F}_\Gamma(z)$ is just 1 as there is no integral to be done,
\begin{equation}
\includegraphics[height=1cm,valign=c]{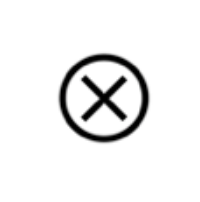}= 1
\end{equation}
For $L=2$ we have a single graph. The integrand form is 
\begin{equation}
\includegraphics[height=1.2cm,valign=c]{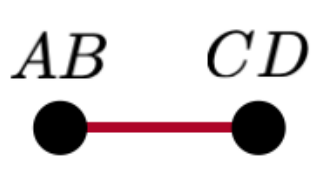} = \frac{d\mu_{AB}d\mu_{CD}\la1234\ra^2n_{12}}{\la AB12\ra\la AB23\ra\la AB34\ra\la AB14\ra\la CD12\ra\la CD23\ra\la CD34\ra\la CD14\ra\la ABCD\ra}
\end{equation}
where the numerator $n_{12}$ is given by
\begin{equation}
 n_{12} = -\la1234\ra\left[\langle AB13\rangle\langle CD24\ra+\langle AB24\rangle\langle CD13\rangle\right] \,. \label{num2}
\end{equation}
After integration over $AB$, renaming $CD\rightarrow AB_0$ we get and factorizing out $N_0$ we get,
\begin{equation}
\includegraphics[height=1.3cm,valign=c]{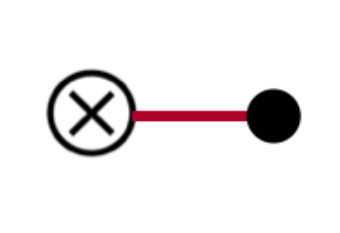} = \log^2z + \pi^2 \,.\label{L2}
\end{equation}

\subsection{Complete three-loop result}

As noted earlier, for $L=3$ we have three different graphs. The associated forms are,
\begin{equation}
\includegraphics[height=2.4cm,valign=c]{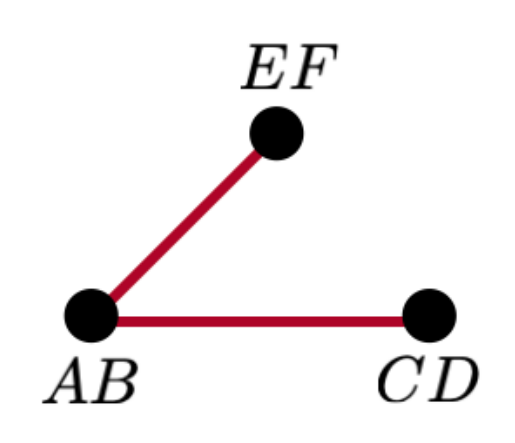}= \frac{d\mu_{AB}d\mu_{CD}d\mu_{EF}\,\la1234\ra^2\, n_{12}n_{13}}{
\begin{array}{c} \la AB12\ra\la AB23\ra\la AB34\ra\la AB14\ra\la CD12\ra\la CD23\ra\la CD34\ra\\ \la CD14\ra\la EF12\ra\la EF23\ra\la EF34\ra\la EF14\ra\la ABCD\ra\la ABEF\ra\end{array}}
\end{equation}
where $n_{12}$ and $n_{13}$ are two-loop numerators (\ref{num2}) where we replace $CD\leftrightarrow EF$ in $n_{12}$ to get $n_{13}$. The second graph is
\begin{equation}
\includegraphics[height=2.4cm,valign=c]{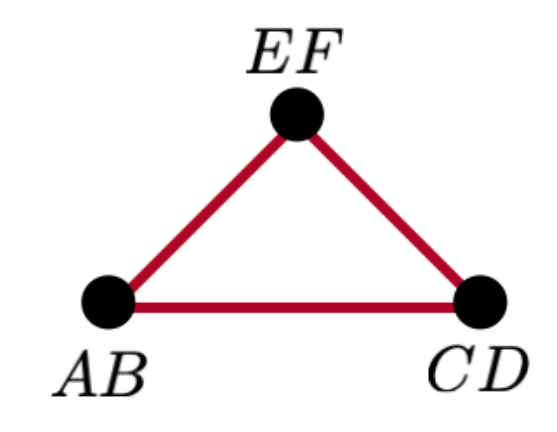}= \frac{d\mu_{AB}d\mu_{CD}d\mu_{EF}\,\la1234\ra^2\,{\cal N}}{
\begin{array}{c} \la AB12\ra\la AB23\ra\la AB34\ra\la AB14\ra\la CD12\ra\la CD23\ra\la CD34\ra\\ \la CD14\ra\la EF12\ra\la EF23\ra\la EF34\ra\la EF14\ra\la ABCD\ra\la ABEF\ra\la CDEF\ra\end{array}}\label{3loop}
\end{equation}
where the numerator ${\cal N}$ can be written as
\begin{equation}
    \mathcal{N} = n_{12}n_{13}n_{23} + {\cal R} \,,\label{L3n}
\end{equation}
where we recognize the product of two-loop numerators between each pair of loops. There is also a ``remainder", ${\cal R}={\cal R}_1 - {\cal R}_2$, which has two terms,
\begin{align}
    {\cal R}_1 &= 4\la AB12\ra\la AB34\ra\la CD12\ra\la CD34\ra\la EF12\ra\la EF34\ra + \mbox{cycl}\\
    {\cal R}_2 &= \la AB12\ra\la AB34\ra  n_{23}[\la CD12\ra\la EF34\ra + \la CD34\ra\la EF12\ra] + \sigma_{AB,CD,EF} + \mbox{cycl}
\end{align}
where $\sigma_{AB,CD,EF}$ denotes the distinct permutation over lines $AB$, $CD$, $EF$, here it is three terms. The symbol ``cycl'' refers to cyclic shift $Z_i\rightarrow Z_{i{+}1}$. 

These forms were not obtained by a direct triangulation, but rather by a different method motivated by \cite{Arkani-Hamed:2014dca}, where we write all poles in the denominator and try to fix the numerator using certain conditions. This happens to be extremely effective in our case as we can always find enough implications of the negativity conditions $\la (AB)_i(AB)_j\ra<0$ which fix the numerator uniquely.

Using the techniques developed in \cite{Henn:2019swt} we find the results for two ``tree" graphs,
\begin{equation}
\includegraphics[height=1.3cm,valign=c]{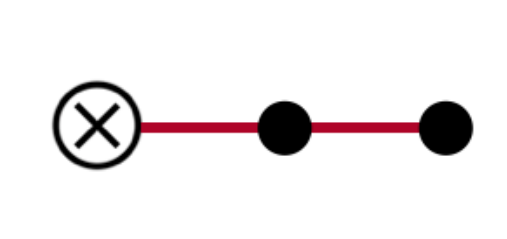}=  \frac{1}{6} \left[ \pi^2 + {\rm log}^2(z) \right] \times \left[ 5 \pi^2 +   \log^2 z \right]\,.
\label{L3a}
\end{equation}
\begin{equation}
\includegraphics[height=2cm,valign=c]{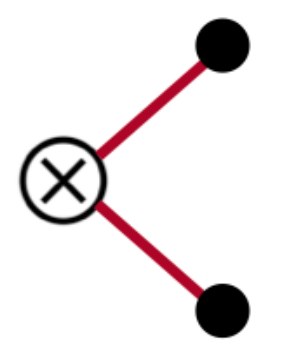}=  \left[ \pi^2 + \log^2 z \right]^2 \,.\label{L3b}
\end{equation}
We will recover these equations in the next section using a differential equations method. Note that (\ref{L3b}) is a square of (\ref{L2}) as there are two independent one-loop integrations.

The expression for the loop graph is more interesting, and we find, 
\begin{equation}
\includegraphics[height=2.2cm,valign=c]{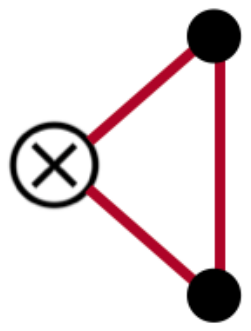} 
\begin{array}{l}
=  8\displaystyle H_{\text{0,0,0,0}}+8 H_{\text{-1,0,0,0}}-16 H_{\text{-1,-1,0,0}}+8 H_{\text{-2,0,0}} - 8 \zeta _3 \left(2 H_{-1}-H_0\right) \\
\hspace{0.5cm} \displaystyle + 4\pi ^2 \left(H_{\text{-1,0}}- H_{\text{-1,-1}}+H_{-2}\right) +\frac{13 \pi ^4}{45}. \end{array}
\end{equation}
Here $H$ are harmonic polylogarithms of argument $z$ \cite{Remiddi:1999ew}. They can be re-expressed in terms of classical polylogarithms of certain arguments, cf. \cite{Alday:2013ip}.
\begin{align}
=& \frac{1}{3} \log^4 z  - 2\log^2 z \left[ -\frac{2}{3} \text{Li}_2\left(\frac{1}{z+1}\right)-\frac{2}{3} \text{Li}_2\left(\frac{z}{z+1}\right)+\frac{\pi ^2}{9} \right] \label{L3c}  \\
&- 2\log z \left[ 4 \text{Li}_3\left(\frac{z}{z+1}\right)-4 \text{Li}_3\left(\frac{1}{z+1}\right) \right] 
+\frac{4}{3} \left[ \text{Li}_2\left(\frac{1}{z+1}\right)+\text{Li}_2\left(\frac{z}{z+1}\right)-\frac{\pi ^2}{6}\right]^2 \nonumber\\
& + \frac{16}{3} \pi ^2 \left[\text{Li}_2\left(\frac{1}{z+1}\right)+\text{Li}_2\left(\frac{z}{z+1}\right)-\frac{\pi ^2}{6}\right]
+16 \text{Li}_4\left(\frac{1}{z+1}\right)+16 \text{Li}_4\left(\frac{z}{z+1}\right)+\frac{\pi ^4}{9}\,. \nonumber
\end{align}
We see that both the forms and the integrals are particularly simple for ``tree'' graphs, while ``loop'' graphs are more involved. In fact, we can find the form for arbitrary tree graphs for any $L$. If we add the contributions (\ref{L3a}), (\ref{L3b}), (\ref{L3c}), dressed with the correct combinatorial factors from (\ref{calF}), we reconstruct the formula for the two-loop function $F^{(2)}(z)$ from section \ref{sec:Wilsonloop}, namely (\ref{Fweak2}). 

There is an interesting question of uniform positivity (or negativity) of expressions we found. This is motivated by \cite{Arkani-Hamed:2014dca,Dixon:2016apl} where the integrands and also the integrated amplitudes were found to be uniformly positive (or negative) for the positive kinematics. It was suggested \cite{Arkani-Hamed:2014dca,Ferro:2015grk,Herrmann:2020qlt} that the positivity is closely related to the existence of the dual Amplituhedron where the amplitudes would correspond to volumes which are naturally positive. Though this relation should only apply to integrands, hence the uniform positivity or negativity of final (integrated) amplitudes is an even more surprising property.

Motivated by this we now inspect the positivity properties of explicit expressions for ${\cal F}_\Gamma$. The kinematics is now represented by a single parameter $z$. Looking at terms (\ref{L3a}), (\ref{L3b}), (\ref{L3c}) we can check that they are all uniformly positive for $z>0$.

Furthermore, if we now look at the function $F(g,z)$ and the perturbative expansion (\ref{Fweak0}), (\ref{Fweak1}), (\ref{Fweak2}), which at each order is given by the sum of expressions for the negative geometries, we see that 
\begin{equation}
    F^{(0)}(z) < 0,\quad F^{(1)}(z) > 0, \quad F^{(2)}(z)<0\qquad \mbox{for $z>0$}
\end{equation}
Based on these (limited) data it is reasonable to conjecture that the $L$-loop function $F^{(L)}(z)>0$ for $L$ odd and $F^{(L)}(z)>0$ for even $L$. This is reminiscent of the manifest positivity (or negativity) in the Amplituhedron triangulations for the integrand. Note that in the expansion (\ref{calF}) for $F^{(2)}(z)$ there are non-trivial cancellations between individual terms as (\ref{L3a}), (\ref{L3b}) contribute with negative sign, while (\ref{L3c}) contributes positively due to the $-1/2$ prefactor.

\subsection{Forms for all tree graphs}

For a general tree graph, the dlog form takes the form
\begin{equation}
    \widetilde{\Omega} = \prod_{k=1}^L \frac{d\mu_k}{D_k} \times \prod_\Gamma N_{ij}\,, \label{tree}
\end{equation}
where the measure for each loop is $d\mu_k = \la (AB)_k\,d^2A_k\ra\la (AB)_k\,d^2B_k\ra$, and we denoted the denominator factor
\begin{equation}
    D_k = \la (AB)_k12\ra\la (AB)_k23\ra\la (AB)_k34\ra\la (AB)_k14\ra\,.
\end{equation}
The last factor is a product of $N_{ij}$ for each edge in the graph, 
\begin{equation}
\includegraphics[height=2.3cm,valign=c]{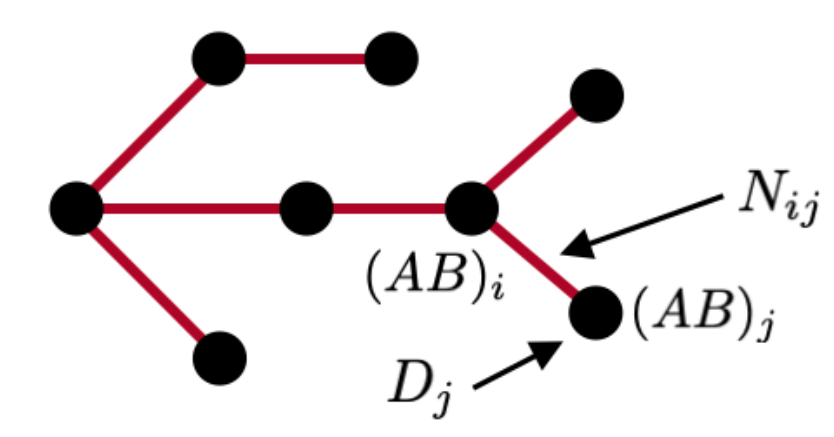}  
\end{equation}
It is given by the two-loop numerator for loops $(AB)_i$ and $(AB)_j$, divided by the propagator $\la (AB)_i(AB)_j\ra$,
\begin{equation}
    N_{ij} = -\frac{(\la (AB)_i13\ra\la (AB)_j24\ra + \la (AB)_i24\ra\la (AB)_k13\ra)}{\la (AB)_i(AB)_j\ra}\,.
\end{equation}
It is surprising that the general $L$-loop dlog form for tree graphs factorizes in this very simple way. 

As an example, we can work out two tree graphs for $L=4$,
\begin{equation}
\includegraphics[height=3cm,valign=c]{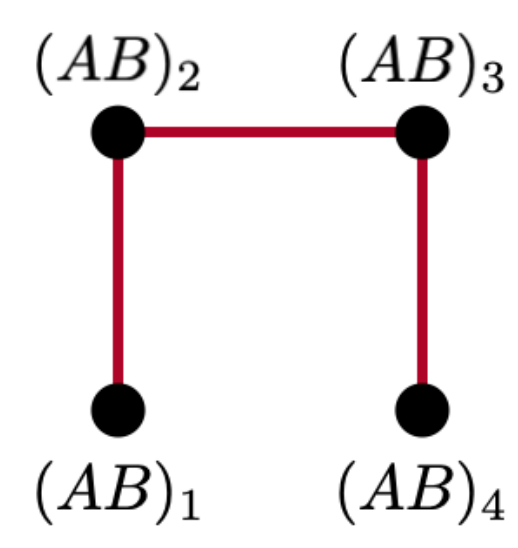}  = \frac{d\mu_1\,d\mu_2\,d\mu_3\,d\mu_4\,N_{12}N_{23}N_{34}}{D_1D_2D_3D_4}
\end{equation}
\begin{equation}
\includegraphics[height=3cm,valign=c]{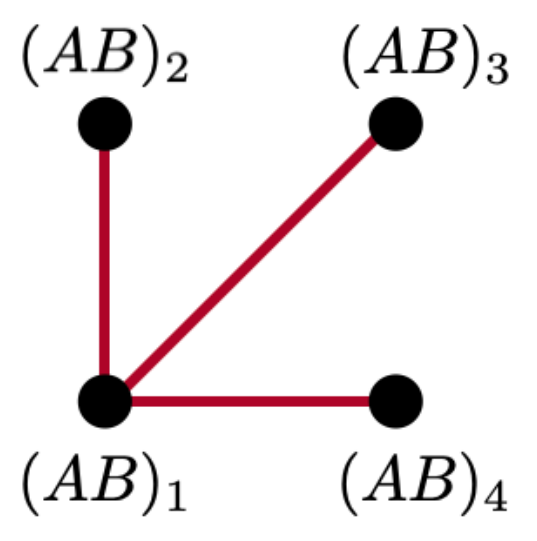}  = \frac{d\mu_1\,d\mu_2\,d\mu_3\,d\mu_4\,N_{12}N_{13}N_{14}}{D_1D_2D_3D_4}
\end{equation}
Finally, let us comment on the derivation and validity of (\ref{tree}). The logic of the expression is simple and it nicely connects to the geometry: thinking about the form $\widetilde{\Omega}$ as the numerator factor over the denominator (set of all poles), the numerator is fixed by two conditions:
\begin{enumerate}
    \item Remove all singularities from the form (cuts) which violate $\la (AB)_i(AB)_j\ra<0$ conditions.
    \item Ensure the form is logarithmic.
\end{enumerate}
The numerator $N_{ij}$ does the first job by construction for a single negative link. Taking the product of all $N_{ij}$ then manifestly removes all the cuts which violate one or more negativity conditions. The second point is more subtle and one has to analyze what type of double (or multiple) poles can appear in the form when evaluated on deeper residues. It is easy to see that all double poles which involve cutting two loops $(AB)_i$ and $(AB)_j$ are removed (again just a consequence that $N_{ij}$ is a two-loop numerator). It is more difficult to see that no additional double pole can appear when we consider cuts involving more than two loops. In this case, it is the tree nature of the graph that prevents these multiple poles to appear. 

In fact, in order to generate such higher poles we need to have a ``closed loop" of negativity links. That is the reason why for $L=3$ the complete graph had more complicated form. If we only consider the first term in (\ref{L3n}), which is the product of two-loop numerators $N_{ij}$, the resulting form has double poles. The additional term ${\cal R}$ in (\ref{L3n}) is there to remove these double poles, while not introducing any forbidden cuts.

\section{Summing ``ladders'' and ``trees''}
\label{sec:resummation}

In this section we focus on tree diagrams and also a special subset of tree diagrams which we call ``ladders" because of some similarities with $\phi^3$ ladder diagrams. 

\subsection{Generating functions}

In section \ref{sec:geometry}, we have seen that $F$, or equivalently ${\cal F}(g,z)$, 
admits a geometric expansion. To each geometry one associates a canonical form, which in turn integrates to a finite function of $z$.
We have seen examples of canonical forms and integrated function in section \ref{section-examples-forms-and-functions}.

Calculating ${\cal F}(g,z)$ for any value of coupling is hard: it requires both finding the forms for general connected graphs, and integrating these forms. Because of the simplifications we saw in the last section, we define a new object ${\cal F}_{\rm tree}(g,z)$ which is a perturbative sum of only tree graphs,
\begin{equation}\label{defFtree}
\includegraphics[width=0.95\columnwidth]{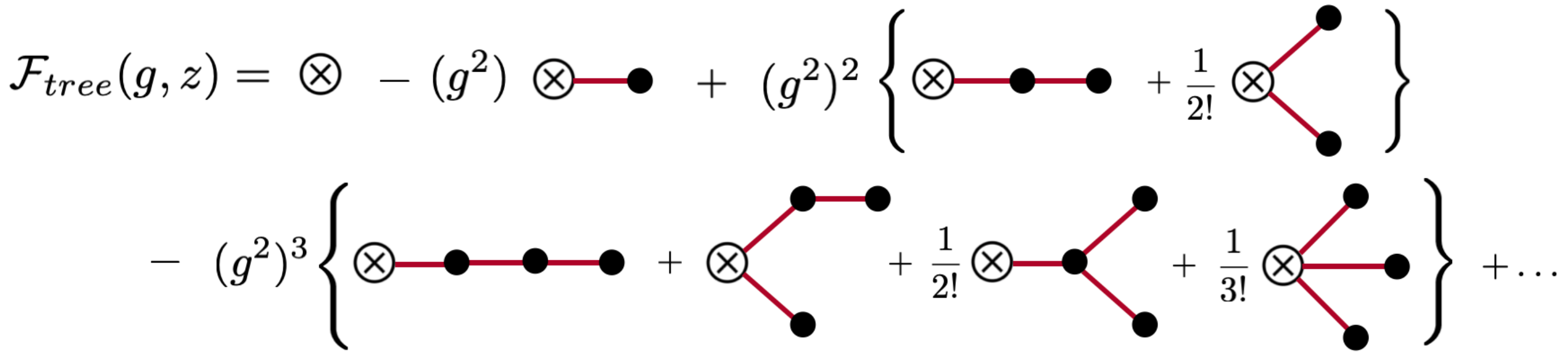}
\end{equation}
It is also useful to define ${\cal H}_{\rm tree}(z)$ to be the generating function for all graphs that begin with the $(AB)_0$ vertex, and attach via a single line to another vertex, before branching out into any tree-structure they like,
\begin{equation}
\includegraphics[width=0.7\columnwidth]{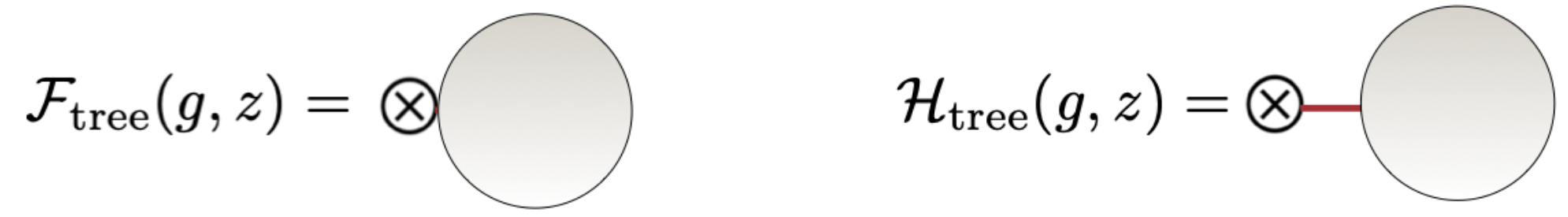}
\end{equation}
The perturbative expansion for ${\cal H}_{\rm tree}(g,z)$ is
\begin{equation}\label{defHtree}
\includegraphics[width=0.57\columnwidth]{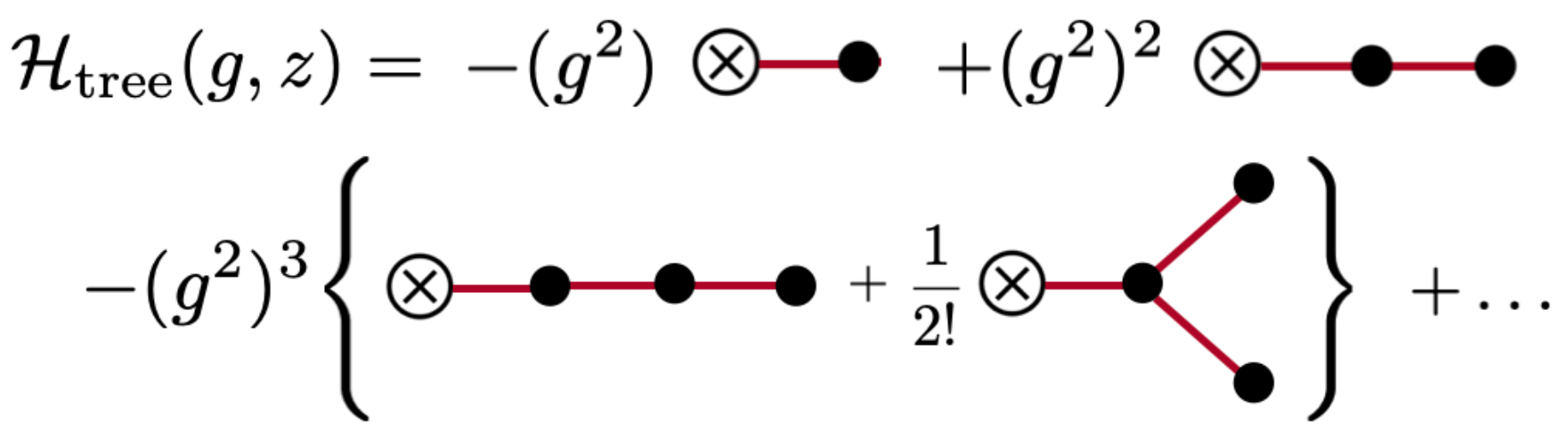}
\end{equation}
Now, there is a very simple relationship between these two generating functions; we have that
\begin{equation}\label{relationFtoH}
{\cal F}_{\rm tree}(g) = e^{{\cal H}_{\rm tree}(g)}\,.
\end{equation}
This relation can be proved by explicitly expanding the exponential and collecting all terms. 

Note that ${\cal F}_{\rm tree}(g,z)$ is a certain approximation of ${\cal F}(g,z)$ when we omit the graphs with internal loops. A priori, it is not clear how good this approximation is as we obviously omit a vast majority of diagrams. We will study this in detail in the following subsections.

Finally, we will define one more quantity ${\cal F}_{\rm ladder}(g,z)$, which is a further simplification of ${\cal F}_{\rm tree}(g,z)$ when only ``ladder'' graphs are considered,
\begin{equation}\label{calFladder}
\includegraphics[width=0.75\columnwidth]{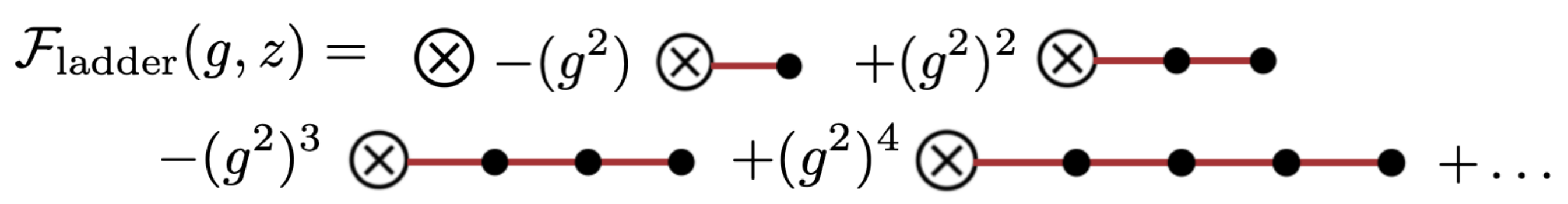}
\end{equation}
We are now ready to turn to resumming all the ``ladder'' and ``tree'' diagrams. Instead of the direct integration of the dlog forms associated with ladder or tree negative geometries, we derive differential equations for these quantities.

\subsection{`Laplace'-type differential equation for ``tree'' and ``ladder'' geometries}

To begin with, we recall that ${\cal F}^{(0)} = 1$.
Now, the simplest non-trivial ``tree''-diagram is at one loop.
As shown in section \ref{sec:geometry}, it is given by
\begin{equation}\label{eq:curlyF1}
{\cal F}^{(1)} = \,\, \includegraphics[width=0.08\columnwidth]{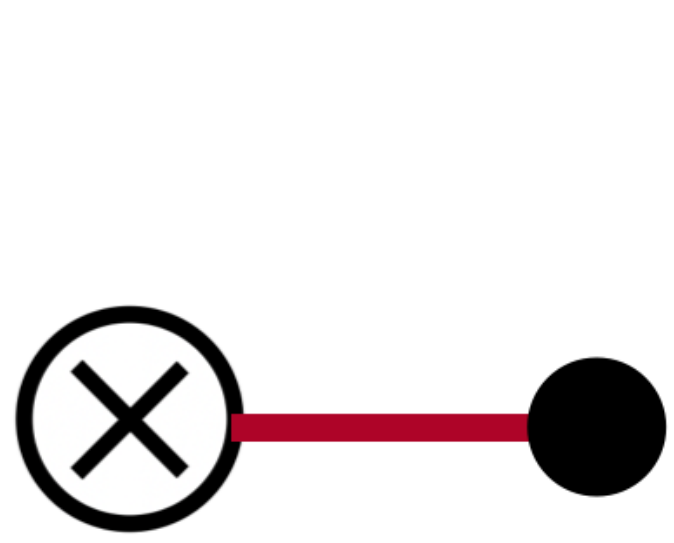}  \,\, = \int_{CD} \frac{\vev{1234} [ \vev{AB 13} \vev{ CD 24} + \vev{AB 24} \vev{CD 13}]}{ \vev{AB CD} \vev{CD 12} \vev{CD 23} \vev{CD 34} \vev{CD 41}}\,.
\end{equation}
We recall that by abuse of notation, we use the same graphical notation for differential forms, and integrated function. The meaning should be clear from the context.

It is easy to see that this graph satisfies a Laplace equation \cite{Drummond:2006rz,Drummond:2010cz}, very similar to the one for the ladder diagrams in $\phi^3$ theory.
The key point is that the dependence on $AB$ is especially simple; with only a linear factor upstairs and a pole downstairs, connecting $AB$ to a single other vertex $CD$. Because of this, it satisfies an extremely simply ``boxing'' equation, which says in effect that if we write 
\begin{align}
{\cal F}^{(1)} = \vev{AB 13} {\cal F}^{(1)}_{24} + \vev{AB 24} {\cal F}^{(1)}_{13}\,,
\end{align}
then acting with the Laplace operator on ${\cal F}^{(1)}_{24}$ or ${\cal F}^{(1)}_{13}$ simply collapses the propagator connecting $AB$ to $CD$, giving us a new graph where $CD$ is replaced with $AB$. In this way, we have, schematically,
\begin{equation}
\includegraphics[width=0.28\columnwidth]{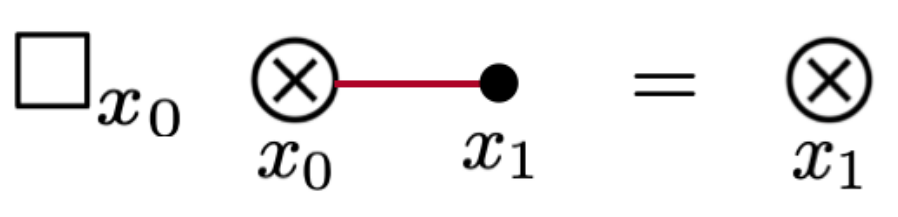}
\end{equation}
Since ${\cal F}$ is a function of a single variable only, we can rewrite this equation in the following concise form,
\begin{align}\label{DEpentagon}
\frac{1}{2} (z \partial_z)^2 {\cal F}^{(1)}  + {\cal F}^{(0)} = 0\,.
\end{align}
Let us see this in more detail. Following \cite{Drummond:2006rz,Drummond:2010cz}, we are considering loop integrals of the form 
\begin{align}\label{eq:Ggeneric}
  G =& \int_{CD} \frac{\langle A B 13 \rangle \langle C D 24 \rangle \langle 1 2 3 4 \rangle}{\langle CD 12 \rangle \langle CD 23 \rangle \langle CD 34 \rangle \langle CD 41 \rangle \langle CD AB \rangle } H\left(  \frac{\vev{12CD} \vev{34CD}}{\vev{14CD}\vev{23CD}} \right)\,.
  \end{align}
In order to use the Laplace equation trick, we switch to dual notation, writing $x\sim CD, x_0 \sim AB$, and $\vev{AB13} = \vev{AB} \vev{13} (x_0 - x_{+})^2$, $\vev{CD24} \equiv \vev{CD} \vev{24} (x-x_{-})^2$. In this way, we have
  \begin{align}
 G =& 
  \frac{(x_0 - x_{+}^2) \langle 13 \rangle \langle 24 \rangle \langle 1234 \rangle}{\langle 12 \rangle \vev{23} \vev{34} \langle 41 \rangle} \times \nonumber \\
  &
  \int \frac{d^4 x}{i \pi^2} \frac{(x-x_{-})^2 }{(x-x_1)^2 \ldots (x-x_4)^2 (x -x_0)^2} H\left( \frac{x_{13}^2 (x-x_2)^2 (x-x_4)^2}{x_{24}^2 (x-x_1)^2 (x-x_3)^2} \right) \,.
\end{align}
Clearly if we divide the above by $(x_0 - x_{13})^2$, acting with $\Box_{x_0}$ will simply collapse the propagator 
\begin{align}
    \Box_{x_0} \frac{1}{(x - x_0)^2} =  -4 i \pi^2 \delta^4(x - x_0)\,.
\end{align} 
In this way we trivially have
\begin{equation}\label{eq:detailbox1}
    (x_0 - x_{+})^2 \Box_{x_0} \left[ \frac{1}{(x_0 - x_{+})^2} {G}(z) \right] = 4 \left( \frac{x_{24}^2}{x_{02}^2 x_{04}^2} + \frac{x_{13}^2}{x_{01}^2 x_{03}^2} \right) H(z)
\end{equation}
We can then simply use the chain rule to simplify the LHS of eq. (\ref{eq:detailbox1}), for any function $G(z)$. The result is 
\begin{align}\label{eq:resultchainrule}
(z \partial_z)^2 { G}(z) = - H(z)\,.
\end{align}
Let us now apply this result to our  function ${\cal F}^{(1)}$ given in eq. (\ref{eq:curlyF1}).
It corresponds to $G$ of eq. (\ref{eq:Ggeneric})  with $H=1$, up to one detail: note that it is the sum of two terms, with the $\langle AB 13 \rangle$ and $\langle AB 24 \rangle$ numerators. However, eq. (\ref{eq:resultchainrule}) holds for each of them separately, and hence for the sum we find eq. (\ref{DEpentagon}), as claimed.
This is our manifestation of the ``boxing trick''. 

Looking at the canonical form for ${\cal F}_{\rm ladder}$, cf. eq. (\ref{calFladder}), we thus see that the Laplace operator trick also works for us, and we obtain
\begin{align}\label{DEladders}
\frac{1}{2}  (z \partial_z)^2  {\cal F}_{\rm ladder} + g^2 {\cal F}_{\rm ladder} = 0\,.
\end{align}
This differential equation can be easily exactly solved; but we need to know two boundary conditions. The first one is obvious: under cyclic action, we have that $z \to 1/z$, and thus we should have an even function of $\log(z)$. The second one is slightly more interesting, it turns out to be ${\cal F}_{\rm ladder}(z \to -1) = 1$. To see this, note that the numerators always have factors of $(AB_0 13)$ or $(AB_0 24)$, and hence the integrals always vanish when $(1 + z) = (AB_013)(AB_024)/(AB_023)(AB_041) \to 0$. 
The solution reads
\begin{align}\label{exactFladder}
{\cal F}_{\rm ladder} = \frac{\cos(\sqrt{2} g \log z)}{\cosh(\sqrt{2} g \pi)} \,.
\end{align}
Now, let us generalize this to  the tree-graphs that are contained in the generating function ${\cal H}_{\rm tree}$.  
Recall that the integrand takes a special form, cf. eqs. (\ref{defFtree}) and (\ref{defHtree}).
Repeating the above steps, it is easy to see that the action of the Laplace operator
simply collapses the propagator connecting $(AB)_0$ to $(AB)_1$, giving us a new graph where $(AB)_1$ is replaced with $(AB)_0$.
For example
\begin{equation}
\includegraphics[width=0.55\columnwidth]{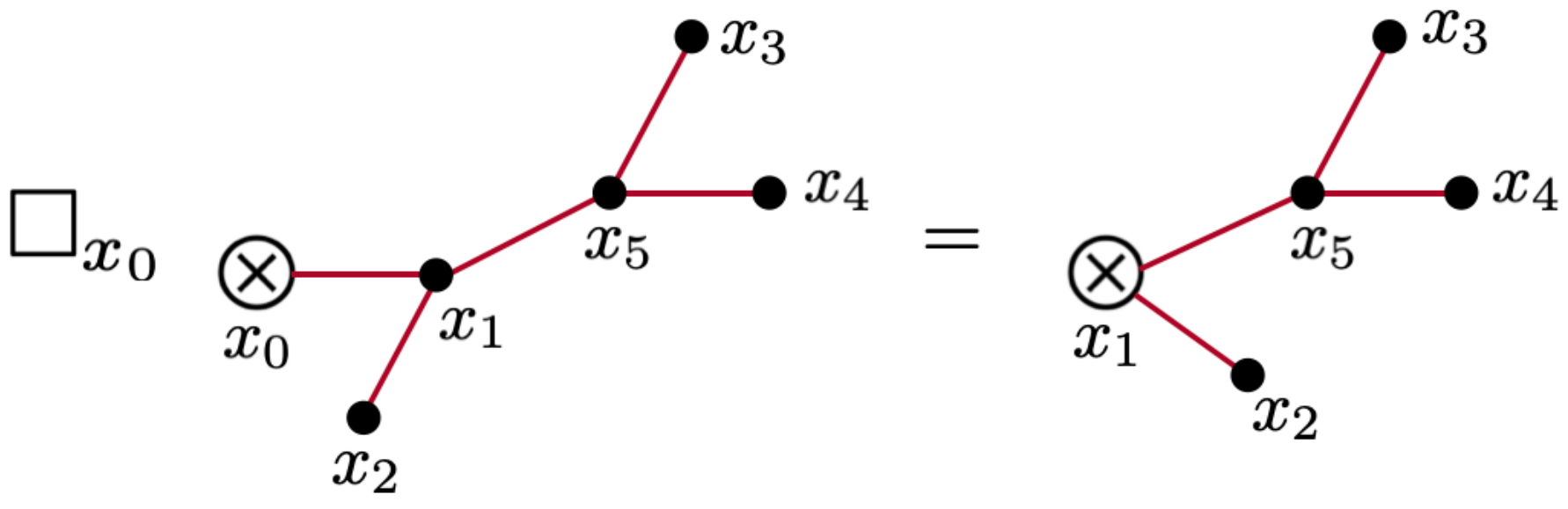}
\end{equation}
and at the level of the generating functions we have, schematically, 
\begin{equation}\label{boxingDE2}
\includegraphics[width=0.5\columnwidth]{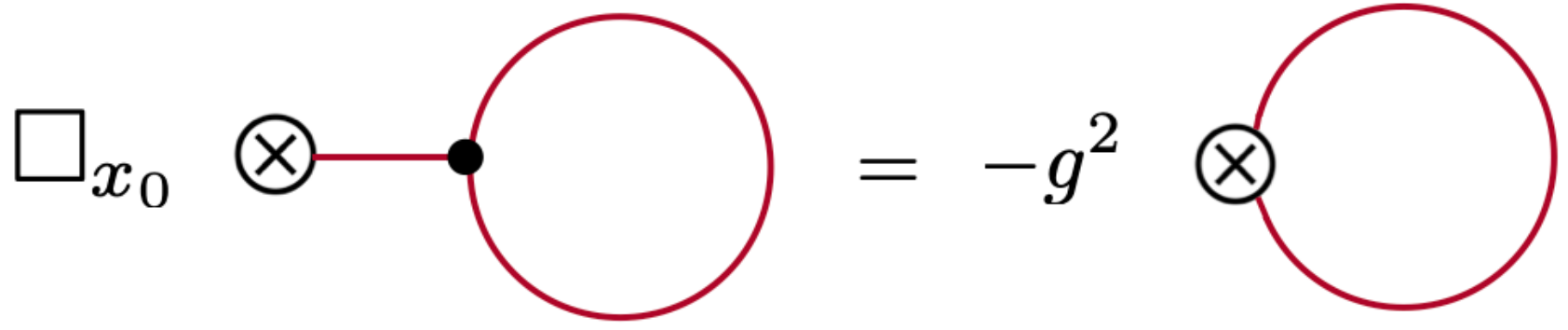}
\end{equation}
Given that all the integrals only depend on the single variable $z$, and recalling eq. (\ref{relationFtoH}), we can rewrite eq. (\ref{boxingDE2}) as
\begin{equation}\label{DEtrees}
\frac{1}{2} (z \partial_z)^2 {\cal H}_{\rm tree} + g^2 e^{{\cal H}_{\rm tree}} = 0\,.
\end{equation}
Note that this equation has some similarity to the equation for resumming the $\phi^3$ ladders, eq. (\ref{DEladdersphi3}), that we began our discussion with, except it is importantly nonlinear, involving $e^{\cal H_{\rm tree}}$. This shows us concretely where the vague ``extra direction'' we alluded to can come from: not from extra kinematic variables, but from a panoply of different ``negative geometric regions'', which at least at ``tree-level'', are easily captured combinatorially by the exponential. We will return later to discuss a bit more what precisely is being captured by this ``extra-direction'', non-linear structure.

The differential equation (\ref{DEtrees}) can be easily exactly solved, with the boundary conditions ${\cal H}_{\rm tree}(z \to -1) = 0$ and ${\cal H}_{\rm tree}(1/z) = {\cal H}_{\rm tree}(z)$. We find
\begin{equation}\label{exactanswertrees_maintext}
{\cal F}_{\rm tree}(g,z) = \frac{A^2}{g^2} \frac{z^A}{(z^A + 1)^2}\,,\quad {\rm where} \,\quad \frac{A}{2 g {\rm cos} \frac{\pi A}{2}} = 1 \,,
\end{equation}
where the integration constant $A$ is determined by imposing the second boundary condition, which implies that ${\cal F}_{\rm tree}(z \to -1)  =1$.

\subsection{Computation of $\Gamma_{\rm ladder}$ and $\Gamma_{\rm tree}$}

Having obtained the result for ${\cal F}_{\rm ladder}$ and ${\cal F}_{\rm tree}$, we turn to computing the corresponding contributions to the cusp anomalous dimenstion. To whit, recall the translation is most easily made when $F(z)$ is expressed as a sum over powers of $z$. Note that 
\begin{align}
{\cal F}_{tree} = A z \partial_z \frac{1}{1 + z^A} =  A \sum_{m=0}^\infty (m A) (-1)^m z^{m A}\,,
\end{align}
and recalling that the rules given in eqs. (\ref{extractgammacusp}) and (\ref{rulefunctional-intro})
we have that 
\begin{align}
g \partial_g  \Gamma_{\rm tree} =& - 2 A \sum_{m=0}^{\infty} (-1)^m (m A) \frac{{\rm sin}(\pi m A)}{\pi m A} \nonumber\\
=& - 2 \frac{A}{\pi} \sum_{m=0}^\infty (-1)^m {\rm sin}(\pi m A)
\end{align}
The infinite sum can be performed as a geometric series by writing sin$(k x) \to \frac{1}{2i} (w^k - w^{-k})$ with $w = e^{i x}$. Note that for real $x$ the sum is evaluated exactly on the radius of convergence $|w|=1$ of $\sum w^{\pm k} = \frac{1}{1 - w^{\pm 1}}$. We find
\begin{align}
\sum_{k = 1}^{\infty} (-1)^{k} {\sin( k x )} {=} - \frac{1}{2} \tan \frac{x}{2} \,
\end{align}
We can of course obtain exactly the same result without expanding in powers of $z^p$, but using the form of ${\cal I}[F]$ given in eq. (\ref{eq:functionalform2}) for a general function $F$. Using that ${\cal F}_{{\rm tree}}$ is a total derivative, the integral over the contour ${\cal C}$ trivializes to a difference between the endpoints at $x=-1$, and we find $\frac{A}{\pi} (\frac{1}{1 + e^{- i \pi A}} - \frac{1}{1 + e^{i \pi A}}) = \frac{A}{\pi} {\rm tan} \frac{A \pi}{2}$.

Employing this formula, 
we finally obtain
\begin{align}\label{Gammatreeexactmain}
g \partial_g \Gamma_{\rm tree}   = \frac{4 A}{\pi} { \tan}(\pi A/2) \,.
\end{align}
We can readily integrate this equation to obtain $\Gamma_{{\rm tree}}(g)$. Using the chain rule we have $\partial_A \Gamma_{{\rm tree}} = 2 {\rm tan} (A \frac{\pi}{2}) (\frac{2}{\pi} + A {\rm tan}(\frac{A \pi}{2}))$. The boundary condition that $\Gamma_{{\rm tree}} \to 0$ as $g \to 0$ also tells us that it vanishes as $A \to 0$, so we can integrate to obtain $\Gamma_{{\rm tree}}$: 
\begin{equation}
    \Gamma_{{\rm tree}} = A \left(\frac{4}{\pi}  {\rm tan} \frac{\pi A}{2} - A \right)\,.
\end{equation}
In a similar way we can associate to $F_{\rm ladder}$ its contribution to $\Gamma_{\rm ladder}$, which has the extremely simple form 
\begin{equation}\label{Gammaladderexact}
\Gamma_{\rm  ladder}(g) = \frac{4}{\pi} {\rm log}\, {\rm cosh} (\sqrt{2} \pi g) \,.
\end{equation}
It is amusing that almost the same expression has occurred as another sort of ``anomalous dimension'', the so-called ``$\Gamma_{\rm octagon}$'' \cite{Coronado:2018cxj,Belitsky:2020qrm} controlling the remainder function for six-particle scattering in a particular limit \cite{Basso:2020xts}, and also appears in scattering amplitudes on the Coulomb branch in a certain limit \cite{Caron-Huot:2021usw}.
The two differ simply by rescaling $g$: 
\begin{equation}
\Gamma_{\rm  ladder}(g) = 2\, \Gamma_{\rm octagon}(\frac{g}{\sqrt{2}})\,.
\end{equation}

\subsection{Analytic structure, weak and strong coupling asymptotics}

Finally, let us analyze the exact formulas we obtained for $F_{\rm  ladder}$, $F_{\rm  tree}$, $\Gamma_{\rm  ladder}$, and $\Gamma_{\rm  tree}$.

It is straightforward to expand ${\cal F}_{\rm  ladder}$, given in eq. (\ref{exactFladder}), at weak coupling,
\begin{align}
{\cal F}_{\rm  ladder}(g;z) = 1 + g^2 (-\pi^2-\log^2 z) +\frac{1}{6} g^4 (\pi^2+ \log^2 z) (5 \pi^2 + \log^2 z) + {\cal O}(g^6) \,.
\end{align}
At strong coupling, we see that, similar to the $\phi^3$ ladders \cite{Broadhurst:2010ds}, the result wildly oscillating but exponentially suppressed,
\begin{align}
|{\cal F}_{\rm  ladder}(g;z)| \le \frac{1}{\cosh(\sqrt{2} g \pi) }  \le 2 e^{-\sqrt{2} g \pi}\,\,\, \stackrel{g \gg 1}{\to}\,\,\, 0 \,.
\end{align}
This is very different from the full strong coupling result, given in eq. (\ref{Fstrong}), which behaves as ${F} \sim g$.

Let us turn analyze ${\cal F}_{\rm  tree}$. The asymptotics of $A$ in eq. (\ref{exactanswertrees_maintext}) for small and large $g$ are readily understood.  We find that $A$ has the following expansions,
\begin{equation}\label{Aasymptotics}
A \to \left\{ \begin{array}{cc} 2 g - \pi^2 g^3 +  \frac{13}{12} \pi^4 g^5 + \cdots & g \ll 1 \\ 1 - 1/(g \pi) + 1/(g \pi)^2 + \cdots & g \gg 1 \end{array} \right. \,.
\end{equation}
It is easy to see why we get an expansion in powers of $g^2$ for small g, but an expansion in powers of $1/g$ for large g. The function $A/\cos(\pi A/2)$ is odd in $A$; but when we expand around $g \to 0$, we are expanding around $A \to 0$, and so this is reflected in a symmetry of the coefficients under $g \to -g$. However, when we expand around large $g \to + \infty$, we are breaking that $g \to -g$ symmetry, and so the expansion of $A$ does not reflect any symmetry in $g$, and has all powers of $1/g$. Note further that while $A$ is completely analytic around the origin, the expansion in power of $1/g$ is asymptotic, with corrections of order $\exp(-g)$.

Given eq. (\ref{Aasymptotics}), we readily find the weak coupling expansion,
\begin{align}
{\cal F}_{\rm  tree}(g;z) = 1 + g^2 (-\pi^2-\log^2 z) +\frac{2}{3} g^4 (\pi^2+ \log^2 z) (2 \pi^2 + \log^2 z) + {\cal O}(g^6) \,.
\end{align}
At strong coupling, we find
\begin{align}
F_{\rm tree} = -\frac{z}{(1+z)^2} + {\cal O}\left(\frac{1}{g} \right) \,.
\end{align}
This result is closer in structure to the exact result than what we found with the ``ladder'' approximation. Unlike the ladder which is exponentially small at strong coupling, it asymptotes to $g^0$, and has corrections that are a series expansion in $1/g$. Of course it still disagrees with the exact result which scales as $g^1$ at strong coupling; clearly the negative geometries with ``loops'' are important for both the correct qualitative and quantitative behaviors of $F(g^2,z)$ at strong coupling.

Interestingly, while $F(g^2,z)$ itself differs significantly between the ``ladders'', ``trees'', and the known exact results, the expression for $\Gamma$ extracted from them are much more similar. Indeed the expression (\ref{Gammatreeexactmain}) has all the correct behavior seen in the exact result for $\Gamma_{\rm tree}$ at both weak and strong coupling, in detail 
\begin{equation}
\Gamma_{\rm tree}(g) \to \left\{ \begin{array}{cc} 4 g^2 - 8 \zeta_2 g^4 + \cdots & g \ll 1 \\ 
 \frac{8}{\pi} g + \frac{1}{ \pi} \frac{1}{g} + \cdots & g \gg 1 \end{array}  \right. \,.
\end{equation}
It is easy to determine the radius of convergence of the expansion for $\Gamma_{tree}(g)$. Note that when $g$ is real we can always find solutions to $A/(2 \cos(\pi A/2)) = g$. But suppose g is pure imaginary, $g = i \bar{g}$. Then we have to have that $A$ is pure imaginary $A = i \bar{A}$, and we must solve 
\begin{equation} 
\frac{\bar{A}}{2 {\rm cosh}(\pi \bar{A}/2)} = \bar{g} \,.
\end{equation}
For small $\bar{g}$ we can always solve this equation and indeed find two solutions; but as a function of $\bar{A}$, the function $\frac{\bar{A}}{2 {\rm cosh}(\pi \bar{A}/2)}$ goes up from the origin and then down at infinity, with a maximum value of $0.21\cdots$ in between. Right when  $\bar{g} =0.21\cdots$ equals this maximum value, the two solutions for $\bar{g}$ collider and for larger $\bar{g}$ we can't find solutions, so $\Gamma_{tree}$, which is a function of $A(g)$, must have a branch point at $g = i \times 0.21 \cdots$.  

The asymptotics of $\Gamma_{\rm  ladder}(g)$ are also  interesting. We have an expansion in powers of $g^2$ for small $g$, and asymptotes to being linear in $g$ at large $g$. But there are no ``string loop'' corrections in powers of $1/g$ at large $g$; instead we only have the non-perturbative corrections of order $\exp(-g)$:
\begin{equation}
\Gamma_{\rm  ladder}(g) \to \left\{ \begin{array}{cc} 4 g^2 - 8 \zeta_2 g^4 + \cdots & g \ll 1 \\  4 \sqrt{2} g - 4 \frac{{\rm log}2}{\pi} + \frac{4}{\pi} e^{-2 \sqrt{2} g \pi} + \cdots & g \gg 1 \end{array} \right. \,.
\end{equation}
The analytic structure in the $g$ plane also differs significantly from the exact result for $\Gamma_{\rm cusp}$, with an infinite series of poles on the imaginary $g$ axis, instead of a branch cut. Finally, the radius of convergence of $\Gamma_{\rm ladder}$ is $|g_*| = \sqrt{2}/4 = 0.35\cdots$. Note this is {\it larger} than the radius of convergence $|g_*|=1/4$ of the exact result; so $\Gamma_{\rm ladder}$ is ``less non-perturbative'' than the exact result, in contrast with $\Gamma_{\rm tree}(g)$ for which as we saw above, $|g_*| = 0.21\cdots$ and is ``more non-perturbative'' than the exact result. 

It is also interesting to ask, how do the coefficients in the perturbative expansions of $\Gamma_{\rm cusp}$, $\Gamma_{\rm ladder}$, and $\Gamma_{\rm tree}$ compare numerically?
Given the series expansions $\Gamma_{a} = \sum_{k \ge 1} g^{2k} c_{a,k}$, we can look at the ratos of coefficients $c_{a,k}$;  
a deviation from $1$ gives us a measure of how close we are from the true answer. 
The ratio of the coefficients between the exact result and the ``ladder'' approximation are

\begin{align}
r_{{\rm ladder}/{\rm cusp}} =  \begin{tabular}{|c|c|c|c|c|c|c|c|}
\hline
$g^2$ & $g^4$ & $g^6$ & $g^8$ & $g^{10}$ & $g^{12}$ & $g^{14}$ & $\cdots$ \\ \hline $1$ & $1$ & $0.73$ & $0.44$ & $0.25$ & $0.14$ & $0.07$ & $\cdots$ \\
\hline
\end{tabular} 
\end{align}

The first two coefficients are unity by construction.
The ``tree'' result appears to be larger than the cusp anomalous dimension, so here we prefer to take the inverse ratio,

\begin{align}
r_{ {\rm cusp}/{\rm tree} } =  \begin{tabular}{|c|c|c|c|c|c|c|c|}
\hline
$g^2$ & $g^4$ & $g^6$ & $g^8$ & $g^{10}$ & $g^{12}$ & $g^{14}$ & $\cdots$ \\ \hline 1 & 1 & 0.92 & 0.83 & 0.74 & 0.63 & 0.53 & $\cdots$ \\
\hline
\end{tabular} 
\end{align}

Just to illustrate what the numbers mean, for example the coefficients at $g^{12}$ numerically are roughly $-128921$ (cusp) and $-203607$ (tree), respectively, so they are of the same order of magnitude. For the ladder case, the same coefficient is approximately $17707$, so it differs already by one order of magnitude.

It is peculiar that the functions ${\cal F}$ which are so different at large $g$ between ``ladders'', ``trees'' and the exact result, can be so similar in extracting the coefficient of the leading IR divergence when integrating over $z$ to extract $\Gamma_{\rm cusp}$! This is presumably due to the fact that the functional form of $F(z)$ differs qualitatively between the examples, which using the algebraic procedure for extracting $\Gamma$ from the expansion of $F$ in powers of $z$ produces relatively similar expressions for $\Gamma$. For instance, from the $\cos(\sqrt{2} g {\rm log} z)$ factor in ${\cal F}_{\rm ladder}(z)$ shows wild oscillation in $z$ at large $g$, which is absent in both $F_{\rm ladder}(z)$ and the exact $F(z)$. Meanwhile $F_{\rm tree}(z)$ has a pole at $z=-1$, which is absent in the exact result which instead has a pole at $z=1$. It is interesting that these significant differences appear to be largely washed out in extracting $\Gamma$ from ${\cal F}$.

\section{Outlook}
\label{sec:outlook}

In this paper we have seen that the positive geometry of the loop-amplituhedron for four-point scattering can naturally be understood as a sum over complementary negative geometries, which in turn can be recognized as the exponential of the sum over ``connected'' negative geometries. Quite beautifully, while the full positive geometry has boundaries corresponding to all the soft and collinear singularities of the amplitude, and hence integrates to something with the usual $1/\epsilon^{2L}$ IR divergences at $L$ loops, the negative geometries only have these boundaries when {\it all} $L$ loops are taken into the soft-collinear region, and hence each connected negative geometry only has the mildest possible $1/\epsilon^2$ divergence. This gives a purely positive-geometric understanding of the exponentiation of IR divergences, and an intrinsic geometric motivation for defining log $M$ as an interesting object of study. Furthermore, if just one loop is left un-integrated, the remaining $(L-1)$ integrations are guaranteed to yield a finite function $F(g,z)$, dually interpreted as the expectation value of the Wilson loop with a single Lagrangian insertion, with a perturbative  expansion in given by polylogarithms in $z$. This provides a long-sought connection between each negative geometry and their canonical ``dlog'' forms in loop momentum space, and {\it finite} polylogs for integated expressions, un-infected by IR divergences. 

The negative geometries do not have a direct association with planar Feynman diagrams; only the complete sum over all geometries is associated with the final physical observable $F(g,z)$. While our starting point was in terms of planar amplitudes, the individual terms in the negative geometry expansion are intrinsically non-planar. This expansion is reminiscent of the non-Abelian exponentiation property for Wilson loops, where the logarithm of the Wilson loop is given directly in terms of certain ``maximally non-Abelian web'' diagrams. In the case of Wilson loops with a non-light-like cusp, each for the web diagrams only has an overall $1/\epsilon_{\rm UV}$ divergences, as opposed to $1/\epsilon^L_{\rm UV}$ poles for a generic $L$-loop diagram. However, there is also a difference: if one takes cusp to be formed of light-like segments, the logarithm of the Wilson loops has a double pole, $1/\epsilon^2_{\rm UV}$, but individual webs typically have higher divergences. As far as we are aware, it is not known in general how to organize the web expansion in terms of manifestly $1/\epsilon^2_{\rm UV}$ divergent terms. (See reference \cite{Erdogan:2011yc} for interesting work in this direction.) Our geometric expansion does provide such a decomposition for the scattering amplitude integrals that are dual to the light-like Wilson loops. It would be fascinating to understand more closely the relationship between webs and our geometric expansion.

We then studied the canonical forms and integrated expressions associated with an infinite class of negative geometries with a ``tree'' structure for the negativity conditions, and showed how this infinite set of contributions can be resummed via a simple non-linear differential equations, at all values of the 't Hooft coupling, leading to non-perturbative expressions for these contributions $F_{{\rm tree}}(g,z)$ and $\Gamma_{{\rm tree}}(g,z)$. 
It is remarkable that an essentially trivial aspect of amplituhedron combinatorics which determines $\Gamma_{{\rm tree}}(g)$ already captures all the qualitative features (and is not terribly far off quantitatively) of the weak $\to$ strong coupling behavior of $\Gamma_{{\rm cusp}}(g)$. This is encouraging, since we also know that the positive geometry of the amplituhedron is not restricted to supersymmetric theories. The geometry of the amplituhedron captures the most complicated universal parts of {\it all} the long-distance/IR singularities of planar amplitudes in any ``adjoint-like'' planar theory.  
The novelty of non-SUSY theories is the presence of {\it new} kinds of singularities, in the UV, that are not present in ${\cal N}=4$ sYM, for instance reflected in the UV running of coupling constants. But we expect that the more universal aspects of amplituhedral geometries, especially the ``negative geometries'' associated with the log of the amplitude, might extend to more general theories, and the sort of weak to strong extrapolation we have found here might be a more robust and general phenomenon. 

On the other hand, the non-perturbative expressions for $F_{{\rm tree}}(g,z)$ itself does {\it not} capture all the correct qualitative behavior at strong coupling--missing the leading $O(g)$ behavior, so clearly the more interesting negative geometries with ``loops of loops'' for negativity conditions, must play an important role in the full physics, but in a way that is interestingly largely washed out in extracting $\Gamma_{\rm cusp}$. It would be interesting to find a more systematic way of thinking about the performing the loop integrations associated with the negative geometries with closed loops, perhaps beginning with the simplest case of ``one-loop'' graphs. Clearly a more systematic way of thinking about the loop integrations is needed. For trees, the ``boxing trick'' we used in this paper shows a simple recursive relationship between the integrated expressions for different negative-geometry-graphs: a simple differential operator on a more complicated tree reduces to a simpler tree by collapsing an edge. Optimistically, one might hope that a recursive structure of this type generalize for any graph, with some simple operation on more complicated graphs being determined by simpler ones. Since all these geometries are associated with finite integrals, at the very least we have a perfectly well-posed mathematical question--closely connected to understanding the cohomology of these negative geometric spaces-- that has a plausible chance of having a natural answer, and which could give the correct generalization for the ``new directions'' we have already seen reflected in our non-linear differential equations arising from the recursive structure of the ``boxing trick'' just for ``tree'' graphs. 

As we have stressed, the truncation to negative geometries that are ``trees'' is natural from the perspective of positive geometries, but does not have an obvious diagrammatic interpretation. It would be interesting to see whether there is a physically natural deformation of ${\cal N} = 4$ sYM amplitudes that might correspond to only keeping these geometries.  
Along these lines, it would be instructive to consider the finite Wilson loops with Lagrangian insertion for higher multiplicity $n$ of particles \cite{progress2}, for which explicit results at $n=5$ are forthcoming \cite{progress3}. This observable also has an expansion in terms of the negative geometries \cite{progress4}, and we again expect a significant simplification of the analysis for ``tree'' negative geometries, so it would be interesting to understand what the strong coupling limit of this truncation to ``trees'' looks like for general $n$, which might tell us more about a possible physical interpretation of this sector of geometries.

\acknowledgments

We thank Fernando Alday, Benjamin Basso, Simon Caron-Huot, Lance Dixon and Matthias Staudacher for useful discussions and comments. N.A-H. is supported by a DOE grant No. SC0009988. J.T. is supported by the DOE grant No. SC0009999. 
This research received funding from the European Research Council (ERC) under the European Union's Horizon 2020 research and innovation programme (grant agreement No 725110), {\it Novel structures in scattering amplitudes}.

\bibliographystyle{JHEP}
\bibliography{refs.bib}

\providecommand{\href}[2]{#2}\begingroup\raggedright\begin{thebibliography}{10}

\bibitem{Maldacena:1997re}
J.~M. Maldacena, \emph{{The Large N limit of superconformal field theories and
  supergravity}}, \href{https://doi.org/10.1023/A:1026654312961}{\emph{Adv.
  Theor. Math. Phys.} {\bfseries 2} (1998) 231}
  [\href{https://arxiv.org/abs/hep-th/9711200}{{\ttfamily hep-th/9711200}}].

\bibitem{Beisert:2006ez}
N.~Beisert, B.~Eden and M.~Staudacher, \emph{{Transcendentality and Crossing}},
  \href{https://doi.org/10.1088/1742-5468/2007/01/P01021}{\emph{J. Stat. Mech.}
  {\bfseries 0701} (2007) P01021}
  [\href{https://arxiv.org/abs/hep-th/0610251}{{\ttfamily hep-th/0610251}}].

\bibitem{Bern:2006ew}
Z.~Bern, M.~Czakon, L.~J. Dixon, D.~A. Kosower and V.~A. Smirnov, \emph{{The
  Four-Loop Planar Amplitude and Cusp Anomalous Dimension in Maximally
  Supersymmetric Yang-Mills Theory}},
  \href{https://doi.org/10.1103/PhysRevD.75.085010}{\emph{Phys. Rev. D}
  {\bfseries 75} (2007) 085010}
  [\href{https://arxiv.org/abs/hep-th/0610248}{{\ttfamily hep-th/0610248}}].

\bibitem{Benna:2006nd}
M.~K. Benna, S.~Benvenuti, I.~R. Klebanov and A.~Scardicchio, \emph{{A Test of
  the AdS/CFT correspondence using high-spin operators}},
  \href{https://doi.org/10.1103/PhysRevLett.98.131603}{\emph{Phys. Rev. Lett.}
  {\bfseries 98} (2007) 131603}
  [\href{https://arxiv.org/abs/hep-th/0611135}{{\ttfamily hep-th/0611135}}].

\bibitem{Basso:2007wd}
B.~Basso, G.~P. Korchemsky and J.~Kotanski, \emph{{Cusp anomalous dimension in
  maximally supersymmetric Yang-Mills theory at strong coupling}},
  \href{https://doi.org/10.1103/PhysRevLett.100.091601}{\emph{Phys. Rev. Lett.}
  {\bfseries 100} (2008) 091601}
  [\href{https://arxiv.org/abs/0708.3933}{{\ttfamily 0708.3933}}].

\bibitem{Kruczenski:2002fb}
M.~Kruczenski, \emph{{A Note on twist two operators in N=4 SYM and Wilson loops
  in Minkowski signature}},
  \href{https://doi.org/10.1088/1126-6708/2002/12/024}{\emph{JHEP} {\bfseries
  12} (2002) 024} [\href{https://arxiv.org/abs/hep-th/0210115}{{\ttfamily
  hep-th/0210115}}].

\bibitem{Roiban:2007dq}
R.~Roiban and A.~A. Tseytlin, \emph{{Strong-coupling expansion of cusp anomaly
  from quantum superstring}},
  \href{https://doi.org/10.1088/1126-6708/2007/11/016}{\emph{JHEP} {\bfseries
  11} (2007) 016} [\href{https://arxiv.org/abs/0709.0681}{{\ttfamily
  0709.0681}}].

\bibitem{Basso:2021omx}
B.~Basso, L.~J. Dixon, D.~A. Kosower, A.~Krajenbrink and D.-l. Zhong,
  \emph{{Fishnet four-point integrals: integrable representations and
  thermodynamic limits}},
  \href{https://doi.org/10.1007/JHEP07(2021)168}{\emph{JHEP} {\bfseries 07}
  (2021) 168} [\href{https://arxiv.org/abs/2105.10514}{{\ttfamily
  2105.10514}}].

\bibitem{Broadhurst:2010ds}
D.~J. Broadhurst and A.~I. Davydychev, \emph{{Exponential suppression with four
  legs and an infinity of loops}},
  \href{https://doi.org/10.1016/j.nuclphysbps.2010.09.014}{\emph{Nucl. Phys. B
  Proc. Suppl.} {\bfseries 205-206} (2010) 326}
  [\href{https://arxiv.org/abs/1007.0237}{{\ttfamily 1007.0237}}].

\bibitem{Drummond:2006rz}
J.~M. Drummond, J.~Henn, V.~A. Smirnov and E.~Sokatchev, \emph{{Magic
  identities for conformal four-point integrals}},
  \href{https://doi.org/10.1088/1126-6708/2007/01/064}{\emph{JHEP} {\bfseries
  01} (2007) 064} [\href{https://arxiv.org/abs/hep-th/0607160}{{\ttfamily
  hep-th/0607160}}].

\bibitem{Drummond:2010cz}
J.~M. Drummond, J.~M. Henn and J.~Trnka, \emph{{New differential equations for
  on-shell loop integrals}},
  \href{https://doi.org/10.1007/JHEP04(2011)083}{\emph{JHEP} {\bfseries 04}
  (2011) 083} [\href{https://arxiv.org/abs/1010.3679}{{\ttfamily 1010.3679}}].

\bibitem{Alday:2011ga}
L.~F. Alday, E.~I. Buchbinder and A.~A. Tseytlin, \emph{{Correlation function
  of null polygonal Wilson loops with local operators}},
  \href{https://doi.org/10.1007/JHEP09(2011)034}{\emph{JHEP} {\bfseries 09}
  (2011) 034} [\href{https://arxiv.org/abs/1107.5702}{{\ttfamily 1107.5702}}].

\bibitem{Engelund:2011fg}
O.~T. Engelund and R.~Roiban, \emph{{On correlation functions of Wilson loops,
  local and non-local operators}},
  \href{https://doi.org/10.1007/JHEP05(2012)158}{\emph{JHEP} {\bfseries 05}
  (2012) 158} [\href{https://arxiv.org/abs/1110.0758}{{\ttfamily 1110.0758}}].

\bibitem{Engelund:2012re}
O.~T. Engelund and R.~Roiban, \emph{{Correlation functions of local composite
  operators from generalized unitarity}},
  \href{https://doi.org/10.1007/JHEP03(2013)172}{\emph{JHEP} {\bfseries 03}
  (2013) 172} [\href{https://arxiv.org/abs/1209.0227}{{\ttfamily 1209.0227}}].

\bibitem{Gatheral:1983cz}
J.~G.~M. Gatheral, \emph{{Exponentiation of Eikonal Cross-sections in
  Nonabelian Gauge Theories}},
  \href{https://doi.org/10.1016/0370-2693(83)90112-0}{\emph{Phys. Lett. B}
  {\bfseries 133} (1983) 90}.

\bibitem{Frenkel:1984pz}
J.~Frenkel and J.~C. Taylor, \emph{{Nonabelian Eikonal Exponentiation}},
  \href{https://doi.org/10.1016/0550-3213(84)90294-3}{\emph{Nucl. Phys. B}
  {\bfseries 246} (1984) 231}.

\bibitem{Korchemskaya:1992je}
I.~A. Korchemskaya and G.~P. Korchemsky, \emph{{On lightlike Wilson loops}},
  \href{https://doi.org/10.1016/0370-2693(92)91895-G}{\emph{Phys. Lett. B}
  {\bfseries 287} (1992) 169}.

\bibitem{Alday:2013ip}
L.~F. Alday, J.~M. Henn and J.~Sikorowski, \emph{{Higher loop mixed correlators
  in N=4 SYM}}, \href{https://doi.org/10.1007/JHEP03(2013)058}{\emph{JHEP}
  {\bfseries 03} (2013) 058} [\href{https://arxiv.org/abs/1301.0149}{{\ttfamily
  1301.0149}}].

\bibitem{Alday:2007hr}
L.~F. Alday and J.~M. Maldacena, \emph{{Gluon scattering amplitudes at strong
  coupling}}, \href{https://doi.org/10.1088/1126-6708/2007/06/064}{\emph{JHEP}
  {\bfseries 06} (2007) 064} [\href{https://arxiv.org/abs/0705.0303}{{\ttfamily
  0705.0303}}].

\bibitem{Drummond:2007aua}
J.~M. Drummond, G.~P. Korchemsky and E.~Sokatchev, \emph{{Conformal properties
  of four-gluon planar amplitudes and Wilson loops}},
  \href{https://doi.org/10.1016/j.nuclphysb.2007.11.041}{\emph{Nucl. Phys. B}
  {\bfseries 795} (2008) 385}
  [\href{https://arxiv.org/abs/0707.0243}{{\ttfamily 0707.0243}}].

\bibitem{Brandhuber:2007yx}
A.~Brandhuber, P.~Heslop and G.~Travaglini, \emph{{MHV amplitudes in N=4 super
  Yang-Mills and Wilson loops}},
  \href{https://doi.org/10.1016/j.nuclphysb.2007.11.002}{\emph{Nucl. Phys. B}
  {\bfseries 794} (2008) 231}
  [\href{https://arxiv.org/abs/0707.1153}{{\ttfamily 0707.1153}}].

\bibitem{Drummond:2007cf}
J.~M. Drummond, J.~Henn, G.~P. Korchemsky and E.~Sokatchev, \emph{{On planar
  gluon amplitudes/Wilson loops duality}},
  \href{https://doi.org/10.1016/j.nuclphysb.2007.11.007}{\emph{Nucl. Phys. B}
  {\bfseries 795} (2008) 52} [\href{https://arxiv.org/abs/0709.2368}{{\ttfamily
  0709.2368}}].

\bibitem{Arkani-Hamed:2013jha}
N.~Arkani-Hamed and J.~Trnka, \emph{{The Amplituhedron}},
  \href{https://doi.org/10.1007/JHEP10(2014)030}{\emph{JHEP} {\bfseries 10}
  (2014) 030} [\href{https://arxiv.org/abs/1312.2007}{{\ttfamily 1312.2007}}].

\bibitem{Arkani-Hamed:2013kca}
N.~Arkani-Hamed and J.~Trnka, \emph{{Into the Amplituhedron}},
  \href{https://doi.org/10.1007/JHEP12(2014)182}{\emph{JHEP} {\bfseries 12}
  (2014) 182} [\href{https://arxiv.org/abs/1312.7878}{{\ttfamily 1312.7878}}].

\bibitem{Franco:2014csa}
S.~Franco, D.~Galloni, A.~Mariotti and J.~Trnka, \emph{{Anatomy of the
  Amplituhedron}}, \href{https://doi.org/10.1007/JHEP03(2015)128}{\emph{JHEP}
  {\bfseries 03} (2015) 128} [\href{https://arxiv.org/abs/1408.3410}{{\ttfamily
  1408.3410}}].

\bibitem{Arkani-Hamed:2017vfh}
N.~Arkani-Hamed, H.~Thomas and J.~Trnka, \emph{{Unwinding the Amplituhedron in
  Binary}}, \href{https://doi.org/10.1007/JHEP01(2018)016}{\emph{JHEP}
  {\bfseries 01} (2018) 016}
  [\href{https://arxiv.org/abs/1704.05069}{{\ttfamily 1704.05069}}].

\bibitem{Arkani-Hamed:2018rsk}
N.~Arkani-Hamed, C.~Langer, A.~Yelleshpur~Srikant and J.~Trnka, \emph{{Deep
  Into the Amplituhedron: Amplitude Singularities at All Loops and Legs}},
  \href{https://doi.org/10.1103/PhysRevLett.122.051601}{\emph{Phys. Rev. Lett.}
  {\bfseries 122} (2019) 051601}
  [\href{https://arxiv.org/abs/1810.08208}{{\ttfamily 1810.08208}}].

\bibitem{Damgaard:2019ztj}
D.~Damgaard, L.~Ferro, T.~Lukowski and M.~Parisi, \emph{{The Momentum
  Amplituhedron}}, \href{https://doi.org/10.1007/JHEP08(2019)042}{\emph{JHEP}
  {\bfseries 08} (2019) 042}
  [\href{https://arxiv.org/abs/1905.04216}{{\ttfamily 1905.04216}}].

\bibitem{Coronado:2018cxj}
F.~Coronado, \emph{{Bootstrapping the Simplest Correlator in Planar $\mathcal N
  = 4$ Supersymmetric Yang-Mills Theory to All Loops}},
  \href{https://doi.org/10.1103/PhysRevLett.124.171601}{\emph{Phys. Rev. Lett.}
  {\bfseries 124} (2020) 171601}
  [\href{https://arxiv.org/abs/1811.03282}{{\ttfamily 1811.03282}}].

\bibitem{Belitsky:2020qrm}
A.~V. Belitsky and G.~P. Korchemsky, \emph{{Octagon at finite coupling}},
  \href{https://doi.org/10.1007/JHEP07(2020)219}{\emph{JHEP} {\bfseries 07}
  (2020) 219} [\href{https://arxiv.org/abs/2003.01121}{{\ttfamily
  2003.01121}}].

\bibitem{Basso:2020xts}
B.~Basso, L.~J. Dixon and G.~Papathanasiou, \emph{{Origin of the Six-Gluon
  Amplitude in Planar $N=4$ Supersymmetric Yang-Mills Theory}},
  \href{https://doi.org/10.1103/PhysRevLett.124.161603}{\emph{Phys. Rev. Lett.}
  {\bfseries 124} (2020) 161603}
  [\href{https://arxiv.org/abs/2001.05460}{{\ttfamily 2001.05460}}].

\bibitem{Caron-Huot:2021usw}
S.~Caron-Huot and F.~Coronado, \emph{{Ten dimensional symmetry of $N$ = 4 SYM
  correlators}},  \href{https://arxiv.org/abs/2106.03892}{{\ttfamily
  2106.03892}}.

\bibitem{Henn:2019swt}
J.~M. Henn, G.~P. Korchemsky and B.~Mistlberger, \emph{{The full four-loop cusp
  anomalous dimension in $\mathcal{N}=4$ super Yang-Mills and QCD}},
  \href{https://doi.org/10.1007/JHEP04(2020)018}{\emph{JHEP} {\bfseries 04}
  (2020) 018} [\href{https://arxiv.org/abs/1911.10174}{{\ttfamily
  1911.10174}}].

\bibitem{Progress1}
N.~Arkani-Hamed, A.~Hillman and S.~Mizera, \emph{{To appear}}, .

\bibitem{Arkani-Hamed:2010zjl}
N.~Arkani-Hamed, J.~L. Bourjaily, F.~Cachazo, S.~Caron-Huot and J.~Trnka,
  \emph{{The All-Loop Integrand For Scattering Amplitudes in Planar N=4 SYM}},
  \href{https://doi.org/10.1007/JHEP01(2011)041}{\emph{JHEP} {\bfseries 01}
  (2011) 041} [\href{https://arxiv.org/abs/1008.2958}{{\ttfamily 1008.2958}}].

\bibitem{Caron-Huot:2010ryg}
S.~Caron-Huot, \emph{{Notes on the scattering amplitude / Wilson loop
  duality}}, \href{https://doi.org/10.1007/JHEP07(2011)058}{\emph{JHEP}
  {\bfseries 07} (2011) 058} [\href{https://arxiv.org/abs/1010.1167}{{\ttfamily
  1010.1167}}].

\bibitem{Eden:2010zz}
B.~Eden, G.~P. Korchemsky and E.~Sokatchev, \emph{{From correlation functions
  to scattering amplitudes}},
  \href{https://doi.org/10.1007/JHEP12(2011)002}{\emph{JHEP} {\bfseries 12}
  (2011) 002} [\href{https://arxiv.org/abs/1007.3246}{{\ttfamily 1007.3246}}].

\bibitem{Alday:2012hy}
L.~F. Alday, P.~Heslop and J.~Sikorowski, \emph{{Perturbative correlation
  functions of null Wilson loops and local operators}},
  \href{https://doi.org/10.1007/JHEP03(2013)074}{\emph{JHEP} {\bfseries 03}
  (2013) 074} [\href{https://arxiv.org/abs/1207.4316}{{\ttfamily 1207.4316}}].

\bibitem{Henn:2020omi}
J.~M. Henn, \emph{{What can we learn about QCD and collider physics from $N$=4
  super Yang-Mills?}},
  \href{https://doi.org/10.1146/annurev-nucl-102819-100428}{\emph{Ann. Rev.
  Nucl. Part. Sci.} {\bfseries 71} (2021) 87}
  [\href{https://arxiv.org/abs/2006.00361}{{\ttfamily 2006.00361}}].

\bibitem{Hodges:2009hk}
A.~Hodges, \emph{{Eliminating spurious poles from gauge-theoretic amplitudes}},
  \href{https://doi.org/10.1007/JHEP05(2013)135}{\emph{JHEP} {\bfseries 05}
  (2013) 135} [\href{https://arxiv.org/abs/0905.1473}{{\ttfamily 0905.1473}}].

\bibitem{Arkani-Hamed:2010pyv}
N.~Arkani-Hamed, J.~L. Bourjaily, F.~Cachazo and J.~Trnka, \emph{{Local
  Integrals for Planar Scattering Amplitudes}},
  \href{https://doi.org/10.1007/JHEP06(2012)125}{\emph{JHEP} {\bfseries 06}
  (2012) 125} [\href{https://arxiv.org/abs/1012.6032}{{\ttfamily 1012.6032}}].

\bibitem{Arkani-Hamed:2012zlh}
N.~Arkani-Hamed, J.~L. Bourjaily, F.~Cachazo, A.~B. Goncharov, A.~Postnikov and
  J.~Trnka, \emph{{Grassmannian Geometry of Scattering Amplitudes}}. Cambridge
  University Press, 4, 2016,
  \href{https://doi.org/10.1017/CBO9781316091548}{10.1017/CBO9781316091548},
  [\href{https://arxiv.org/abs/1212.5605}{{\ttfamily 1212.5605}}].

\bibitem{Bourjaily:2013mma}
J.~L. Bourjaily, S.~Caron-Huot and J.~Trnka, \emph{{Dual-Conformal
  Regularization of Infrared Loop Divergences and the Chiral Box Expansion}},
  \href{https://doi.org/10.1007/JHEP01(2015)001}{\emph{JHEP} {\bfseries 01}
  (2015) 001} [\href{https://arxiv.org/abs/1303.4734}{{\ttfamily 1303.4734}}].

\bibitem{Arkani-Hamed:2014via}
N.~Arkani-Hamed, J.~L. Bourjaily, F.~Cachazo and J.~Trnka, \emph{{Singularity
  Structure of Maximally Supersymmetric Scattering Amplitudes}},
  \href{https://doi.org/10.1103/PhysRevLett.113.261603}{\emph{Phys. Rev. Lett.}
  {\bfseries 113} (2014) 261603}
  [\href{https://arxiv.org/abs/1410.0354}{{\ttfamily 1410.0354}}].

\bibitem{Bourjaily:2015jna}
J.~L. Bourjaily and J.~Trnka, \emph{{Local Integrand Representations of All
  Two-Loop Amplitudes in Planar SYM}},
  \href{https://doi.org/10.1007/JHEP08(2015)119}{\emph{JHEP} {\bfseries 08}
  (2015) 119} [\href{https://arxiv.org/abs/1505.05886}{{\ttfamily
  1505.05886}}].

\bibitem{Herrmann:2020qlt}
E.~Herrmann, C.~Langer, J.~Trnka and M.~Zheng, \emph{{Positive geometry, local
  triangulations, and the dual of the Amplituhedron}},
  \href{https://doi.org/10.1007/JHEP01(2021)035}{\emph{JHEP} {\bfseries 01}
  (2021) 035} [\href{https://arxiv.org/abs/2009.05607}{{\ttfamily
  2009.05607}}].

\bibitem{Herrmann:2020oud}
E.~Herrmann, C.~Langer, J.~Trnka and M.~Zheng, \emph{{Positive Geometries for
  One-Loop Chiral Octagons}},
  \href{https://arxiv.org/abs/2007.12191}{{\ttfamily 2007.12191}}.

\bibitem{Henn:2020lye}
J.~Henn, B.~Mistlberger, V.~A. Smirnov and P.~Wasser, \emph{{Constructing d-log
  integrands and computing master integrals for three-loop four-particle
  scattering}}, \href{https://doi.org/10.1007/JHEP04(2020)167}{\emph{JHEP}
  {\bfseries 04} (2020) 167}
  [\href{https://arxiv.org/abs/2002.09492}{{\ttfamily 2002.09492}}].

\bibitem{He:2020uxy}
S.~He, Z.~Li, Y.~Tang and Q.~Yang, \emph{{The Wilson-loop $d$ log
  representation for Feynman integrals}},
  \href{https://doi.org/10.1007/JHEP05(2021)052}{\emph{JHEP} {\bfseries 05}
  (2021) 052} [\href{https://arxiv.org/abs/2012.13094}{{\ttfamily
  2012.13094}}].

\bibitem{He:2018okq}
S.~He and C.~Zhang, \emph{{Notes on Scattering Amplitudes as Differential
  Forms}}, \href{https://doi.org/10.1007/JHEP10(2018)054}{\emph{JHEP}
  {\bfseries 10} (2018) 054}
  [\href{https://arxiv.org/abs/1807.11051}{{\ttfamily 1807.11051}}].

\bibitem{Arkani-Hamed:2017tmz}
N.~Arkani-Hamed, Y.~Bai and T.~Lam, \emph{{Positive Geometries and Canonical
  Forms}}, \href{https://doi.org/10.1007/JHEP11(2017)039}{\emph{JHEP}
  {\bfseries 11} (2017) 039}
  [\href{https://arxiv.org/abs/1703.04541}{{\ttfamily 1703.04541}}].

\bibitem{Arkani-Hamed:2014dca}
N.~Arkani-Hamed, A.~Hodges and J.~Trnka, \emph{{Positive Amplitudes In The
  Amplituhedron}}, \href{https://doi.org/10.1007/JHEP08(2015)030}{\emph{JHEP}
  {\bfseries 08} (2015) 030} [\href{https://arxiv.org/abs/1412.8478}{{\ttfamily
  1412.8478}}].

\bibitem{Remiddi:1999ew}
E.~Remiddi and J.~A.~M. Vermaseren, \emph{{Harmonic polylogarithms}},
  \href{https://doi.org/10.1142/S0217751X00000367}{\emph{Int. J. Mod. Phys. A}
  {\bfseries 15} (2000) 725}
  [\href{https://arxiv.org/abs/hep-ph/9905237}{{\ttfamily hep-ph/9905237}}].

\bibitem{Dixon:2016apl}
L.~J. Dixon, M.~von Hippel, A.~J. McLeod and J.~Trnka, \emph{{Multi-loop
  positivity of the planar $ \mathcal{N} $ = 4 SYM six-point amplitude}},
  \href{https://doi.org/10.1007/JHEP02(2017)112}{\emph{JHEP} {\bfseries 02}
  (2017) 112} [\href{https://arxiv.org/abs/1611.08325}{{\ttfamily
  1611.08325}}].

\bibitem{Ferro:2015grk}
L.~Ferro, T.~Lukowski, A.~Orta and M.~Parisi, \emph{{Towards the Amplituhedron
  Volume}}, \href{https://doi.org/10.1007/JHEP03(2016)014}{\emph{JHEP}
  {\bfseries 03} (2016) 014}
  [\href{https://arxiv.org/abs/1512.04954}{{\ttfamily 1512.04954}}].

\bibitem{Erdogan:2011yc}
O.~Erdo\u{g}an and G.~Sterman, \emph{{Gauge Theory Webs and Surfaces}},
  \href{https://doi.org/10.1103/PhysRevD.91.016003}{\emph{Phys. Rev. D}
  {\bfseries 91} (2015) 016003}
  [\href{https://arxiv.org/abs/1112.4564}{{\ttfamily 1112.4564}}].

\bibitem{progress2}
D.~Chicherin and J.~Henn, \emph{{Symmetry properties of Wilson loops with a
  Langrangian insertion, to appear}}, .

\bibitem{progress3}
N.~Arkani-Hamed, D.~Chicherin, J.~Henn and J.~Trnka, \emph{{In progress}}, .

\bibitem{progress4}
D.~Chicherin and J.~Henn, \emph{{To appear}}, .

\end{thebibliography}\endgroup

\end{document}